\documentclass[prd,aps,superscriptaddress,preprintnumbers,notitlepage,nofootinbib]{revtex4}
\usepackage{epsf}
\usepackage{epsfig}
\usepackage[utf8]{inputenc}
\usepackage{graphics}
\usepackage{graphicx}
\usepackage{amsmath}
\usepackage{amssymb}
\usepackage{bbold}
\usepackage{hyperref} 
\usepackage{xcolor}
\usepackage{adjustbox}
\usepackage{slashed}
\usepackage{multirow}
\usepackage{hhline}
\usepackage{caption2} 
\usepackage[T1]{fontenc}
\usepackage[utf8]{inputenc}
\usepackage{tabularx,ragged2e,booktabs,caption2}
\newcolumntype{C}[1]{>{\Centering}m{#1}}

\newcommand{\cmmnt}[1]{}

\newcommand{\beq}{\begin {equation}}  
\newcommand{\eeq}{\end   {equation}} 
\newcommand{\bea}{\begin {eqnarray}} 
\newcommand{\eea}{\end   {eqnarray}}  
\newcommand{\baa}{\begin {array}   } 
\newcommand{\eaa}{\end   {array}   }     
\newcommand{\bit}{\begin {itemize} }
\newcommand{\eit}{\end   {itemize} }
\newcommand{\be }{\begin {equation}} 
\newcommand{\ee }{\end   {equation}}

\newcommand{\longeq}[1]{\overset{\mathrm{#1}}{=\joinrel=\joinrel=\joinrel=}}

\newcommand{\gsim}{\buildrel > \over {_\sim}}

\begin{document}
\raggedbottom

\preprint{ACFI-T18-18}

\title{ \large {\color{black} Type-II Seesaw Scalar Triplet Model at a 100\,TeV $pp$ Collider:\\ Discovery and Higgs Portal Coupling Determination}}

\author {Yong Du\footnote{Email: yongdu@umass.edu}}
\affiliation{Amherst Center for Fundamental Interactions, Physics Department, University of Massachusetts Amherst, Amherst, MA 01003 USA}
\author{Aaron Dunbrack\footnote{Email: aaron.dunbrack@stonybrook.edu}}
\affiliation{Department of Physics and Astronomy, Stony Brook University, Stony Brook, NY 11790, USA}
\author{Michael J. Ramsey-Musolf\footnote{Email: mjrm@physics.umass.edu}}
\affiliation{Amherst Center for Fundamental Interactions, Physics Department, University of Massachusetts Amherst, Amherst, MA 01003, USA}
\affiliation{Kellogg Radiation Laboratory, California Institute of Technology, Pasadena, CA 91125 USA}
\author {Jiang-Hao Yu\footnote{Email: jhyu@itp.ac.cn}}
\affiliation{Amherst Center for Fundamental Interactions, Physics Department, University of Massachusetts Amherst, Amherst, MA 01003, USA}
\affiliation{CAS Key Laboratory of Theoretical Physics, Institute of Theoretical Physics, Chinese Academy of Sciences, Beijing 100190, P. R. China}
\affiliation{School of Physical Sciences, University of Chinese Academy of Sciences, No.19A Yuquan Road, Beijing 100049, P.R. China}

\date{\today}

\begin{abstract}
{\color{black} We investigate the collider phenomenology of the scalar triplet  particles in the Type-II seesaw model at a 100\,TeV $pp$ collider. Depending on triplet vacuum expectation value $v_\Delta$, the dominant discovery channels could be $H^{++} H^{--}$ and $H^{\pm\pm}H^\mp$. We find the $H^{\pm\pm}H^\mp\to W^\pm W^\pm h W^\mp/\ell^\pm \ell^\pm h W^\mp$ channels are promising for both model discovery at relatively large $v_\Delta$ and determination of the Higgs portal couplings $\lambda_4$ and $\lambda_5$. We also find that these two channels are complementary to indirect determination of $\lambda_4$ from future precise measurements on $h\to\gamma\gamma$ decay rate. Together with pair production of the doubly-charged Higgs subsequently decaying into same-sign di-leptons, the $H^{\pm\pm}H^\mp$ channels have the potential to cover a significant portion of the parameter space of the Type-II seesaw complex scalar triplet model. }
\end{abstract}
\maketitle

\section{Introduction}
\label{sec:intro}
Explaining the origin of neutrino masses is a key open problem in particle physics. The significant difference in magnitudes between the masses of the charged and neutral leptons suggests that the dynamics responsible for the observed light neutrino masses, generically denoted here as $m_\nu$, are different than those of the Standard Model (SM) Higgs mechanism. Among the most widely considered is the seesaw mechanism. Its theoretical attractiveness rests in part on the idea that the suppression of $m_\nu$ results from a ratio of physical scales rather than the appearance of tiny dimensionless Yukawa couplings in the Lagrangian. Several variants of the seesaw mechanism have been studied over the years, with perhaps the types I, II, and III models\,\cite{Minkowski:1977sc,Ramond:1979py,GellMann:1980vs,Yanagida:1979as,Mohapatra:1979ia,Schechter:1980gr,Schechter:1981cv,Konetschny:1977bn,Cheng:1980qt,Lazarides:1980nt,Magg:1980ut,Foot:1988aq,Witten:1985bz,Mohapatra:1986aw,Mohapatra:1986bd,Val86,Barr:2003nn,Mohapatra:1980yp} the most thoroughly considered. 

It remains to be seen which, if any of these scenarios, is realized in nature. In the conventional type-I model\,\cite{Minkowski:1977sc,Yanagida:1979as,GellMann:1980vs,Mohapatra:1979ia,Ramond:1979py}, the scale of the heavy, right-handed (RH) Majorana neutrinos, $M_N$, lies well above the energies directly accessible in the laboratory, making a direct probe of this scenario infeasible. Theorists have considered lower scale variants with $M_N$ at the TeV scale or below, a possibility that allows for more direct experimental tests, including the observation of the RH neutrinos in high energy collider searches or beam dump experiments.  In this case, the scale of the relevant Yukawa couplings need not be too different from those of the charged leptons. 

In this study, we consider the type-II scenario\,\cite{Konetschny:1977bn,Magg:1980ut,Schechter:1980gr,Cheng:1980qt,Lazarides:1980nt,Mohapatra:1980yp}, wherein the scale of $m_\nu$ is governed by the product of Yukawa couplings $h_\nu$ and the vacuum expectation value (vev) $v_\Delta$ of the neutral component of a complex triplet $\Delta$ that transforms as (1,3,2) under the SM gauge groups. Constraints from electroweak precision tests require that $v_\Delta$ be no larger than a few GeV, though it could be considerably smaller. Consequently, the Yukawa couplings $h_\nu$ may be as large as $\mathcal{O}(1)$. As in the case of low scale type I models, the mass scale of the $\Delta$ may lie at the TeV scale or below without introducing new naturalness issues beyond those already present in the SM Higgs sector. It is, then, interesting to ask under what conditions one may discover the new degrees of freedom essential to the type II scenario and to what extent its interactions determined. 

In this study, we focus on these questions, paying particular attention to the $\Delta$ interactions in the scalar sector. With the discovery of the SM-like Higgs boson\cite{Aad:2012tfa,Chatrchyan:2012xdj}, it is timely to consider the scalar sector potential in more detail. In general, the presence of additional scalar degrees of freedom that interact with the Higgs doublet $\Phi$ may enhance stability of the potential, as has been noted in the case of the $\Delta$ in Refs.{{\,\cite{Chao:2012mx,Haba:2016zbu,Chun:2012jw,Bonilla:2015eha}}}. In addition, $\Delta$-$\Phi$ interactions may allow for a strong first order electroweak phase transition (SFOEWPT), thereby providing the needed conditions for generation of the cosmic baryon asymmetry through electroweak baryogenesis\footnote{The electroweak symmetry-breaking transition in the SM is of a crossover type\cite{Aoki:1999fi,Rummukainen:1998as,Csikor:1998eu,Laine:1998jb,Gurtler:1997hr,Kajantie:1995dw}.}. In both cases, knowledge of the Higgs portal couplings $\lambda_4$ and $\lambda_5$ (defined below) is essential. This study represents our first effort to provide a roadmap for discovery of the $\Delta$ and determination of its scalar sector couplings, building on the results of earlier studies that focus on the collider phenomenology of the $\Delta$ at LEP and the LHC\footnote{Note that the triplet $\Delta$ also exists in the left-right symmetric model (LRSM), see Ref.\,\cite{Pati:1974yy,Mohapatra:1974gc,Senjanovic:1975rk,Maiezza:2016ybz,Cao:2012ng,Hsieh:2010zr,Dev:2016dja,Jung:2008pz,Barenboim:2000nn,Huitu:1997vh,Cvetic:1991kh,Grifols:1988ag} and reference therein for related works.} as well as its impact contributions on the SM Higgs  di-photon decay rate\,\cite{Quintero:2012jy, Kanemura:2013vxa, Chun:2013vma, Yagyu:2014aaa, Kanemura:2014goa, Muhlleitner:2003me, Chun:2012zu, Kanemura:2014ipa, Chun:2013fya, Chiang:2012dk, Ghosh:2017pxl, Mitra:2016wpr, Haba:2016zbu, Chen:2013dh, Dev:2013ff, Akeroyd:2011zza, Yue:2010zu, Akeroyd:2005gt, Akeroyd:2012ms, Akeroyd:2011ir, Ong:2011fx, Akeroyd:2009hb, Huitu:2017cpc, Biswas:2017tnw, Das:2016bir, Kikuchi:2013kya, Aoki:2012jj, Aoki:2012yt, Arhrib:2012vp, Kanemura:2012rs, Aoki:2011pz, Rodejohann:2010bv, Chen:2010uc, Fukuyama:2009xk, Nishiura:2009yd, Nishiura:2009jn, Akeroyd:2009nu, Petcov:2009zr, Godbole:1994np, Gogoladze:2008gf, Akeroyd:2007zv, Garayoa:2007fw, Ma:2006mr, deS.Pires:2005au, Kakizaki:2003jk, Alanakian:1997ii, CoarasaPerez:1995wa, Arhrib:2014nya, Arbabifar:2012bd, Han:2015hba, Akeroyd:2010je, Akeroyd:2012nd, Akeroyd:2012rg, delAguila:2013mia, Cao:2014nta, Shen:2014rpa, Han:2015sca, Bi:2015fra, Shen:2015ora, Shen:2015bna, Shen:2015pih, Cao:2016hvg, Melfo:2011nx, Xing:2015wzz, Yagyu:2011hh, kang:2014jia, Bonilla:2015jdf, Perez:2008ha, Barger:1982cy, Gunion:1989in, Han:2007bk, Huitu:1996su, Dion:1998pw, Dev:2017ouk, Sui:2017qra, Agrawal:2018pci, Cai:2017mow, Li:2018jns}.

{\color{black} Searches for the complex triplet scalars -- including doubly charged $H^{\pm\pm}$, singly charged $H^\pm$, and neutral Higgs particles $H$ and $A$ -- have been carried out at the LHC. 
A smoking gun for the CTHM has conventionally been the presence of the  $H^{\pm\pm}$ decaying into a same-sign di-lepton final state and has been intensively investigated by the ATLAS and CMS collaboration\,\cite{Aad:2012cg, Aaltonen:2011rta, ATLAS:2012mn, ATLAS:2012hi, Aad:2014hja, ATLAS:2014kca, Aad:2015oga, Sirunyan:2017ret, Aaboud:2017qph}. For other channels related to CTHM discovery, there are also many studies have been done at the LHC, see Appendix\,\ref{app:expcon} for a detailed summary.}

In what follows, we explore the potential for both discovery of the $\Delta$ and determination of its scalar sector couplings at a prospective future 100\,TeV proton-proton collider, such as the Super Proton Proton Collider (SppC) under consideration in China and the CERN Future Circular Collider (FCC-hh). Given the higher center of mass energy and prospective integrated luminosity, a 100 TeV pp collider will provide coverage for a considerably larger portion of model parameter space than is feasible with the Large Hadron Collider (LHC). In this context, there exist two distinct mass spectra for the $\Delta$ (governed by the model parameters), as discussed in detail in Sec.\,\ref{subsec:input}. By working in the \lq\lq normal mass hierarchy", where $m_h\le m_{H/A}\approx m_\Delta\le m_{H^\pm}\le m_{H^{\pm\pm}}$ with $m_\Delta$ the mass scale of the model, we find that:
\begin{itemize}
\item The future 100\,TeV $pp$ collider with an integrated luminosity of $30\,\rm ab^{-1}$ can discover the triplet model up to $m_\Delta\lesssim4.5$\,TeV for $v_\Delta\le10^{-4}$\,GeV and $m_\Delta\lesssim1$\,TeV for $v_\Delta\gtrsim10^{-4}$\,GeV. Our result is shown in Fig.\,\ref{bdtdis}.
\item Upon discovery, the Higgs portal parameter $\lambda_5$ can be determined from the mass spectrum of $H^{\pm\pm}$ and $H^\pm$ for $m_\Delta\lesssim1$\,TeV, while $\lambda_4$ is determined by the branching ratio (BR) of $H^\pm\to hW^\pm$. The $h\to\gamma\gamma$ decay rate also provides a complementary probes of the related parameter space, as we discuss below in relation to Fig.\,\ref{haa}.
\end{itemize}

In our analysis leading to these conclusions, we first study the same-sign di-lepton decay channel for $pp\to H^{++}H^{--}$, whose production cross section at $\sqrt{s}=100$ TeV is the largest among all triplet scalar channels. We find that this channel is only suitable for the triplet model discovery at small $v_\Delta$, where the corresponding Yukawa couplings $h_\nu$ that govern the $H^{\pm\pm}$ decay rate can be relatively large and still consistent with the scale of $m_\nu$. For relatively large $v_\Delta$,  we find that there exist other promising discovery channels, particularly $pp\to H^{\pm\pm}H^\mp$ with $H^{\pm\pm} \to W^\pm W^\pm/\ell^\pm\ell^\pm$ and $H^\mp \to h W^\mp$. Considering these channels at both small and large $v_\Delta$ will allow for discovery over the entire range of $v_\Delta$ parameter space for triplet mass up to $\sim 4.5$ TeV ($\sim 1$ TeV) as can be seen from Fig.\,\ref{bdtdis}. 

Assuming discovery, the next question we ask is: How does one determine the Higgs portal couplings? We find that measurement of the rate for $pp\to H^{\pm\pm}H^\mp$ with  $H^{\pm\pm}H^\mp\to W^\pm W^\pm h W^\mp/\ell^\pm\ell^\pm h W^\mp$  $W^\mp$ decaying leptonically will be advantageous. These two channels probe a signifiant portion of the relevant entire parameter space as can be seen from Fig.\,\ref{haa}. The presence of the charged triplet scalars with masses and couplings in the same range could also lead to an observable deviation of  the $h\to\gamma\gamma$ signal strength compared to Standard Model expectations. For triplet scalar masses below roughly one TeV, the prospective future collider (circular $e^+e^-$ and $pp$) measurements of the Higgs di-photon decay rate could yield significant constraints on the values or the Higgs portal coupling needed for discovery of the $H^{\pm\pm}H^\mp\to W^\pm W^\pm h W^\mp/\ell^\pm\ell^\pm h W^\mp$ modes. For heavier triplet masses, the discovery potential for these modes would be relatively unconstrained.   

{{The structure of this paper is as follows: In Sec.\,\ref{sec:model}, we set up the complex triplet Higgs model and discuss its key features and various model constraints. We also discuss neutrino mass generation from the type-II seesaw mechanism as well as experimental constraints on the neutrino masses. In Sec.\,\ref{paramdeter}, we focus on how to determine the model parameters from future collider measurements, and in Sec.\,\ref{sec:decayprod}, we study production cross sections and decay patterns of the triplet Higgs particles. Sec.\,\ref{sec:modeldis} presents our result for model discovery at the 100\,TeV collider, and Sec.\,\ref{sec:lam45} discusses a strategy for the determination of $\lambda_4$. Sec.\,\ref{sec:conclusion} is our conclusion, and we summarize the details in the Appendices.}}

\section{The Complex Triplet Higgs Model}
In this section, we will discuss setup of the triplet model and various model constraints. We will also discuss key features of the model in Sec.\,\ref{subsec:mdlkey} and close this section by illustrating how neutrino masses are generated through a Type-II seesaw mechanism and by discussing current constraints on the neutrino masses.
\label{sec:model}
\subsection{Model setup}\label{model:setup}
The type-II seesaw model contains the SM Higgs doublet $\Phi$ with hypercharge $Y_\Phi=1$ and the complex triplet Higgs field $\Delta$ with hypercharge $Y_\Delta=2$\,\cite{Konetschny:1977bn} written in a matrix form \,\cite{Mohapatra:1979ia, Cheng:1980qt, Lazarides:1980nt, Schechter:1980gr}
\begin{eqnarray}\label{basis}
\Phi=\left[
\begin{array}{c}
\varphi^+\\
\frac{1}{\sqrt{2}}(\varphi+v_\Phi+i\chi)
\end{array}\right], \quad
\Delta =
\left[
\begin{array}{cc}
\frac{\Delta^+}{\sqrt{2}} & H^{++}\\
\frac{1}{\sqrt{2}}(\delta+v_\Delta+i\eta) & -\frac{\Delta^+}{\sqrt{2}} 
\end{array}\right],
\end{eqnarray}
where $v_\Phi$ denotes the doublet vev satisfying $\sqrt{v_\Phi^2 + v_\Delta^2}\equiv v\approx246$\,GeV, which is the scale of electroweak spontaneous symmetry breaking (EWSB). And as will be discussed below, $v_\Delta$ will be strongly constrained by the $\rho$ parameter. This scalar extension extension of the SM is also know as the complex triplet Higgs model (CTHM).

The kinetic Lagrangian is
\begin{align}
\mathcal{L}_{\rm{kin}}&=(D_\mu \Phi)^\dagger (D^\mu \Phi)+\rm{Tr}[(D_\mu \Delta)^\dagger (D^\mu \Delta)],
\end{align}
with the covariant derivatives 
\begin{equation}
D_\mu \Phi=\left(\partial_\mu+i\frac{g}{2}\tau^aW_\mu^a+i\frac{g'Y_{\Phi}}{2}B_\mu\right)\Phi, \quad
D_\mu \Delta=\partial_\mu \Delta+i\frac{g}{2}[\tau^aW_\mu^a,\Delta]+i\frac{g'Y_{\Delta}}{2}B_\mu\Delta,
\end{equation}
where $g'$ and $g$ are the U(1)$_Y$ and SU(2)$_L$ gauge couplings, respectively. The second term in $D_\mu \Delta$ introduces new interactions between the electroweak gauge bosons and the triplet, which contributes to the masses of the former  when the triplet gets a nonzero vev.

We write the general CTHM potential as
\begin{align}
V(\Phi,\Delta)&= - m^2\Phi^\dagger\Phi + M^2\rm{Tr}(\Delta^\dagger\Delta)+\left[\mu \Phi^Ti\tau_2\Delta^\dagger \Phi+\rm{h.c.}\right]+\lambda_1(\Phi^\dagger\Phi)^2 \nonumber\\
&~~~~+\lambda_2\left[\rm{Tr}(\Delta^\dagger\Delta)\right]^2 +\lambda_3\rm{Tr}[ \Delta^\dagger\Delta \Delta^\dagger\Delta]
+\lambda_4(\Phi^\dagger\Phi)\rm{Tr}(\Delta^\dagger\Delta)+\lambda_5\Phi^\dagger\Delta\Delta^\dagger\Phi,
\end{align}
where $m$ and $M$ are the mass parameters and $\lambda_i$ (i=1,$\ldots$, 5) are the dimensionless quartic scalar couplings, which are all real due to hermiticity of the Lagrangian. The $\mu$ parameter, however, is in general complex and, thus, a possible source of CP violation (CPV). But as discussed in Ref.\,\cite{Arhrib:2011uy,Dey:2008jm}, the CPV phase from $\mu$ is in fact unphysical and can always be absorbed by a redefinition of the triplet field.

After EWSB, the minimization conditions 
\be
\frac{\partial V}{\partial \Phi_j} = 0, \qquad \frac{\partial V}{\partial \Delta_j} = 0
\ee
imply that 
\bea
m^2 &=& \lambda_1 v_\Phi^2 + \frac{\lambda_{45}v_\Delta^2}{2} - \sqrt2 \mu v_\Delta, \\
M^2 &=&  \frac{ \mu v_\Phi^2}{\sqrt{2}v_\Delta} - \lambda_{23} v_\Delta^2 - \frac{\lambda_{45}v_\Phi^2}{2},
\eea
with 
\begin{align}
\lambda_{ij}\equiv\lambda_i+\lambda_j.
\end{align}
We will use the same notation below. 

The scalar states are, in general, mixtures of the field components that carry the same electric charge: ($\varphi$, $\delta$, $\chi$, $\eta$); ($\varphi^{\pm}$, $\Delta^\pm$); and $H^{\pm\pm}$, which is already in its mass eigenstate. The absence of a CPV phase in the potential implies that the real and imaginary parts of the neutral doublet and triplet fields cannot mix with each other. To diagonalize the corresponding mass matrices, we introduce the following matrices to rotate them into their mass eigenstates $G^0$, $A$, $h$, $H$, $G^\pm$ and $H^\pm$:
\begin{eqnarray}
\left(\begin{array}{c}\varphi\\\delta\end{array}\right)&=&\left(\begin{array}{cc}\cos \alpha & -\sin\alpha \\\sin\alpha   & \cos\alpha\end{array}\right)\left(\begin{array}{c}h\\H\end{array}\right),\label{mhpp}
\quad 
\left(\begin{array}{c}\varphi^\pm\\\Delta^\pm\end{array}\right)=\left(\begin{array}{cc}\cos \beta_\pm & -\sin\beta_\pm \\\sin\beta_\pm   & \cos\beta_\pm\end{array}\right)\left(\begin{array}{c} G^\pm\\H^\pm\end{array}\right),\nonumber\\
\left(\begin{array}{c}\chi\\\eta\end{array}\right)&=&\left(\begin{array}{cc}\cos \beta_0 & -\sin\beta_0 \\\sin\beta_0   & \cos\beta_0\end{array}\right)\left(\begin{array}{c} G^0\\ A\end{array}\right),
\end{eqnarray}
with the mixing angles given by
\begin{eqnarray}
\cos\beta_\pm&=&\frac{v_\Phi}{\sqrt{v_\Phi^2+2v_\Delta^2}},\quad \sin\beta_\pm=\frac{\sqrt{2}v_\Delta}{\sqrt{v_\Phi^2+2v_\Delta^2}},\quad \tan\beta_\pm=\frac{\sqrt{2}v_\Delta}{v_\Phi},\label{betapm} \\
\cos\beta_0&=&\frac{v_\Phi}{\sqrt{v_\Phi^2+4v_\Delta^2}},\quad \sin\beta_0=\frac{2v_\Delta}{\sqrt{v_\Phi^2+4v_\Delta^2}},\quad \tan\beta_0=\frac{2v_\Delta}{v_\Phi}, \label{beta0}\\
\tan2\alpha &=&\frac{v_\Delta}{v_\Phi}\cdot\frac{2v_\Phi  \lambda_{45}-\frac{2 \sqrt2 \mu v_\Phi}{v_\Delta} }{2v_\Phi\lambda_1-\frac{v_\Phi\mu}{\sqrt{2}v_\Delta}-\frac{2 v_\Delta^2 \lambda_{23}}{v_\Phi}}. \label{tan2a}
\end{eqnarray}  
Here $G^0$ and $G^\pm$ are the would-be Goldstone bosons that become the longitudinal components of the $Z$ and $W^\pm$. Among the remaining scalars,  $A$ is the pseudoscalar; $h$ is the CP-even Higgs, which is recognized as the SM Higgs particle; $H$ is the other CP-even Higgs particle with a heavier mass compared with $h$; and $H^\pm$ and $H^{\pm\pm}$ are the singly- and doubly-charged Higgs particles respectively. 

It is useful to express the corresponding mass eigenvalues in terms of the parameters in the potential, vevs, and mixing angles:
\begin{eqnarray}
&&m_{H^{\pm\pm}}^2=m_\Delta^2-v_\Delta^2\lambda_3-\frac{\lambda_5}{2}v_\Phi^2,\label{mhpp}\\
&&m_{H^\pm}^2=\left(m_\Delta^2-\frac{\lambda_5}{4}v_\Phi^2\right)\left(1+\frac{2v_\Delta^2}{v_\Phi^2}\right),\label{mhp}\\
&&m_A^2 =m_\Delta^2\left(1+\frac{4v_\Delta^2}{v_\Phi^2}\right), \label{mA}\\
&&m_h^2= 2v_\Phi^2\lambda_1\cos^2\alpha+\left( m_\Delta^2+2\lambda_{23}v_\Delta^2\right) \sin^2\alpha + \left( \lambda_{45} v_\Phi v_\Delta - \frac{2v_\Delta}{v_\Phi}m_\Delta^2\right) \sin2\alpha,\label{mh}\\
&&m_H^2=2v_\Phi^2\lambda_1\sin^2\alpha+ \left( m_\Delta^2+2\lambda_{23}v_\Delta^2 \right) \cos^2\alpha -  \left( \lambda_{45} v_\Phi v_\Delta - \frac{2v_\Delta}{v_\Phi}m_\Delta^2 \right) \sin2\alpha,\label{mH}
\end{eqnarray}
where 
\begin{align}
m_\Delta^2\equiv \frac{v_\Phi^2\mu}{\sqrt{2}v_\Delta}.
\end{align}
As will be discussed below, experimental constraints on the $\rho$ parameter require $v_\Delta\ll v_{\Phi}$, which in turn results in a small $\sin\alpha$ in general as can be seen from Eq.\eqref{tan2a}. Taking the small $v_\Delta$ and $\sin\alpha$ limit, we see that, from the mass expressions above, $m_\Delta$ basically determines the mass scale of the CTHM. We will discuss this in more detail in Sec.\,\ref{subsec:input}.

Since we seek to gain information about the potential parameters from measurements of the scalar boson properties, it is also useful to express the potential parameters in terms of the masses, vevs, and mixing angles:

\begin{eqnarray}
\mu&=&\frac{\sqrt{2}v_\Delta^2}{v_\Phi^2}m_\Delta^2 =\frac{\sqrt{2}v_\Delta}{v_\Phi^2+4v_\Delta^2}m_A^2,\\
\lambda_1 & = &\frac{1}{2v_\Phi^2}(m_h^2\cos^2\alpha+m_H^2\sin^2\alpha),\\
\lambda_2 & =& \frac{1}{2v_\Delta^2}\left[2m_{H^{\pm\pm}}^2+v_\Phi^2\left(\frac{m_A^2}{v_\Phi^2+4v_\Delta^2}-\frac{4m_{H^{\pm}}^2}{v_\Phi^2+2v_\Delta^2}\right)+m_H^2\cos^2\alpha+m_h^2\sin^2\alpha\right],\label{lam2}\\
\lambda_3 & =& \frac{v_\Phi^2}{v_\Delta^2}\left(\frac{2m_{H^{\pm}}^2}{v_\Phi^2+2v_\Delta^2}-\frac{m_{H^{\pm\pm}}^2}{v_\Phi^2}-\frac{m_A^2}{v_\Phi^2+4v_\Delta^2}\right), \label{lam3}
\end{eqnarray}
\begin{eqnarray}
\lambda_4 & =& \frac{4m_{H^{\pm}}^2}{v_\Phi^2+2v_\Delta^2}-\frac{2m_A^2}{v_\Phi^2+4v_\Delta^2}+\frac{m_h^2-m_H^2}{2v_\Phi v_\Delta}\sin2\alpha,  \\
\lambda_5 & =& 4\left(\frac{m_A^2}{v_\Phi^2+4v_\Delta^2}-\frac{m_{H^{\pm}}^2}{v_\Phi^2+2v_\Delta^2}\right).
\end{eqnarray}
From Eq.\,\eqref{lam2}-\eqref{lam3}, we observe that $v_\Delta$ appears in the denominators. Thus, if we take the physical masses as our model input, then in the small $v_\Delta$ limit, we may need to fine tune the masses in order to maintain perturbative values for the couplings $\lambda_{2,3}$. {{Consequently, we will use $\lambda_{2,3}$ as independent input parameters for simulation.}}
\vspace{-0.5cm}
\subsection{Model constraints}
\label{model:const}
\subsubsection{Constraint on $v_\Delta$ from the $\rho$ parameter}\label{vdcons}
After the EWSB, the electroweak gauge boson masses receive contributions from both the doublet and triplet vevs. At tree level, one has
\begin{eqnarray}
m_W^2 = \frac{g^2}{4}(v_\Phi^2+2v_\Delta^2), \quad m_Z^2 =\frac{g^2}{4\cos^2\theta_W}(v_\Phi^2+4v_\Delta^2),
\end{eqnarray}
with $\theta_W$ the weak mixing angle. The ratio between $m_W$ and $m_Z$ is strongly constrained through the $\rho$ parameter which is defined as
\begin{eqnarray}
\rho \equiv \frac{m_W^2}{m_Z^2\cos^2\theta_W}\longeq{CTHM}\frac{1+\frac{2v_\Delta^2}{v_\Phi^2}}{1+\frac{4v_\Delta^2}{v_\Phi^2}}.\label{mwz}
\end{eqnarray}
The SM predicts $\rho=1$ exactly at tree level, which has been confirmed experimentally to high precision. One therefore expects $v_\Delta$ to be much smaller than $v_\Phi$ from Eq.\,\eqref{mwz} in the CTHM, and in small $v_\Delta$ limit,
\begin{eqnarray}
\rho \simeq 1 - \frac{2v_\Delta^2}{v_\Phi^2}.
\end{eqnarray}
Electroweak precision tests\cite{Agashe:2014kda} gives the $1\sigma$ result $\rho=1.0006\pm0.0009$, which leads to 
\bea
0\le v_\Delta \lesssim 3.0 {\rm ~\,GeV}
\eea
and thus $v_\Delta\ll v_\Phi$.

\subsubsection{Constraint from stability, perturbative unitarity, and perturbativity}\label{secallconst}
\begin{figure}[thb!]
\captionstyle{flushleft}
\begin{tabular}{cc}
\includegraphics[scale=0.32]{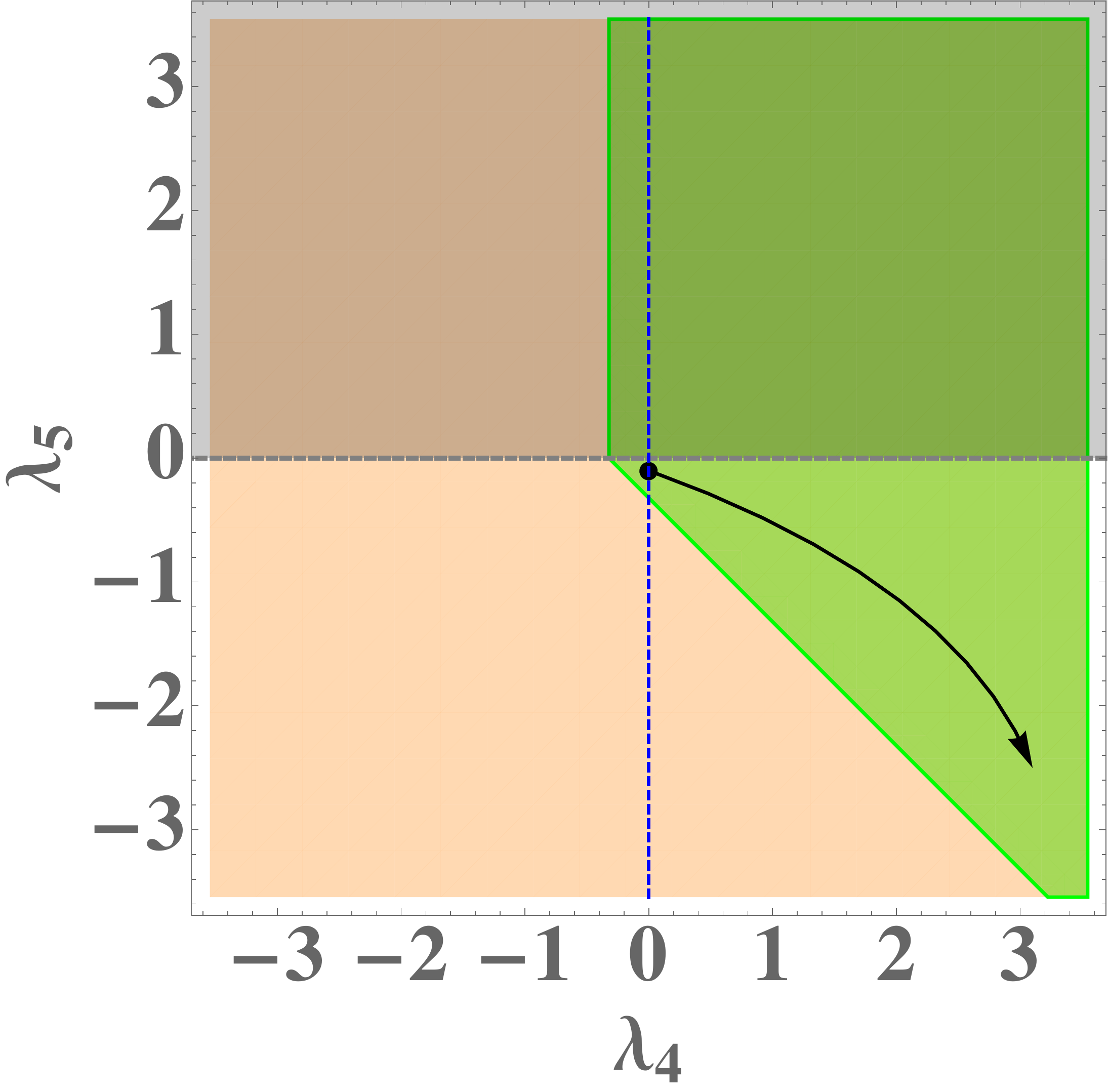} ~& ~ \includegraphics[width=75mm,height=75mm]{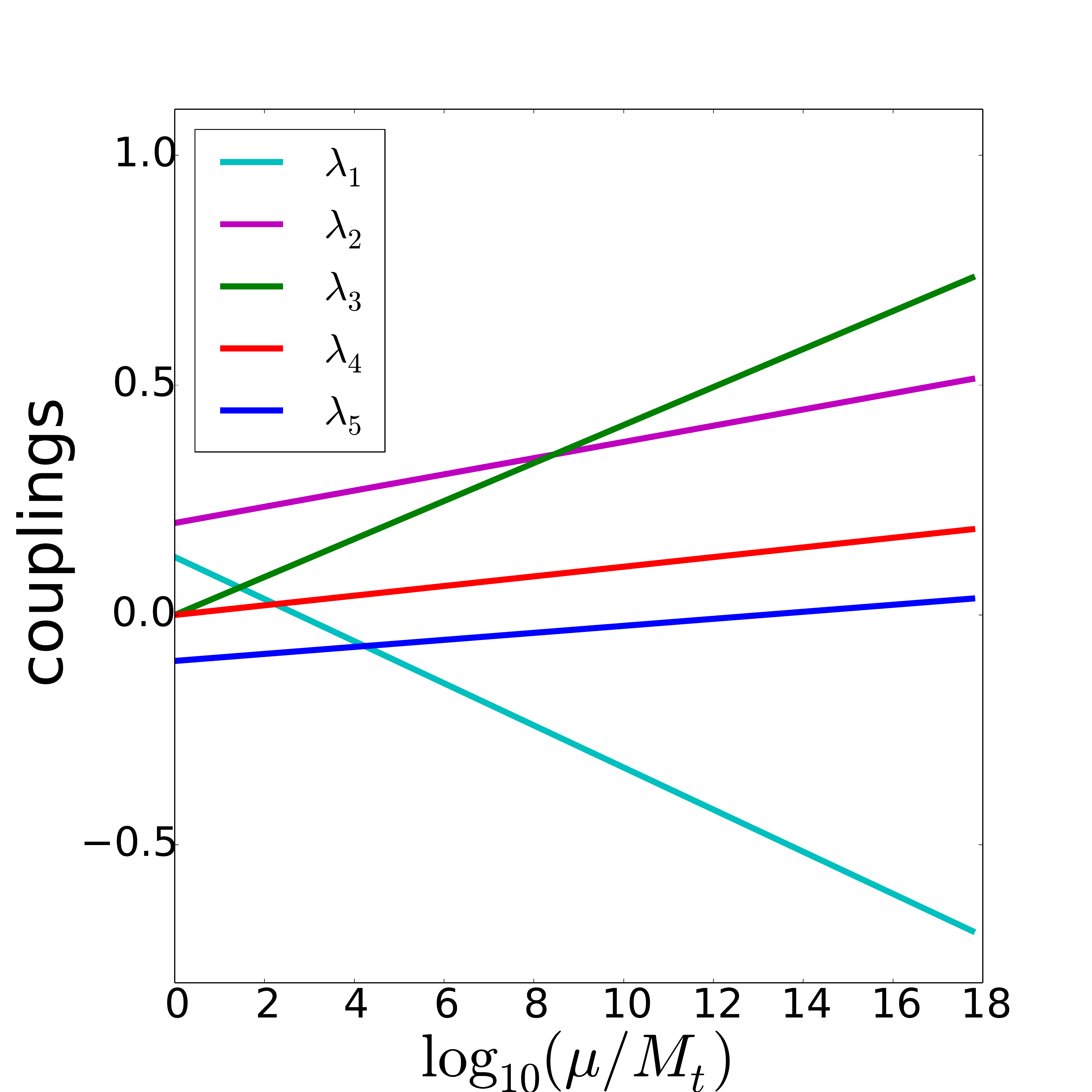}
\end{tabular}
\caption{{{Left panel: Tree-level vacuum stability (green region) and perturbative unitarity (orange region) constraints on the $\lambda_4$-$\lambda_5$ plane with $\lambda_2=0.2$ and $\lambda_3=0$. Right panel: One-loop running of the Higgs quartic couplings at $\lambda_2=0.2$, $\lambda_3=0$, $\lambda_4=0$ and $\lambda_5=-0.1$ with $M_t=173.1$\,GeV being our input scale. The black arrow in the left figure corresponds to regions in which vacuum stability is stable up to a higher scale.}}}\label{stability}
\end{figure}
    
{{Constraints from vacuum stability, perturbative unitarity, and perturbativity have  been studied in\,\cite{Arhrib:2011uy,Haba:2016zbu,Chao:2006ye,Schmidt:2007nq, Bonilla:2015eha,Machacek:1983tz,Machacek:1983fi,Machacek:1984zw,Ford:1992pn,Arason:1991ic,Barger:1992ac,Luo:2002ey,Chao:2012mx,Chun:2012jw} and are summarized below in our notation:}}
\begin{itemize}
\item Vacuum stability (VS)\footnote{Here and below, ``\&'' means the logical conjunction ``and''.}:
\begin{align}
\lambda_1\ge0 ~\&~ \lambda_2+\text{Min}\left\{\lambda_3,\frac{\lambda_3}{2}\right\}\ge0  ~\&~ \nonumber\\
\lambda_4+\text{Min}\left\{0,\lambda_5\right\}+2\text{Min}\left\{\sqrt{\lambda_1\lambda_{23}},\sqrt{\lambda_1(\lambda_2+\frac{\lambda_3}{2})}\right\}\ge0.
\end{align}
\item Perturbative unitarity (PU):
\begin{align}
|\lambda_{45}|\le\kappa\pi ~\&~ |\lambda_4|\le\kappa\pi ~\&~ |2\lambda_4+3\lambda_5|\le2\kappa\pi~\&~2|\lambda_1|\le\kappa\pi~\&~2|\lambda_2|\le\kappa\pi  ~\&~ \nonumber\\
2|\lambda_{23}|\le\kappa\pi~\&~|\lambda_4-\frac{\lambda_5}{2}|\le\kappa\pi~\&~|2\lambda_2-\lambda_3|\le\kappa\pi  ~\&~ \nonumber\\
|\lambda_{12}+2\lambda_3\pm\sqrt{(\lambda_1-\lambda_2-2\lambda_3)^2+\lambda_5^2}|\le\kappa\pi  ~\&~ \nonumber\\
|3\lambda_{13}+4\lambda_2\pm\sqrt{(3\lambda_1-4\lambda_2-3\lambda_3)^2+\frac{3}{2}(2\lambda_4+\lambda_5)^2}|\le\kappa\pi,\label{peruni}
\end{align}
where $\kappa=8$ or $16$ depending on one's choice on the partial wave amplitude of an elastic scalar scattering from the consideration of S-matrix unitarity. For detailed discussion, see Ref.\,\cite{Arhrib:2011uy}. 
\item Perturbativity: {Keeping only the top Yukawa coupling, gauge interactions, and scalar potential couplings, the one-loop renormalization group equations (RGEs) rewritten in our notation are\footnote{Two-loop RGEs for the Higgs portal parameters have been studied in Ref.\,\cite{Chao:2012mx}.}}
\begin{align}
\left(4\pi\right)^{2}\frac{dg_{i}}{dt} & =b_{i}g_{i}^{3}\textrm{ with }b_{i}=\left(\frac{47}{10},-\frac{5}{2},-7\right)\,,\\
\left(4\pi\right)^{2}\frac{dy_t}{dt} & =y_t\left[\frac{9}{2}y_t^2-\left(\frac{17}{20}g_1^2+\frac{9}{4}g_2^2+8g_3^2\right)\right]\,,\\
\left(4\pi\right)^{2}\frac{d\lambda_{1}}{dt} & =\frac{27}{200}g_{1}^{4}+\frac{9}{20}g_{1}^{2}g_{2}^{2}+\frac{9}{8}g_{2}^{4}-\left(\frac{9}{5}g_{1}^{2}+9g_{2}^{2}\right)\lambda_{1}+24\lambda_{1}^{2}+3\lambda_{4}^{2}+3\lambda_{4}\lambda_{5}+\frac{5}{4}{\lambda_{5}}^{2}\nonumber \\
 & +12\lambda_{1}y_{t}^{2}-6y_{t}^{4}\,,\\
 \left(4\pi\right)^{2}\frac{d\lambda_{2}}{dt} & =\frac{54}{25}g_{1}^{4}-\frac{36}{5}g_{1}^{2}g_{2}^{2}+15g_{2}^{4}-\left(\frac{36}{5}g_{1}^{2}+24g_{2}^{2}\right)\lambda_{2}+2\lambda_{4}^{2}+2\lambda_{4}\lambda_{5}\nonumber \\
 & +28\lambda_{2}^{2}+24\lambda_{2}\lambda_{3}+6{\lambda_{3}}^{2}\,,
 \end{align}
\begin{align}
\left(4\pi\right)^{2}\frac{d\lambda_{3}}{dt} & =\frac{72}{5}g_{1}^{2}g_{2}^{2}-6g_{2}^{4}+{\lambda_{5}}^{2}-\left(\frac{36}{5}g_{1}^{2}+24g_{2}^{2}\right)\lambda_{3}+24\lambda_{2}\lambda_{3}+18{\lambda_{3}}^{2}\,,\\
\left(4\pi\right)^{2}\frac{d\lambda_{4}}{dt} & =\frac{27}{25}g_{1}^{4}-\frac{18}{5}g_{1}^{2}g_{2}^{2}+6g_{2}^{4}-\left(\frac{9}{2}g_{1}^{2}+\frac{33}{2}g_{2}^{2}\right)\lambda_{4}+12\lambda_{1}\lambda_{4}+4\lambda_{1}\lambda_{5}+4\lambda_{4}^{2}\nonumber \\
 & +16\lambda_{2}\lambda_{4}+12\lambda_{3}\lambda_{4}+{\lambda_{5}}^{2}+6\lambda_{2}\lambda_{5}+2\lambda_{3}\lambda_{5}+6\lambda_{4}y_{t}^{2}\,,\\
\left(4\pi\right)^{2}\frac{d\lambda_{5}}{dt} & =\frac{36}{5}g_{1}^{2}g_{2}^{2}-\left(\frac{9}{2}g_{1}^{2}+\frac{33}{2}g_{2}^{2}\right)\lambda_{5}+4\lambda_{1}\lambda_{5}+8\lambda_{4}\lambda_{5}+4{\lambda_{5}}^{2}+4\lambda_{2}\lambda_{5}\nonumber \\
 & +8\lambda_{3}\lambda_{5}+6\lambda_{5}y_{t}^{2}\,.
\end{align}
with $t\equiv\ln(\mu/m_t)$. For perturbativity, we require a similar approximate condition on the quartic Higgs couplings as in Ref.\,\cite{Gonderinger:2012rd}, which is based on the work of Ref.\,\cite{Riesselmann:1996is} i.e., 
\begin{align}
\lambda_i(\mu)\lesssim\lambda_{\rm FP}/3, \quad \forall\ m_Z\leq\mu\leq\Lambda,
\end{align}
where $\lambda_{\rm FP}\simeq12$ in the renormalization of Ref.\,\cite{Hambye:1996wb} and $\Lambda$ is the cutoff scale of the theory.
\end{itemize}

{\color{black}{Fig.~\ref{stability} gives constraints from VS (green region) and PU (orange region) at tree-level. The black dot  corresponds to our benchmark point discussed in Sec.\,\ref{subsec:BMandS}, {\it i.e.}, 
\begin{equation}
\lambda_2=0.2\, , \quad \lambda_3 = \lambda_4 = 0\, , \quad \lambda_5 = -0.1\,.
\end{equation}
After solving the above mentioned RGEs, one finds that that VS and perturbativity up to the Planck scale impose stringent constraints on $\lambda_i$'s\,\cite{Chao:2012mx}. For our benchmark point as input at the scale $\mu=m_t$, the resulting running couplings are shown in Fig.\,\ref{stability}.
From the right panel of Fig.\,\ref{stability}, it is clear that the CTHM stays perturbative even at the Planck scale. We also find that the potential develops a second minimum at $\mathcal{O}(10^{5}\text{-}10^{6}\rm \,GeV)$. The presence of this second minimum implies that the SM vacuum may become either unstable or metastable above this scale. In principle, stability could be preserved to higher scales with the presence of additional contributions to the RGEs associated with particles heavier than this threshold. A detailed investigation of the possible U.V. embedding of the CTHM goes beyond the scope of the present study. We observe, however, that the stability region for our benchmark point lies well above the range of triplet scalar masses that we consider below. Moreover, 
 one may also increase the scale at which the potential may develop a second minimum by increasing $\lambda_4$ while preserving perturbativity, which is indicated by the black arrow in the left panel of Fig.\,\ref{stability}. We will discuss this point further in Sec.\,\ref{subsec:disbdt}.}}

\subsection{Key features of the CTHM}
\label{subsec:mdlkey}
 Since $v_\Delta\ll v_\Phi$ due to the $\rho$ parameter constraint, we expect, in general, $\tan2\alpha$ (and thus $\sin\alpha$) to be small. In this case,  we have from  from Eq.\,\eqref{tan2a},
\begin{align}
\tan2\alpha\approx
\frac{v_\Delta}{v_\Phi}
\cdot
\frac{2v_\Phi^2\lambda_{45}-4m_\Delta^2}{{2\lambda_1 v_\Phi^2}-m_\Delta^2}
{{\approx\frac{v_\Delta}{v_\Phi}\cdot\frac{2v_\Phi^2\lambda_{45}-4m_\Delta^2}{m_h^2-m_\Delta^2} }},\label{t2aapprox}
\end{align}
{{Then in this small $\sin\alpha$ limit, the expressions for the masses given in Eq.\,(\ref{mhpp}-\ref{mH}) can be simplified to}} 
\begin{align}\label{hierarchy}
m_h^2\simeq2v_\Phi^2\lambda_1\simeq2v^2\lambda_1,~~ m_H \simeq m_\Delta \simeq m_A,~~ m_{H^\pm}^2\simeq m_\Delta^2-\frac{\lambda_5}{4}v_\Phi^2,~~ m_{H^{\pm\pm}}^2\simeq m_\Delta^2-\frac{\lambda_5}{2}v_\Phi^2.
\end{align}
We see that $m_\Delta$ sets the overall  mass scale of the triplet scalars whereas  $\lambda_1$ is basically determined by $m_h$ and $v$. Moreover, in the large $m_\Delta$ limit, the mass splitting is 

\begin{align}
\Delta m=|m_{H^{\pm\pm}}-m_{H^\pm}|\approx|m_{H^\pm}-m_{H,A}|\approx\frac{|\lambda_5|v_\Phi^2}{8m_\Delta}\approx\frac{|\lambda_5|v^2}{8m_\Delta},\label{massspec}
\end{align}
which depends only on $\lambda_5$, $m_\Delta$, and $v$. Thus, by measuring the masses of any two triplet scalars of differing charges, one could determine both $m_\Delta$ and the Higgs portal coupling $\lambda_5$. A practical corollary is  in the large $m_\Delta$ limit, once one of the triplet Higgs particles is discovered, the relatively small mass splitting (compared to $m_\Delta$) would provide guidance as to the mass region for discovery of the other triplet Higgs scalars.


\subsection{Neutrino masses from a type-II seesaw mechanism}
\label{subsec:seesaw}

In the CTHM, the neutrino masses are generated through a type-II seesaw mechanism via the Yukawa Lagrangian\,\cite{Mohapatra:1979ia, Cheng:1980qt, Lazarides:1980nt, Schechter:1980gr}
\begin{align}
\mathcal{L}_Y= & (h_\nu)_{ij}\overline{L^{ic}}i\tau_2\Delta L^j+\rm{h.c.}.\label{nvlag}
\end{align}
Here, $L=(v_L,e_L)^T$ is the l SU(2)$_L$ doublet; $h_\nu$ is the neutrino Yukawa matrix, which is a $3\times3$ complex and symmetric matrix as has been shown for a general case in Ref.\,\cite{Bilenky:1987ty}. After the EWSB with $v_\Delta\neq0$, neutrinos of different flavors mix through $h_\nu$, as implied by neutrino oscillations. The mass matrix $h_\nu v_\Delta$ also breaks the lepton number explicitly\footnote{In principle, one could assign a lepton number of -2 to $\Delta$ so that the overall Lagrangian conserves lepton number before EWSB. The third term in $V(\Phi, \Delta)$ would then explicitly break lepton number conservation. The coefficient of the the dimension five lepton number violating mass term ${\bar{L^C}} H^T H L$ is then proportional to $\mu/M^2$.}, implying that neutrinos are of the Majorana type with their masses being
\begin{align}
(m_\nu)_{ij}=\sqrt{2}(h_\nu)_{ij} v_\Delta. \label{numass}
\end{align}
{{Experimentally, sum of neutrino masses is constrained to be $\sum_i m_i<0.23$\,eV by the Planck Collaboration via assuming the existence of three light massive neutrinos, the validity of the $\Lambda$ Cold Dark Matter ($\Lambda$CDM) model and using the supernovae and the Baryon Acoustic Oscillations data\,\cite{Agashe:2014kda, Ade:2015xua}. Given this constraint, we choose $m_\nu=0.01\rm\,eV$ for each of the three light neutrinos throughout the paper. In principle, one can choose a larger (smaller) value for the neutrino masses while still satisfying the experimental constraints. Larger (smaller) neutrino masses will correspond to a larger (smaller) $h_\nu$ for fixed $v_\Delta$, which will in turn affect the same-sign di-lepton decay BRs of $H^{\pm\pm}$. The BRs will then affect the parameter space relevant for model discovery. We will discuss effects from smaller/larger $m_\nu$ in Sec.\,\ref{subsec:disbdt}.}}

\section{Model parameter determination}\label{paramdeter}
The model parameters for the CTHM are, na\"ively, $\{g, g', v_\Phi, v_\Delta, \mu, \lambda_i, h_{\nu}\}$(i=1,2,3,4,5), or in the mass eigenstates after the EWSB, $\{\alpha_{E.M.}, G_F, m_Z, m_h, m_H, m_A, m_{H^\pm},$ $m_{H^{\pm\pm}}, v_\Delta, \sin\alpha, m_\nu\}$. $\alpha_{E.M.}, G_F, m_Z, m_h$ are already well-known from electroweak precision and Higgs mass measurements, and in order to further determine other parameters of the CTHM, we will need discovery of the new particles to know their masses and the measurement of the mixing angle $\sin\alpha$ as well. Therefore, in the following sub-sections, we will discuss how to experimentally determine the other parameters of the CTHM. In the end of this section, we will also discuss how to determine the input model parameters from consideration of perturbativity, which is essential for our collider study in Sec.\,\ref{sec:modeldis} and Sec.\,\ref{sec:lam45}.
\subsection{Mass spectrum and determination of $\lambda_1$ and $\lambda_5$}
\label{subsec:input}
From Sec.\,\ref{subsec:mdlkey}, we conclude that $\sin\alpha$ is in general small and in this small $\sin\alpha$ limit, we have Eq.\eqref{hierarchy}, i.e.,
\begin{align}
m_h^2\simeq2v_\Phi^2\lambda_1,~~ m_H \simeq m_\Delta \simeq m_A,~~ m_{H^\pm}^2\simeq m_\Delta^2-\frac{\lambda_5}{4}v_\Phi^2,~~ m_{H^{\pm\pm}}^2\simeq m_\Delta^2-\frac{\lambda_5}{2}v_\Phi^2,\tag{\ref{hierarchy}}
\end{align}
we see that: (a) When $\lambda_5\le0$, $m_h<m_H\simeq m_A\le m_{H^\pm}\le m_{H^{\pm\pm}}$, we call this the Normal Mass Hierarchy (NMH); (b) while when $\lambda_5\ge0$, $m_{H^{\pm\pm}} \le m_{H^{\pm}}\le m_A\simeq m_H$ and $m_h< m_H$, we call this the Reversed Mass Hierarchy (RMH). For the NMH, SM $h$ is the lightest particle and $H^{\pm\pm}$ is the heaviest one, the order of the mass spectra is unique. While for the RMH, $A$ or equivalently $H$ is the heaviest particle, but the mass order between $h$ and ($H^\pm$, $H^{\pm\pm}$) is unclear and will generally depend on our model input.

On the other hand, from $m_h^2\simeq2v^2\lambda_1$, we conclude that $\lambda_1\approx\frac{m_h^2}{2v^2}\approx0.129$. While to determine $\lambda_5$, one can use the mass splitting $\Delta m\approx\frac{|\lambda_5|v^2}{8m_\Delta}$ as defined in Eq.\,\eqref{massspec} upon discovery.

\subsection{Measurement of the mixing angle $\sin\alpha$ for determination of $\lambda_4$}\label{lam4determ}
\begin{figure}[thb!]
\captionstyle{flushleft}
\begin{tabular}{cc}
   \includegraphics[width=80mm,height=65mm]{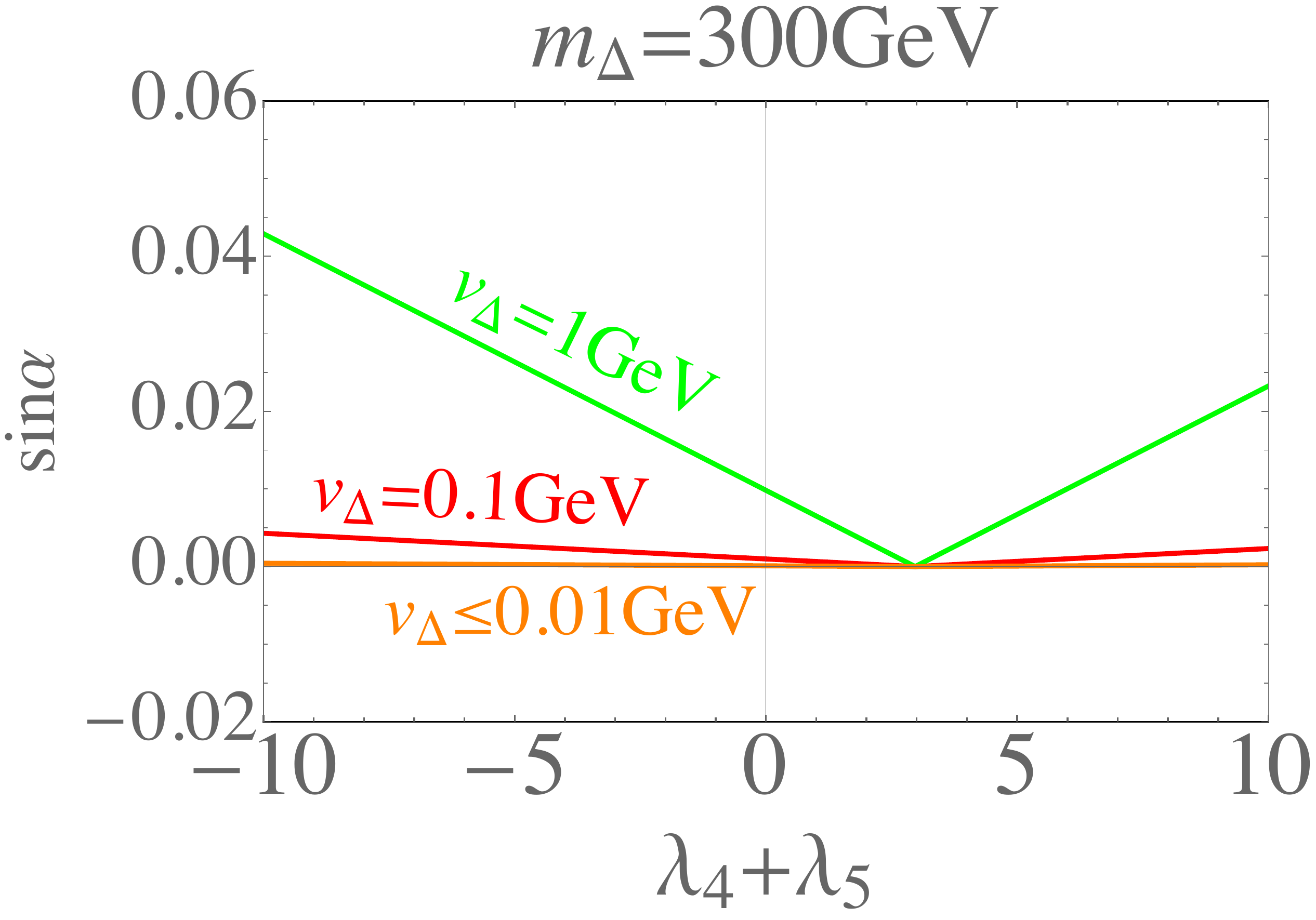} &  \includegraphics[width=80mm,height=65mm]{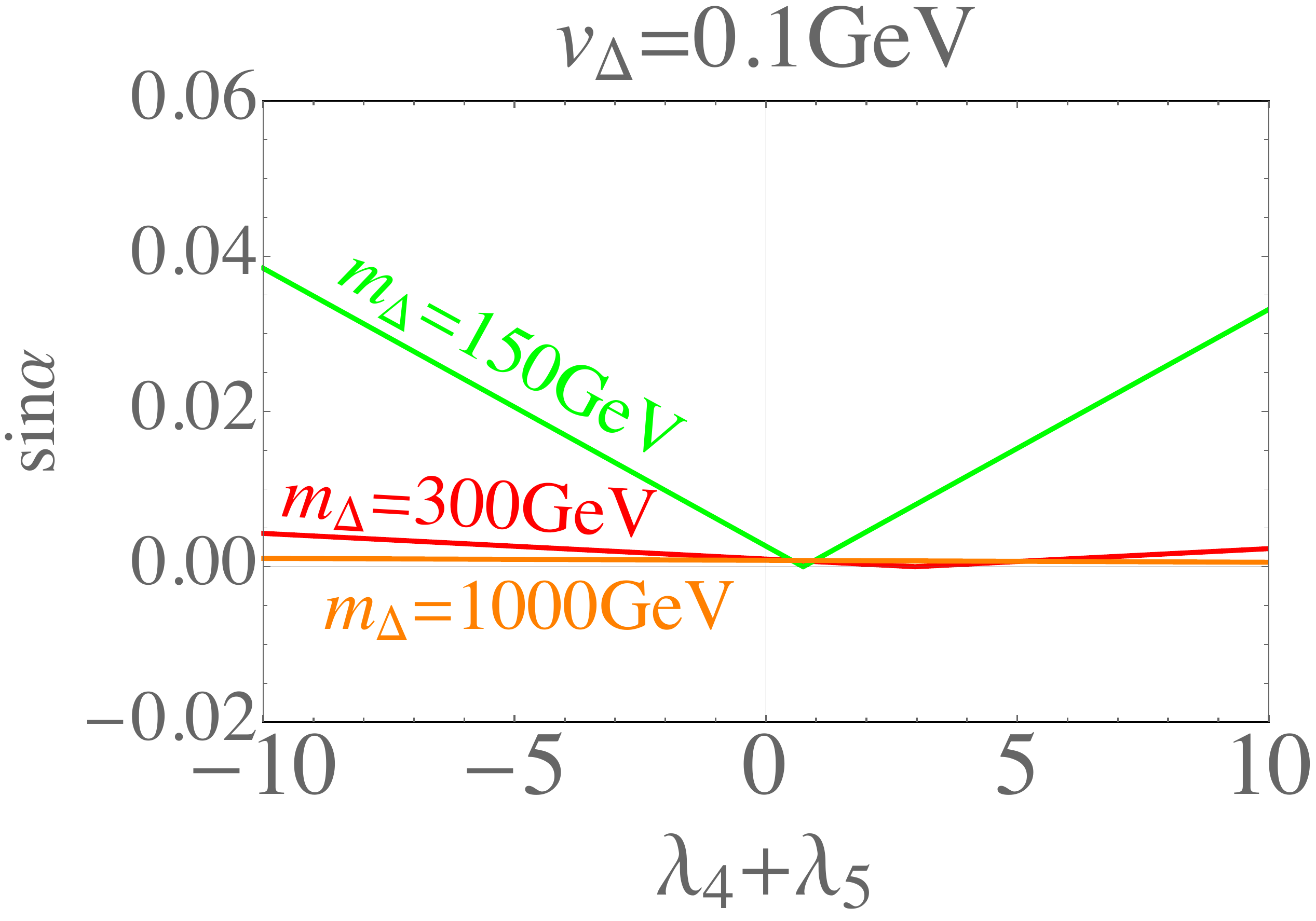}
\end{tabular}
\caption{The dependence of $\sin\alpha$ on $\lambda_{23}$ is negligible due to the smallness of $v_\Delta$, and $\lambda_1\approx m_h^2/(2v^2)\approx0.129$, such that $\sin\alpha$ is approximately a function of $\lambda_{45}$, $m_\Delta$ and $v_\Delta$. On the left (right) panel we fix $m_\Delta=300$\,GeV ($v_\Delta=0.1$\,GeV) and plot $\sin\alpha$ with respect to $\lambda_{45}$ with different $v_\Delta$'s ($m_\Delta$'s). One observes that $\sin\alpha$ becomes sufficiently small for increasing $m_\Delta$ and/or decreasing $v_\Delta$.}\label{saplot}
\end{figure}

To determine $\lambda_4$, we note that from Eq.\,\eqref{t2aapprox}, we can solve for $\alpha$: 
\begin{align}
\alpha\approx\left\{\begin{array}{cc}\frac{1}{2}\arctan\left(\frac{v_\Delta}{v_\Phi}\cdot\frac{2v_\Phi^2\lambda_{45}-4m_\Delta^2}{m_h^2-m_\Delta^2}\right), & \text{if~} \frac{2v_\Phi^2\lambda_{45}-4m_\Delta^2}{m_h^2-m_\Delta^2}\ge0\\ \pi+\frac{1}{2}\arctan\left(\frac{v_\Delta}{v_\Phi}\cdot\frac{2v_\Phi^2\lambda_{45}-4m_\Delta^2}{m_h^2-m_\Delta^2}\right), & \text{if~} \frac{2v_\Phi^2\lambda_{45}-4m_\Delta^2}{m_h^2-m_\Delta^2}<0\end{array}\right.,\label{alphaeq}
\end{align}
which implies $\sin\alpha$ is in general a two-to-one function. This feature of $\sin\alpha$ is graphically reflected in Fig.\,\ref{saplot}. In addition, from Fig.\,\ref{saplot}, we see that $\sin\alpha$ indeed decreases with increasing $m_\Delta$ and/or decreasing $v_\Delta$. For example, when $m_\Delta\gtrsim300$\,GeV and/or $v_\Delta\lesssim0.1$\,GeV, {{$\sin\alpha\lesssim0.01$.}} 

\begin{center}
\begin{minipage}{\linewidth}
\centering
\captionof{table}{Three-point vertices related to the determination of $\lambda_{4,5}$. $\lambda_5$ is determined through mass splitting, $\lambda_4$ is determined through the mixing angle $\sin\alpha$, which is sensitive to $\lambda_{45}$.} \label{tab:1} 
\begin{tabular}{ C{1.25in} C{3.85in} }\toprule[1.5pt]
\bf Vertex & \bf Coupling \\\midrule
$hAZ$ & $-\frac{g}{2\cos\theta_W}(\cos\alpha\sin\beta_0-2\sin\alpha \cos\beta_0)$ \\\midrule
$HZZ$ & $\frac{2ie m_Z}{\sin2\theta_W}(2\sin\beta_0\cos\alpha-\cos\beta_0\sin\alpha)$ \\\midrule
$HW^+W^-$ & $igm_Z\cos\theta_W(\sin\beta_0\cos\alpha - \cos\beta_0\sin\alpha)$ \\\midrule
$hH^-W^+$ & $\frac{ig}{2}(\sin\beta_\pm\cos\alpha - \sqrt{2}\cos\beta_\pm\sin\alpha)$ \\
\bottomrule[1.25pt]
\end {tabular}\par
\bigskip
\end{minipage}
\end{center}

On the other hand, the variation of $\sin\alpha$ with $\lambda_{45}$ can also be used to determine $\lambda_{45}$ through various gauge boson-Higgs couplings. {We focus on gauge boson-Higgs vertices as electroweak production of the triplet Higgs particles is the dominant production mechanism in the CTHM. After a careful investigation of all the triple vertices listed in Appendix\,\ref{frapp}, we find that only four of the gauge boson-Higgs couplings are linearly dependent on $\sin\alpha$.}\footnote{{Some of the non gauge boson-Higgs type vertices are also $\sin\alpha$ linearly dependent as can be seen from the $hH^{++}H^{--}$ vertex in Appendix\,\ref{frapp}, but the corresponding production cross section is smaller compared with the dominant electroweak production.}} {These couplings will eventually affect the decay BRs of the BSM particles. Thus, after their discovery, one could determine $\lambda_5$ from the mass splitting and $\lambda_4$ from the triplet Higgs decay BRs} \footnote{{Here we remind the reader that the Higgs portal parameters $\lambda_{4,5}$ are of particular interest as they may allow a SFOEWPT to explain the baryon asymmetry of the universe (BAU). In this paper, however, we will not discuss the effects on phase transition or baryogenesis from the CTHM but rather leave it for future work.}}.



\subsection{$\lambda_2$ and $\lambda_3$ determination}
Different from the determination of $\lambda_4$ and $\lambda_5$, however, $\lambda_2$ and $\lambda_3$ are in general very difficult or even impossible to measure as they are always suppressed by $v_\Delta^2$ (for mass terms) or by $v_\Delta$ (for three-body interactions). One possible way to measure them is through the quartic triplet Higgs interactions, but the production cross section will again be suppressed by the smallness of $v_\Delta$ in general. Note that since $\lambda_2$ and $\lambda_3$ are irrelevant to electroweak phase transition, it is unnecessary to pay too much attention to their determination.

\subsection{Choice of input model parameters}
\label{sec:input}
As discussed in last three sub-sections, experimentally, one can use the SM Higgs mass, the mass difference and the mixing angle to determine $\lambda_1$, $\lambda_4$ and $\lambda_5$. But recall that, in Sec.\,\ref{vdcons}, the $\rho$ parameter requires $v_\Delta$ to be negligible compared with $v_\Phi$ or $v$, which is about the same order as the Higgs masses. The ratio of the Higgs masses and $v_\Delta$ will then lead to very large $\lambda_{2,3}$ by referring back to Eq.\,(\ref{lam2}-\ref{lam3}), thus to preserve perturbativity of the CTHM, one will have to ``fine-tune'' the Higgs masses to obtain reasonable values for $\lambda_{2,3}$. To avoid the ``fine tuning'', we choose $\lambda_{2,3}$ instead as our input in our theoretical study. As also discussed in Sec.\,\ref{subsec:input} and Sec.\,\ref{lam4determ}, (a) Since we know the Higgs mass exactly, we choose $m_h$ instead of $\lambda_1$ as our model input; (b) we choose $m_\Delta$ and $\lambda_5$ as our model input as they determine the mass spectrum; (c) $\sin\alpha$ is negligible at small $v_\Delta$, thus to avoid ``fine tuning'' $\lambda_4$, we choose $\lambda_4$ instead of $\sin\alpha$ as our model input. Another reason for choosing $\lambda_4$ as our model input is that it {frequently} always appears in pair with $\lambda_5$ such that one can infer $\lambda_4$ from the combination once we know $\lambda_5$. {At the same time, relevant quantities may depend separately on $\lambda_4$ and $\lambda_5$, {\it e.g.}, $H^\pm$ decay BRs.}
To summarize, our model input parameters are $\{\alpha_{E.M.}, G_F, m_Z, m_h, m_\Delta, v_\Delta, \lambda_2, \lambda_3, \lambda_4, \lambda_5, m_\nu\}$.

{Here we emphasize that the input parameters need to be carefully chosen to avoid fine tuning the masses or to preserve the validity of perturbation theory from $\lambda_{2,3}$, otherwise one may easily fall into the region where perturbation theory is invalid. For example, for the plots in the second row of Fig.\,2 in Ref.\,\cite{Aoki:2011pz}, the authors used the scalar masses as their input. We find that using their input, only when $v_\Delta\gtrsim1$\,GeV will the value of $\lambda_3$ respect perturbativity, whereas for smaller $v_\Delta$'s, $\lambda_3$ can be as large as $10^{21}$.}

\section{Production and Decay Rates of the Scalars in the CTHM}
\label{sec:decayprod}
As discussed in last section, the mass ordering of the RMH will in general depend on our model input. For simplicity, we will work in the NMH throughout the paper, in which framework the production and decay rates of the BSM Higgs particles are studied in detail below. {{While we want to point out that, in the RMH, though the decay patterns, the decay BRs and thus our Fig.\,\ref{bdtdis} and Fig.\,\ref{haa} will change, the same channels studied in this paper can still be used for model discovery and Higgs portal parameter determination.}}
\subsection{Production cross section of the Higgs particles in the CTHM}\label{subsec:sg7bg}
\label{subsec:prod}
In SM, the Higgs boson can be produced via gluon fusion or vector boson fusion (VBF), but in the CTHM, single production of the triplet Higgs particles via gluon fusion or VBF is highly suppressed by small $v_\Delta$\footnote{SM $h$ production via gluon fusion, however, does not suffer from suppression from the smallness of $v_\Delta$.}. 
Therefore, single production of the triplet Higgs particles through gluon fusion or VBF will not be considered in this paper. For double scalar production, a pair of triplet scalars can { be produced through electoweak Drell-Yan processes or gluon fusion}.
As in the single Higgs production case, however, double scalar Higgs particle production via an intermediate $H$ or $A$, which is produced through gluon fusion, is again highly suppressed by small $v_\Delta$. {No such suppression occurs for electroweak pair production. Consequently, we focus on the latter.}

To  study quantitively the production cross sections of the triplet Higgs particles, we first use {\tt Mathematica} and {\tt FeynRules}\,2.3.13\,\cite{Alloul:2013bka,Christensen:2008py} to generate the Universal FeynRules Output (UFO) model file\,\cite{Degrande:2011ua} of the CTHM, then we use {\tt MadGraph}\,2.3.3\,\cite{Alwall:2014hca} to implement the CTHM UFO file to obtain the production cross sections at $\sqrt{s}=14$\,TeV and $\sqrt{s}=100$\,TeV. However, we find that for the channels we are going to study in this paper, the number of events at $\sqrt{s}=14$\,TeV and $\mathcal{L}=3\,\rm ab^{-1}$ is too few even without considering the corresponding backgrounds, so we only list the cross section result at $\sqrt{s}=100$\,TeV here.

\begin{figure}[thb!]
\captionstyle{flushleft}
\begin{tabular}{cc}
   \includegraphics[width=90mm,height=75mm]{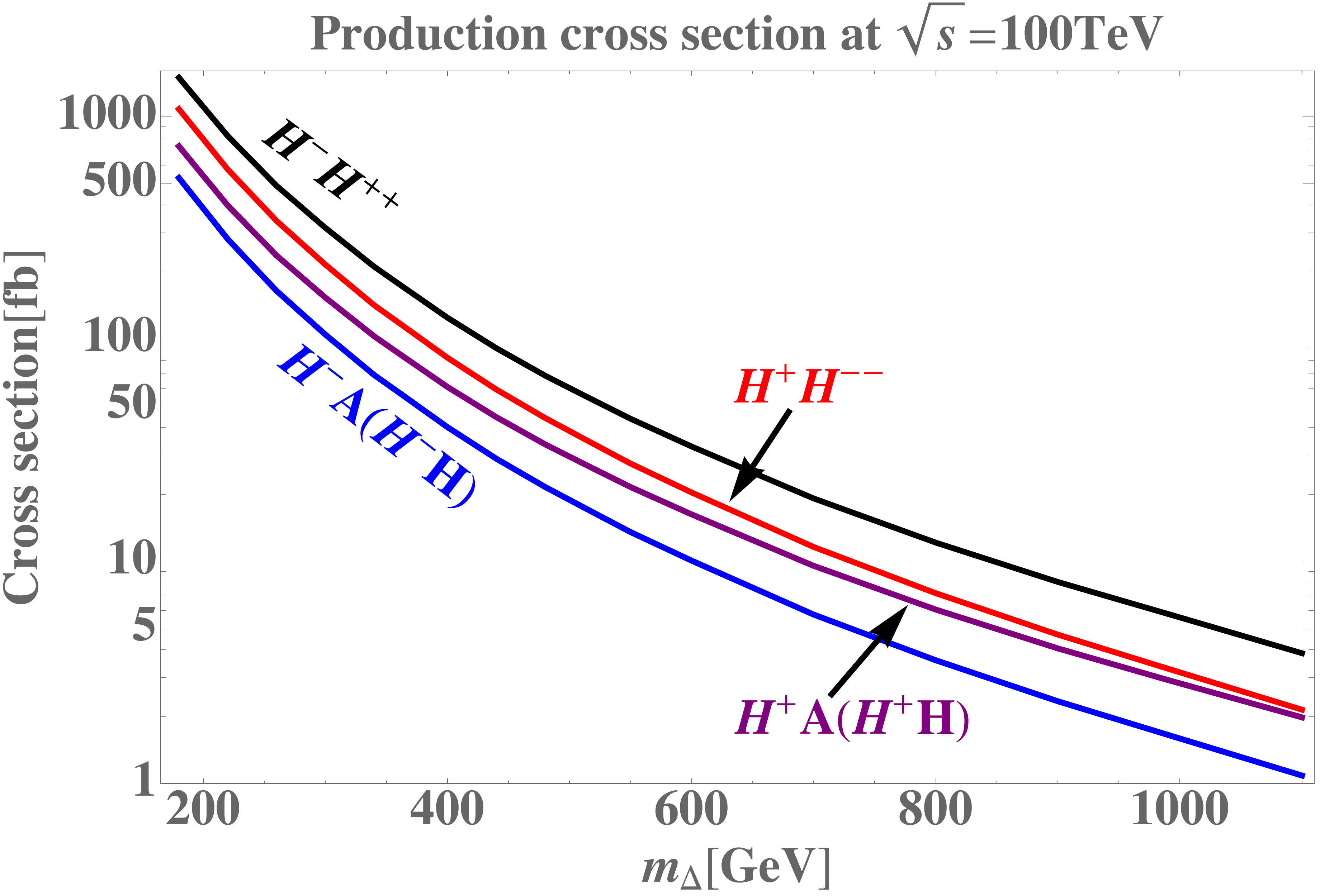} &  \includegraphics[width=90mm,height=75mm]{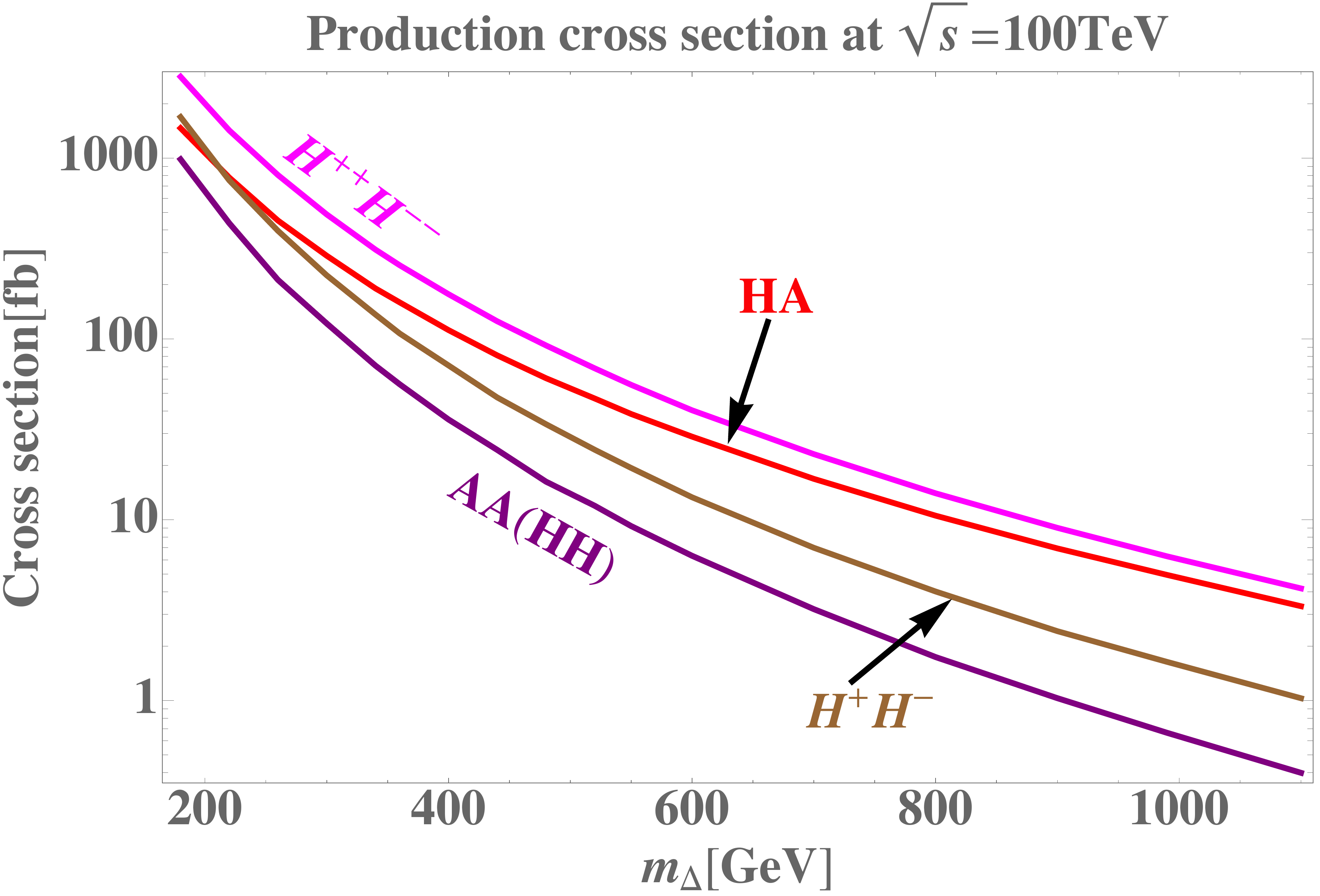}
\end{tabular}
\caption{
Production cross section as a function of $m_\Delta$ at $\sqrt{s}=100$\,TeV with $v_\Delta=10^{-3}$\,GeV. We set $\lambda_2=0.2$, $\lambda_3=0$, $\lambda_4=0$ and $\lambda_5=-0.1$, which correspond to the black dot in the left panel of Fig.\,\ref{stability} in order to be consistent with the NMH framework and to satisfy the model constraints discussed in Sec.\,\ref{secallconst}. The left panel is for associated Higgs production channels while the right one is for pair production except the $HA$ channel. Since the production cross section of $HA$ is very close to $H^-H^{++}$, we include it in the right panel to make the plots more readable.}\label{higgsprod}
\end{figure}



{The pair production cross sections depend on the couplings of the electroweak gauge bosons to the scalars and on the scalar masses. In what follows, we cast these dependences in terms of our independent parameters.} {{Note that $\lambda_1$ is basically fixed by $v$ and $m_h$, while the effects of  $\lambda_{2,3}$ are suppressed by small $v_\Delta$. In short, the production cross sections will be largely insensitive to $\lambda_{2,3}$ but will  depend significantly on $\lambda_{4,5}$.} To be consistent with the NMH, which requires a negative $\lambda_5$, and to satisfy the constraints discussed in Sec.\,\ref{model:const}, we choose $\lambda_2=0.2$, $\lambda_3=0$, $\lambda_4=0$ and $\lambda_5=-0.1$.} As an example, we fix $v_\Delta=10^{-3}$\,GeV and obtain the production cross sections given in Fig.\,\ref{higgsprod}, from which we see that pair production of $H^{++}H^{--}$ has the largest production cross section followed by $H^{++}H^-$. On the other hand, $H^+H^{--}$ will always be produced simultaneously with $H^-H^{++}$. We therefore expect an enhancement of the cross section from the combination of $H^-H^{++}$ and $H^+H^{--}$ channels. 

The hierarchy of the various production cross sections is briefly explained below: (a) {Besides a factor of four enhancement from the electric charge of $H^{\pm\pm}$, $H^{++}H^{--}$ pair has a larger cross section than $H^+H^-$ because it is constructively produced through $s$-channel $\gamma$ and $Z$ exchange. In contrast, the $H^+H^-$ pair production is suppressed due to destructive interference\,\cite{Akeroyd:2011zza}}.{ Note that even though $m_{H^{\pm\pm}}>m_{H^{\pm}}$, the mass splitting is not large due to our choice of $\lambda_5$; therefore, the lighter $H^\pm$ mass does not compensate for the aforementioned factors. (b) $H^{++}H^{-}$ has a larger cross section than $H^{--}H^{+}$ because the former is dominantly produced through a $W^+$ while the latter is through a $W^-$. (c) {{$HH$ and $AA$ channels, or $H^\pm A$ and $H^\pm H$ channels, have the same production cross sections due to mass degeneracy of $H$ and $A$.}} (d) {{$H^\pm A/H^\pm H$ has a smaller cross section than $HH/AA$, and $HA$ has a smaller cross section than $H^{++}H^{--}/H^{++}H^{-}$, because of the couplings.}}} (e) In the NMH, $m_{H^\pm}>m_{H/A}$, but the couplings involved for $H^+H^-$ is larger than those for $H^+A/H^+H$, the phase space and the couplings will compete such that at small $m_\Delta$, $H^+H^-$ has larger cross section while at large $m_\Delta$, $H^+A/H^+H$ has a larger cross section. This is also true for $HA$ and $H^+H^-$ channels.


In order to study the collider signatures of the triplet Higgs particles, it is natural to focus on $H^{\pm\pm}H^{\mp\mp}$ and $H^{\pm\pm}H^{\mp}$ channels since they have the largest production cross sections compared with other channels. To determine the final states, we will study their dominant decay channels in next sub-section.

\subsection{Decay rates of the scalar Higgs particles in the CTHM}\label{moddec}
To further determine the dominant decay modes of the triplet Higgs particles in the CTHM for collider simulation, we calculate their decay rates by taking $h_\nu=\mathbb{1}_{3\times3}$ for simplicity. All our decay formulas agree with those in  Appendix A of Ref.\,\cite{Aoki:2011pz} if one also takes the unit matrix limit there. 

{\color{black} In order to illustrate the potential parameter-dependence of various decay channels, we show in Fig.~\ref{slicedecay}  
the BRs for the charged and neutral triplet states as functions of the relevant combinations of $\lambda_4$ and $\lambda_5$ for representative values of 
$m_\Delta=400$\,GeV and $v_\Delta=10^{-4}$\,GeV.

\begin{figure}[thb!]
\captionstyle{flushleft}
\begin{tabular}{cc}
\includegraphics[width=75mm]{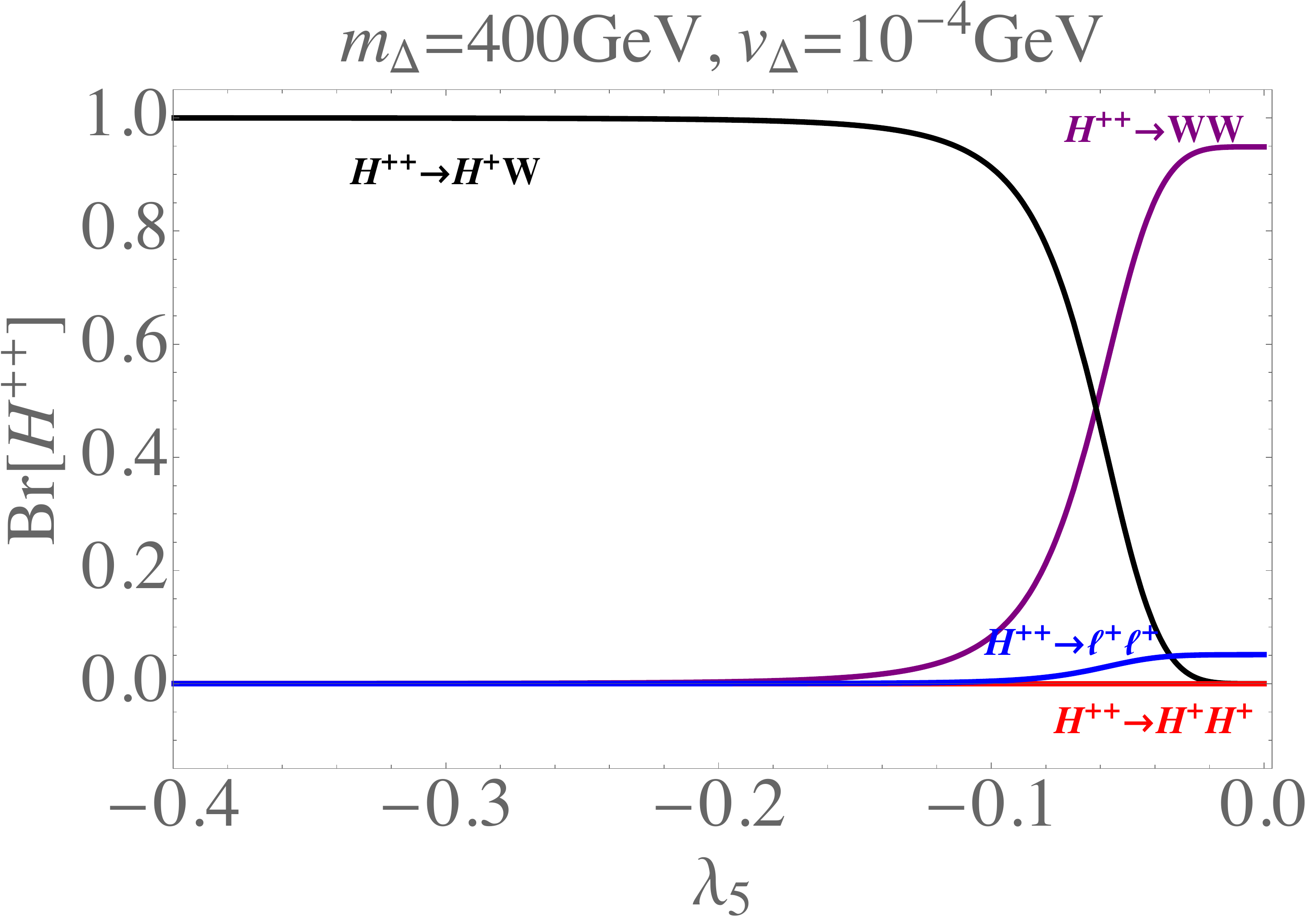} & \includegraphics[width=75mm]{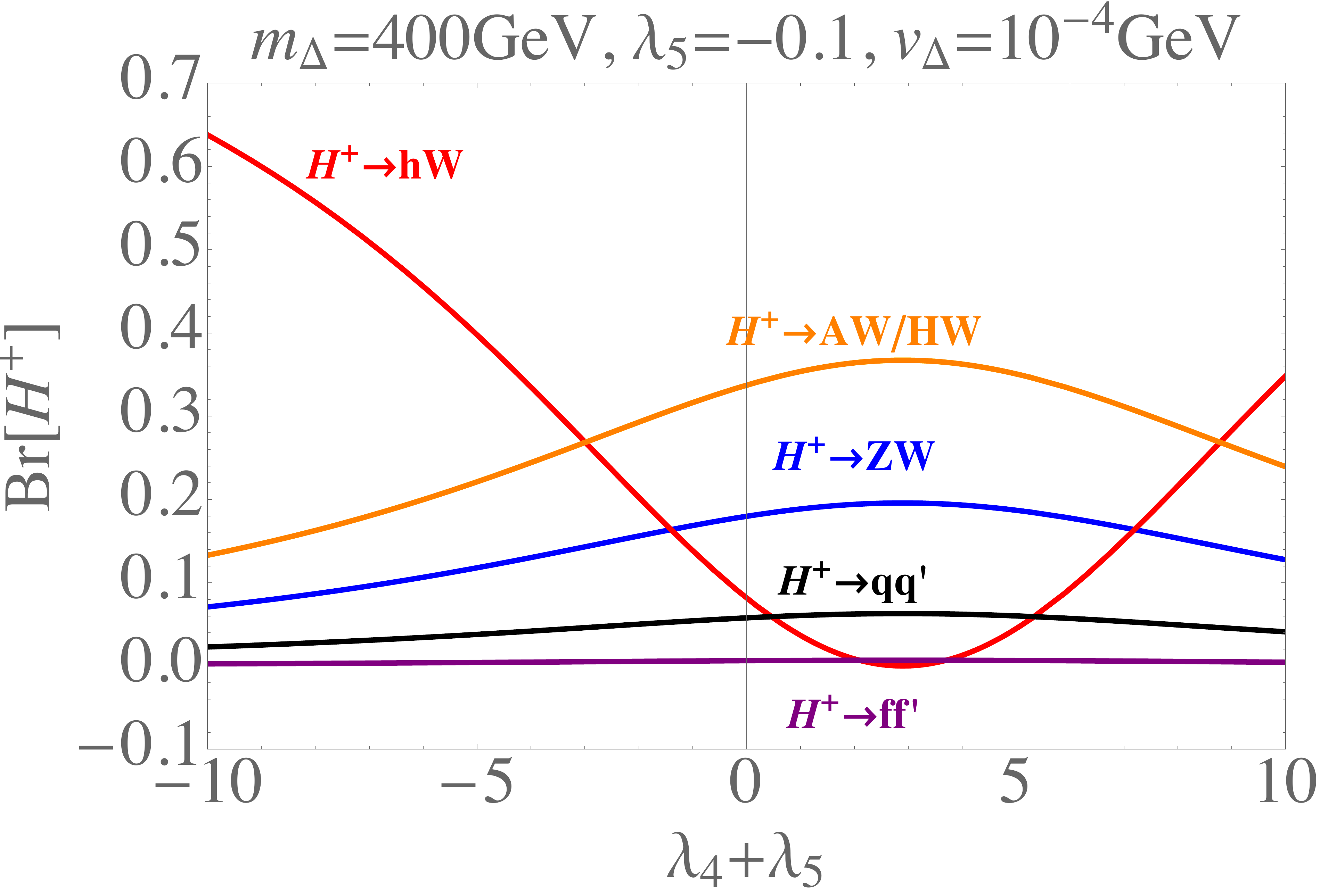}   \\
\includegraphics[width=75mm]{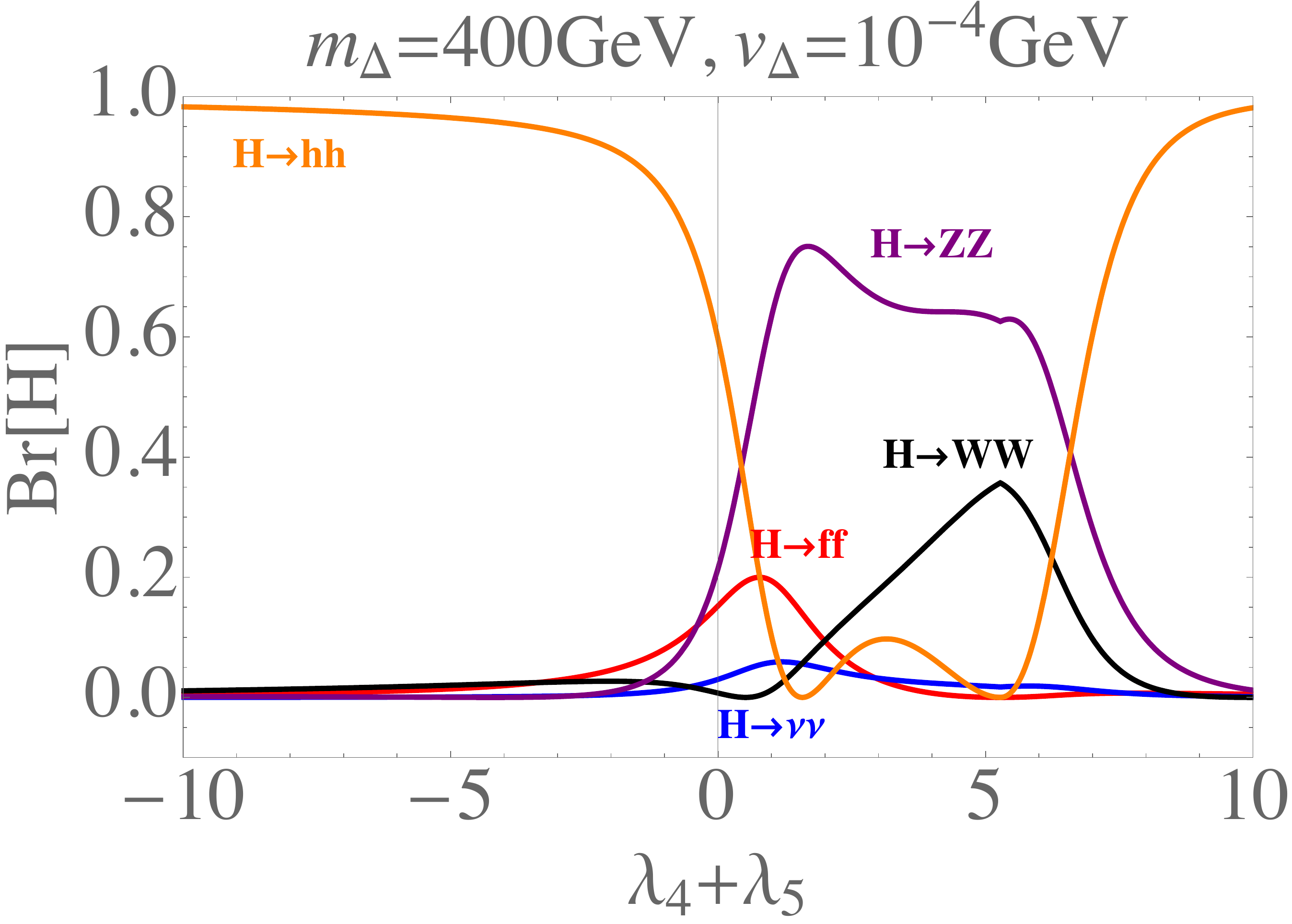} &   \includegraphics[width=75mm]{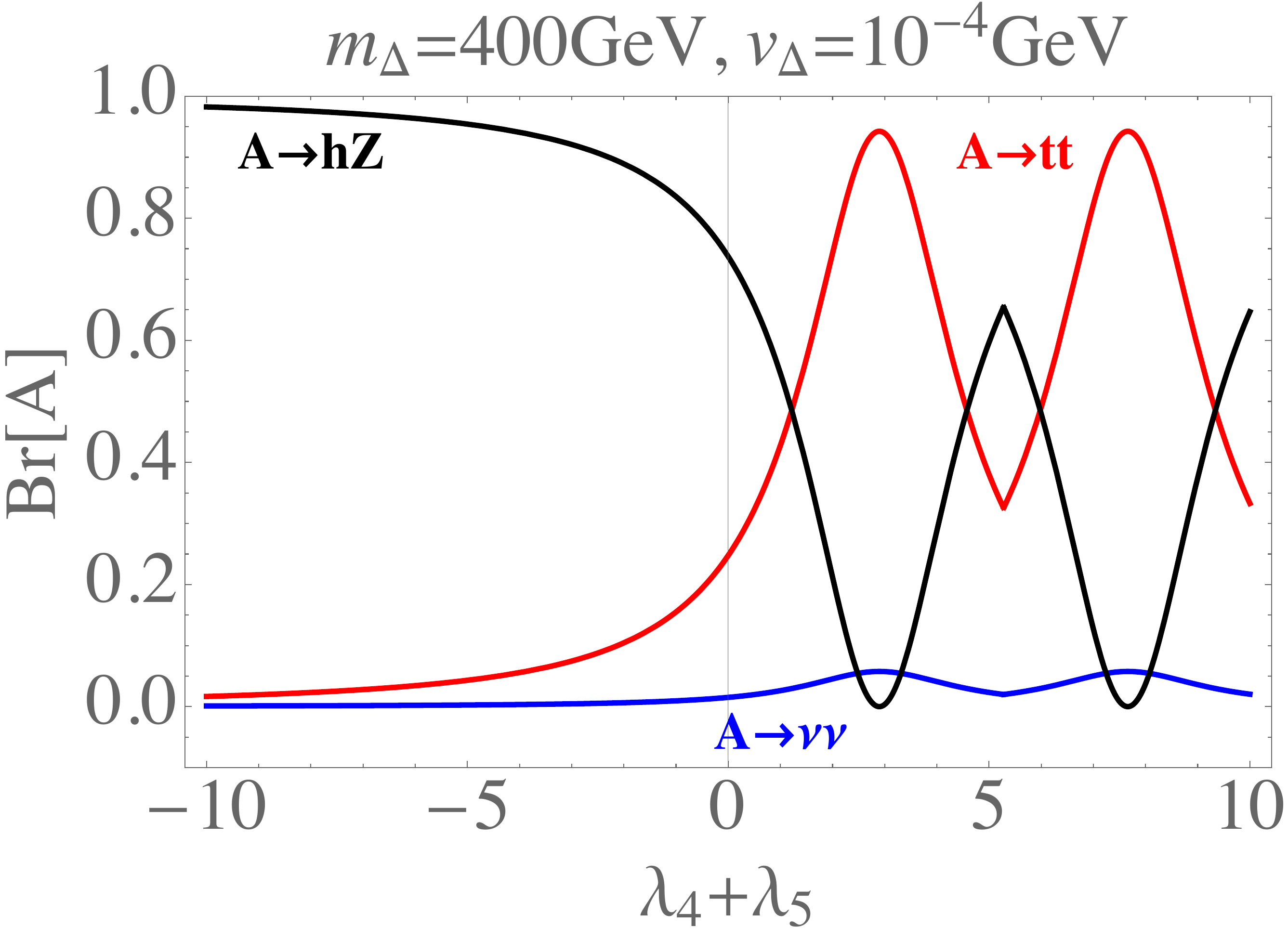}   \\
\end{tabular}
\caption{Decay BRs for $H$, $A$, $H^{\pm\pm}$ and $H^\pm$ as a function of $\lambda_4$ and $\lambda_5$ for representative values of $m_\Delta=400$\,GeV and $v_\Delta=10^{-4}$\,GeV. For a detailed discussion on the decay features, one can refer to the main text in Sec.\,\ref{moddec}.}\label{slicedecay}
\end{figure}
In this study, we will focus on the NMH with $\lambda_5<0$. From the top left panel of Fig.~\ref{slicedecay}, we observe that the $H^{\pm\pm}$ BRs to $H^\pm W^\pm$ and $W^\pm W^\pm$ depend strongly on this parameter in the vicinity of our benchmark point value: $\lambda_5=-0.1$. From the top right plot, we also observe that the $\mathrm{BR}(H^\pm \to h W^\pm)$ also depends strongly on $\lambda_4+\lambda_5$.  Even though in the vicinity of our benchmark point with $\lambda_4+\lambda_5=-0.1$ the $h W^\pm$ mode is subdominant, the corresponding BR depends more strongly on $\lambda_4+\lambda_5$ than do the other modes. Consequently, we will focus on this channel for the decay of the singly-charged scalar. The bottom two panels give the neutral scalar BRs. Though we will not utilize this information in the present study, we include them here for completeness and for future reference. 

It is also useful to determine how the $H^{\pm\pm}$ BRs vary with $m_\Delta$ and $v_\Delta$. To that end, }
{{in Fig.\,\ref{decayregionplotHPP}, we show the regions of parameter space where the BR to various final states is greater than 40\% for $H^{\pm\pm}$. In the left panel of Fig.\,\ref{decayregionplotHPP}, we consider the ($v_\Delta$, $\lambda_5$) plane for fixed $m_\Delta$, while the right panel gives the ($m_\Delta$, $\lambda_5$) plane for fixed $v_\Delta$. Note that $H^{\pm\pm}$ decay BRs are independent on $\lambda_4$ and for the NMH, one has $\lambda_5 < 0$. 

\begin{figure}[thb!]
\captionstyle{flushleft}
\begin{tabular}{cc}
   \includegraphics[width=75mm,height=65mm]{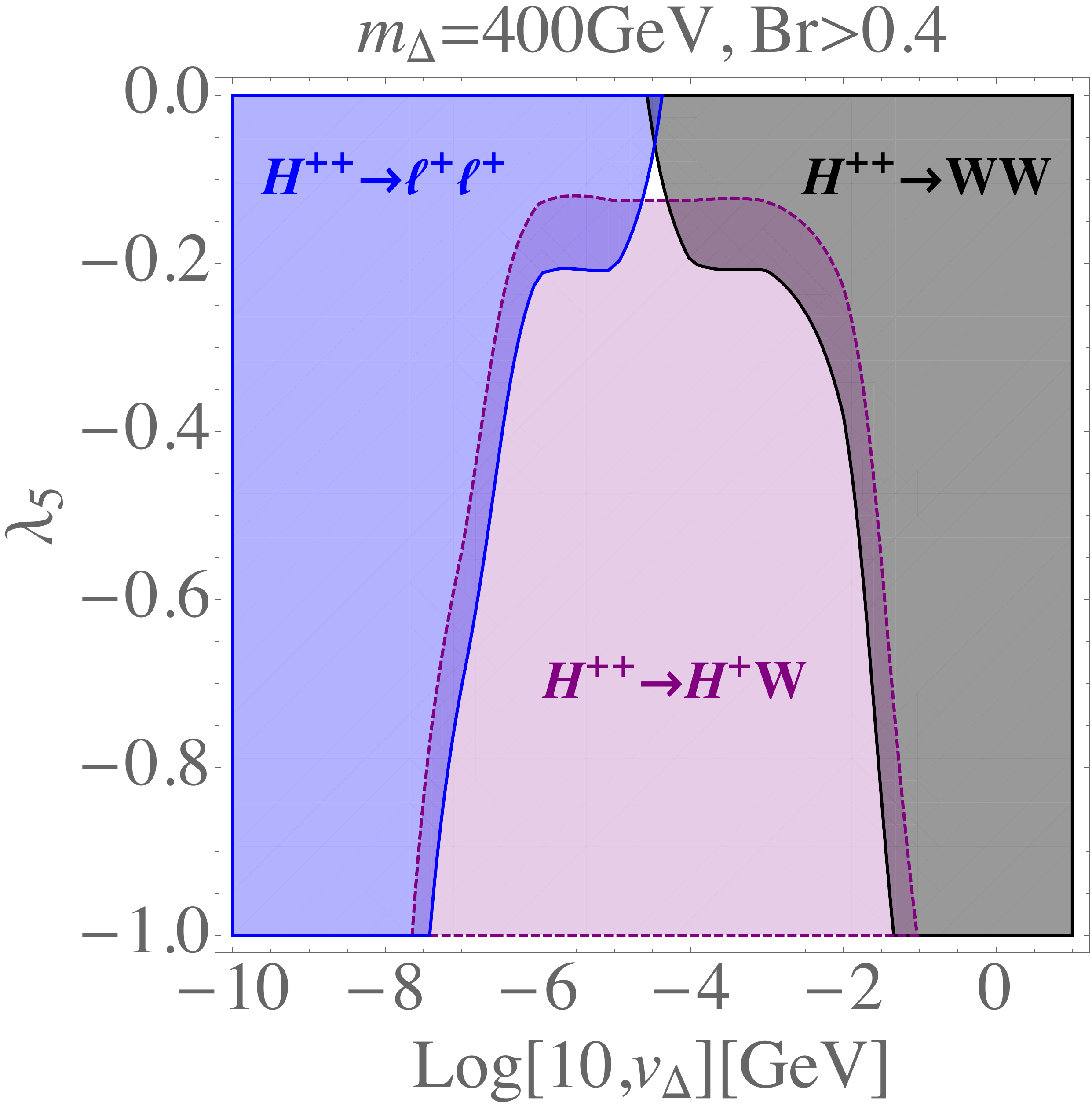} &  \includegraphics[width=75mm,height=65mm]{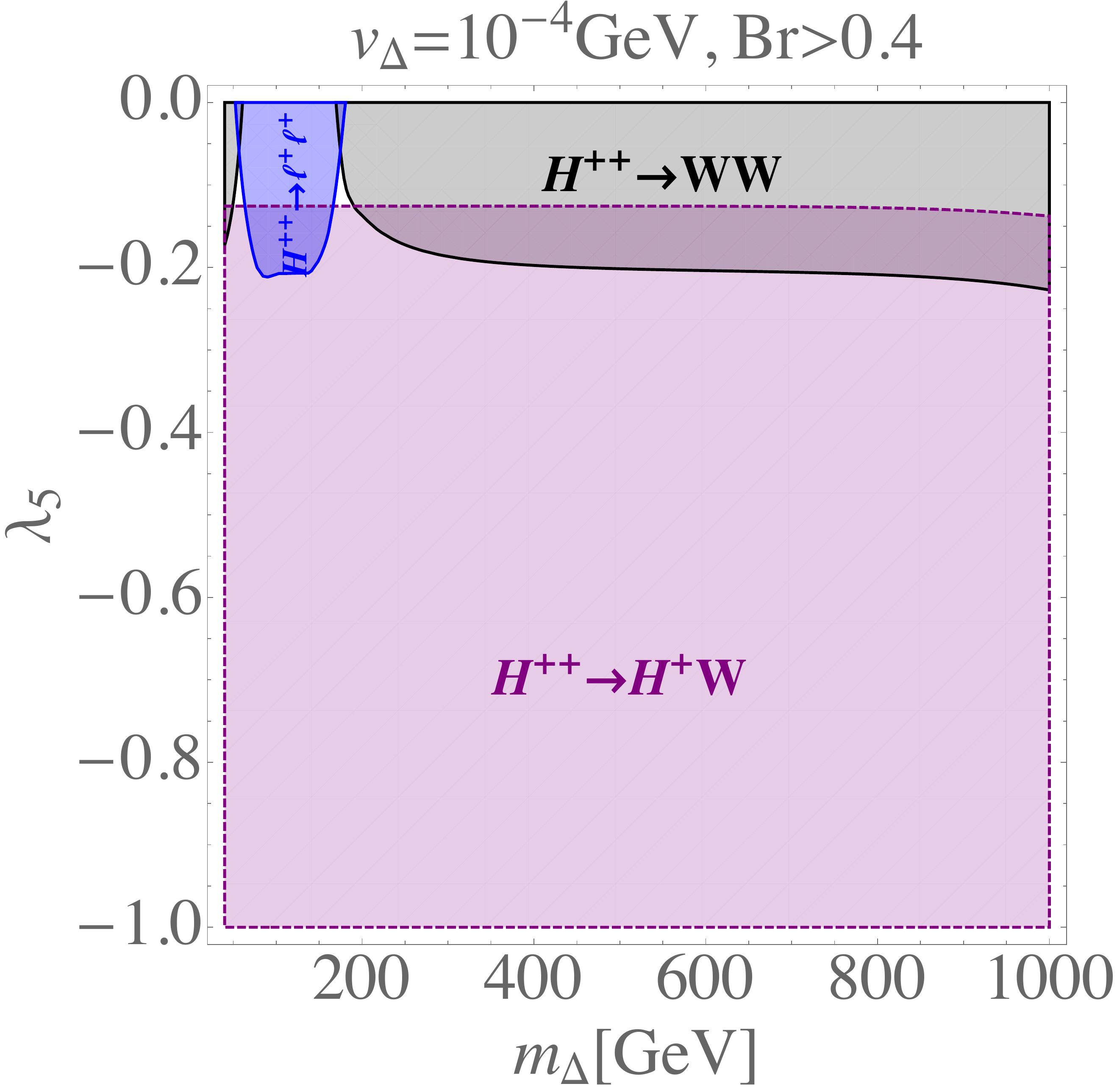} \\
\end{tabular}
\caption{Decay region plots for $H^{\pm\pm}$ with BR$\ge40\%$. Left panel is with $m_\Delta=400\rm\,GeV$ and right panel is with $v_\Delta=10^{-4}\rm\,GeV$. Purple region is the $H^\pm W^\pm$ channel, black is the same-sign di-W boson channel and blue is the same-sign di-lepton channel. $\lambda_5$ is in the negative region to be consistent with the NMH framework.}\label{decayregionplotHPP}
\end{figure}

\begin{figure}[thb!]
\captionstyle{flushleft}
\begin{tabular}{cc}
   \includegraphics[width=75mm,height=65mm]{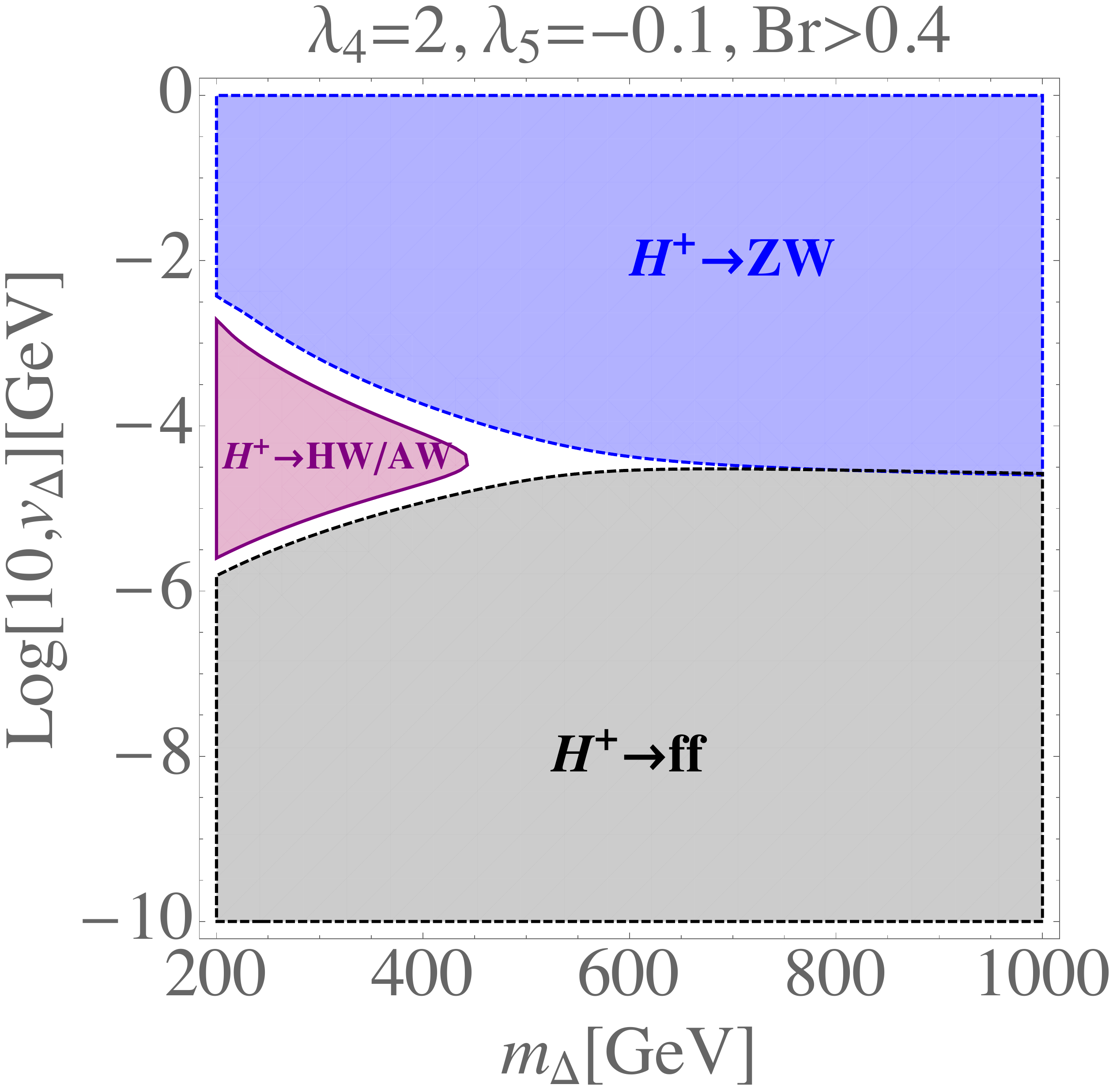} &   \includegraphics[width=75mm,height=65mm]{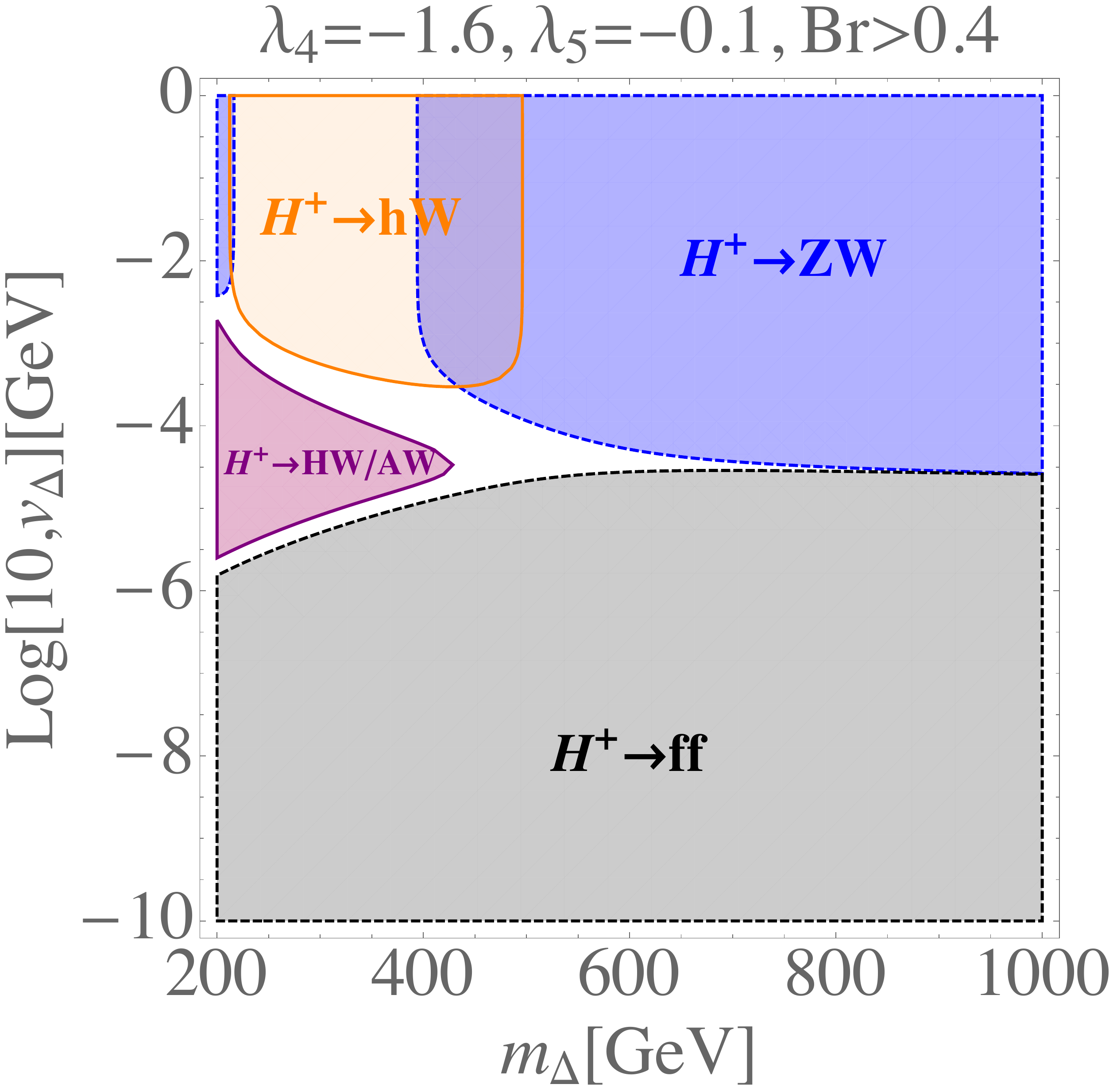} \\
  \includegraphics[width=75mm,height=65mm]{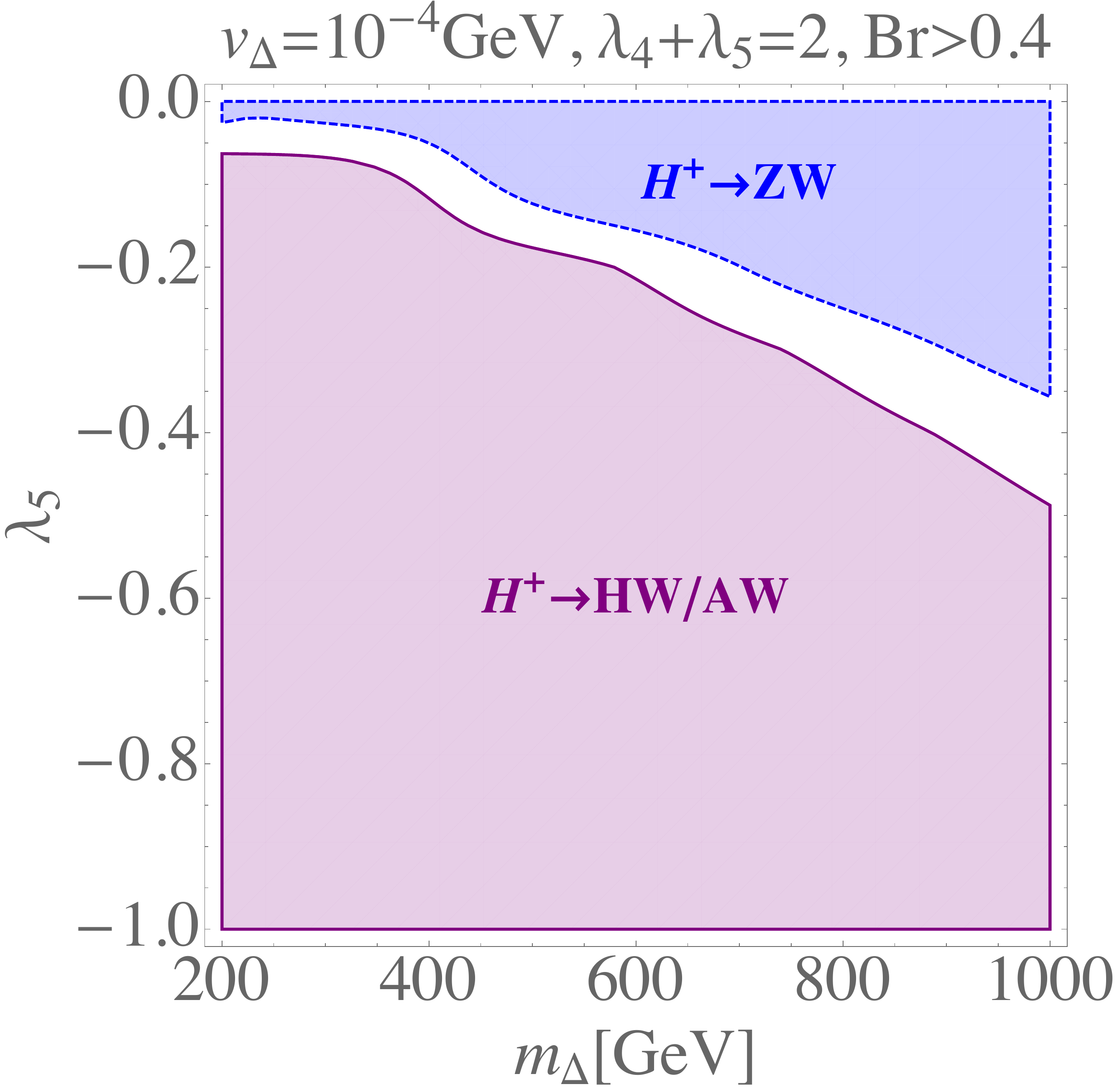} &   \includegraphics[width=75mm,height=65mm]{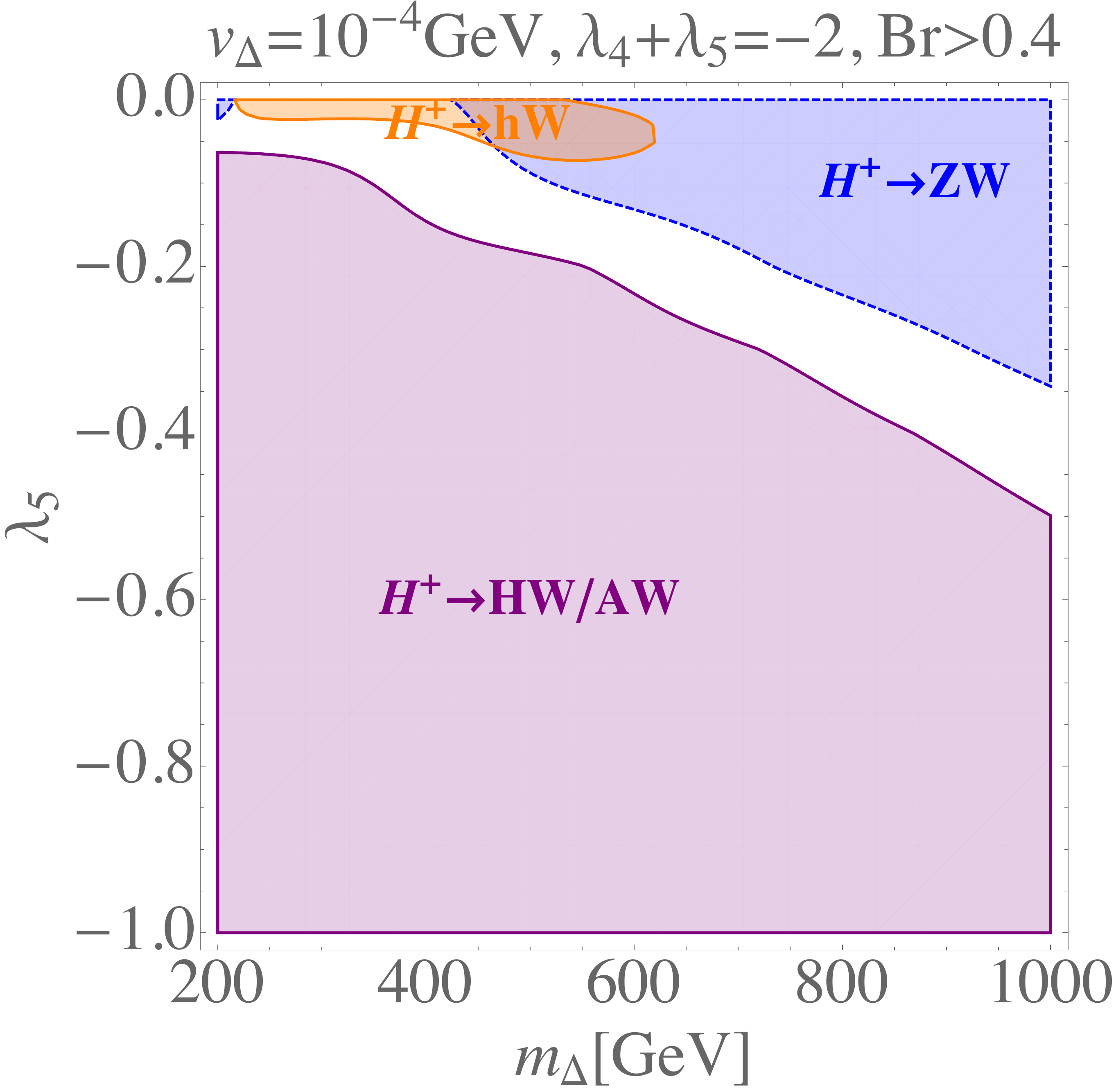}   \\
  \includegraphics[width=75mm,height=65mm]{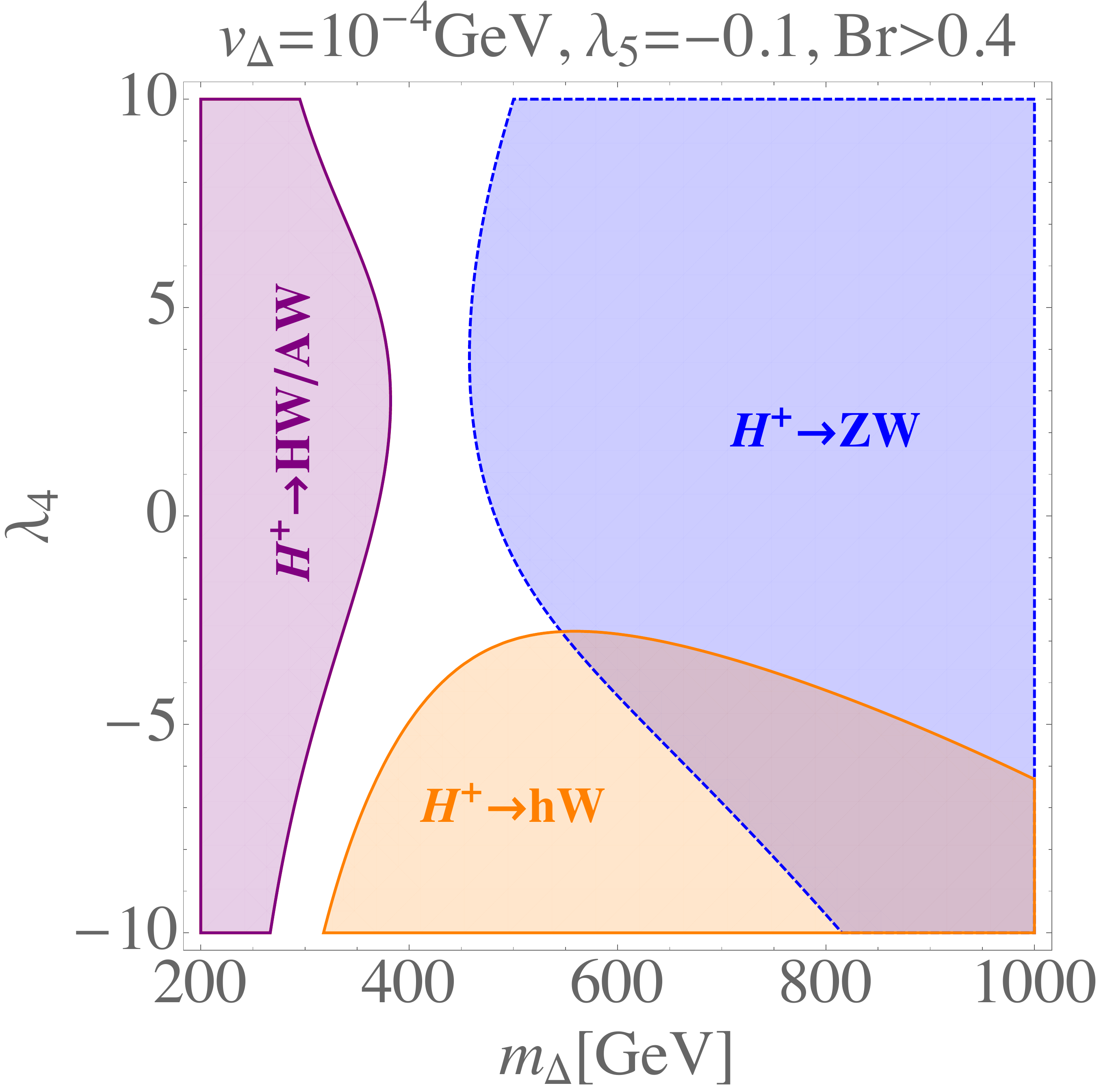} &   \includegraphics[width=75mm,height=65mm]{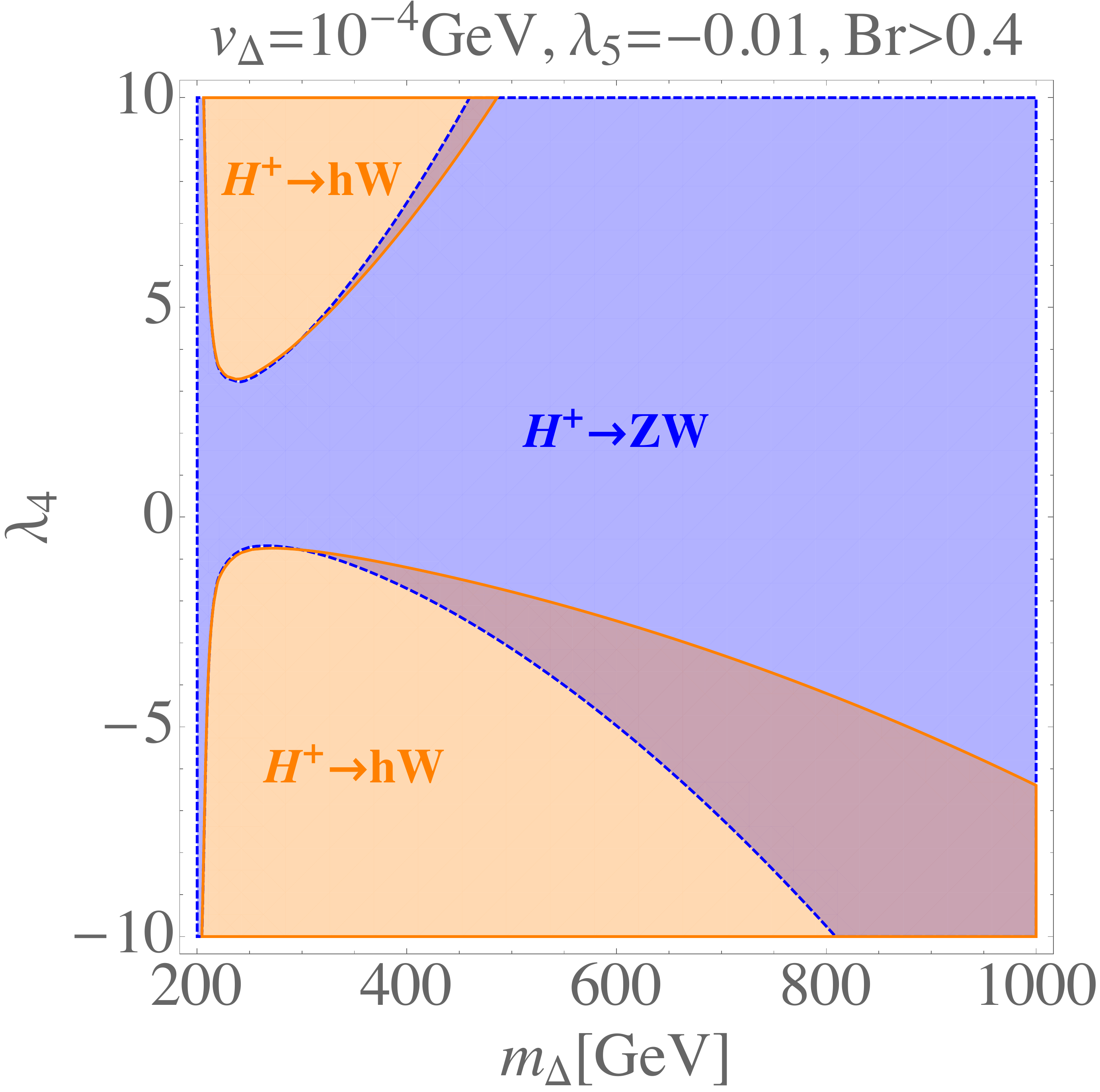} \\
\end{tabular}
\caption{Decay region plots for $H^{\pm}$ with $\rm BR\ge40\%$. Purple region is for $HW$ and $AW$, blue for $ZW$, orange for $hW$ and black for the lepton final state. The first row is with the same $\lambda_5$ but opposite-sign $\lambda_4$; the second row is with the same $v_\Delta$ but opposite-sign $\lambda_{45}$ and the third row is with the same $v_\Delta$ but different $\lambda_5$. From those plots we conclude that $H^\pm\rightarrow hW^\pm$ channel prefers $\lambda_{45}<0$ in general. For $\lambda_5=-0.01$,  $H^\pm\rightarrow hW^\pm$ also gains a large branching ratio when $\lambda_4$ goes from negative to positive as can be seen from the last graph.}\label{decayregionplotHP}
\end{figure}

From Fig.\,\ref{decayregionplotHPP}, we observe that for $H^{\pm\pm}$, the dominant decay channels are $H^{\pm\pm}\rightarrow \ell^\pm\ell^\pm~(W^\pm W^\pm)$ at small (large) $v_\Delta$ when $m_\Delta=400\rm\,GeV$.  For intermediate values of the triplet vev, {\it e.g.} $v_\Delta=10^{-4}\rm\,GeV$, those two channels dominate when $\lambda_5\ge-0.2$. Besides the large BRs in the corresponding regions of $v_\Delta$, additional advantages for these channels are: (1) Clean final states: Leptons in the final states are relatively easy to identify and analyze experimentally; (2) Absence of cascade decay: The $H^{\pm}W^\pm$ decay mode will introduce extra decay chains, {{making the final state more complicated. We emphasize, however, that even though the same-sign di-$W$ boson (di-lepton) channel dominates for large (small) $v_\Delta$, one may still probe the intermediate $v_\Delta$ region using the $\ell^\pm\ell^\pm$ and $W^\pm W^\pm$ channels. Although these channels have relatively small BRs in this $v_\Delta$ region,  we find that by combining these channels with information from other triplet Higgses, one could still explore this region without resorting to the $H^{\pm\pm}\to W^\pm H^\pm$ channel. This feature will become more apparent in our main discovery reach plot Fig.~\ref{bdtdis} and attendant discussion.}}


We also note in passing that at small $v_\Delta$, same-sign di-lepton channel dominates and actually has a $100\%$ decay BR. For those regions where the same-sign di-lepton channel has a $100\%$ decay BR, experimental constraints are  strong. We will discuss this point in detail in Sec.\,\ref{subsec:disbdt}.}

{{In Fig.\,\ref{decayregionplotHP}, we show the regions of parameter space where the $H^\pm$ decay BR to various final states is greater than 40\%. Since the BR functions for $H^\pm$ depend on $v_\Delta$, $m_\Delta$, $\lambda_4$ and $\lambda_5$ individually, the decay region plots for $H^\pm$ are more complicated than those for the doubly charged scalars. We thus plot the dominant decay channels in different planes: In the first row of Fig.\,\ref{decayregionplotHP}, we consider the ($v_\Delta$, $m_\Delta$) plane with varying $\lambda_{45}$, while in the second (third) row, we consider the ($v_\Delta$, $\lambda_{5(4)}$) plane with fixed $\lambda_{4(5)}$ and $v_\Delta$.}} Recall that from Table\,\ref{tab:1}, only the $H^\pm\rightarrow hW^\pm$ channel is related to the determination of $\lambda_4$ {{through the mixing angle $\sin\alpha$ as discussed in Sec.\,\ref{lam4determ}}}. {\color{black} We observe that $\lambda_{45}<0$ generally leads to a large BR for the $H^\pm\rightarrow hW^\pm$ channel, though there also exist some regions giving a large BR$(H^\pm\to hW^\pm)$ for $\lambda_{45}>0$.}

With the foregoing observations in mind, we will next study the following channels for model discovery: {{$p p \to H^{++}H^{--}$ and $p p \to H^{\pm\pm}H^{\mp}$ with $H^{\pm\pm}\rightarrow \ell^\pm\ell^\pm~(W^\pm W^\pm)$ and $H^{\mp}\rightarrow hW^\mp$.}}

\subsection{Present experimental constraints}
\label{subsec:sum}
Present experimental constraints on the charged Higgs particles we study here already exclude some portions of the CTHM  parameter space especially from studies on the $p p \to H^{++}H^{--} \to \ell^+\ell^-\ell'^-\ell'^-$ ($\ell=e,\mu$) process. Thus, before moving to the detailed collider study of some specific channels, we review the current direct LHC experimental constraints. A detailed summary can be found in Appendix \ref{app:expcon}, with the most stringent ones given below:
\begin{enumerate}
\item For $H^{\pm\pm}$: By assuming a 100\% di-lepton decay BR, the lower limit on $m_{H^{\pm\pm}}$ is constrained to be 870\,GeV\,\cite{Aaboud:2017qph} for a $\mu^\pm \mu^\pm$ final state. In Ref.\,\cite{ATLAS:2012mn}, an upper limit on the cross section with the $\ell^\pm\ell^\pm$ {{($\ell=e,\mu$)}} final state is set to be between 1.7\,fb and 67\,fb. While by assuming $H^{\pm\pm}$ is long-lived\footnote{As explained in the footnote of Ref.\,\cite{Aad:2015oga}, ``long-lived'' means a particle that does not decay within the full depth of the ATLAS detector.}, $m_{H^{\pm\pm}}\in[50,600]$\,GeV is excluded\,\cite{Aad:2015oga}.
\item For $H^\pm$: $\sigma(pp\rightarrow H^\pm t[b])\times\text{BR}(H^\pm\rightarrow \tau\nu)<\text{1.9\text{\,fb}-15}$\,fb for $m_H^\pm\in(200,2000)$\,GeV\,\cite{Aaboud:2016dig}, while for a VBF produced $H^\pm$, $\sigma(pp\rightarrow H^\pm+X)\times\text{BR}(H^\pm\rightarrow W^\pm Z)<\text{36\text{\,fb}-573}$\,fb for $m_H^\pm\in(200,2000)$\,GeV\,\cite{Sirunyan:2017sbn}. {{Here, a larger mass corresponds to a smaller upper bound on the product of the production cross section and the BR. A similar meaning is implied in the following.}}
\item For $H$ and $A$: In Ref.\,\cite{Khachatryan:2016are}, the upper limit on $\sigma(pp\rightarrow S' \rightarrow SZ)\times\text{BR}(S\rightarrow b\bar{b}(\tau^+\tau^-))\times\text{BR}(Z\rightarrow\ell^+\ell^-)$ ($S',~S$ are $H$ or $A$ with $m_{S'}>m_S$) is constrained to be 5\,fb-10\,fb for $\ell^+\ell^-\tau^+\tau^-$ final state with $m_{H/A}\in(500,1000)$\,GeV and $m_{A/H}\in(90,400)$\,GeV; while for $\ell^+\ell^-b\bar{b}$ final state, the upper limit is 1\,fb-100\,fb with $m_H\in[300,100000]$\,GeV. For the degenerate case, i.e., $m_A=m_H$, which is true in our case, the parameter space remains unexplored.
\end{enumerate}

For the charged Higgs particles, we will recast constraints from the charged Higgs particles to the parameter space of the CTHM in Sec.\,\ref{subsec:disbdt}, in which we show the part of the parameter space that is already ruled out by current experimental constraints for the benchmark point we choose.

\section{Model discovery}
\label{sec:modeldis}
As discussed in last section, $H^{++}H^{--}$ has the largest production cross section and will be the dominant discovery channel for the triplet model; $H^{\pm}H^{\mp\mp}$ has the second largest production cross section and is directly related to the determination of $\lambda_{4,5}$. In addition, since the same-sign di-lepton decay channel of the $H^{\pm\pm}$ particle is dominant only at small $v_\Delta$ from left panel of Fig.\,\ref{decayregionplotHPP} and the $H^\pm\rightarrow hW^\pm$ decay channel dominates at large $v_\Delta$ from first row of Fig.\,\ref{decayregionplotHP}, we expect these two channels to be complementary to each other to cover most of the model parameter space. Therefore, in this section, we will study in detail the discovery of the triplet model through these two channels, i.e., $pp\rightarrow H^{++}H^{--}$ and $pp\rightarrow H^{\pm\pm}H^{\mp}$ with $H^{\pm\pm}\rightarrow\ell^\pm\ell^\pm/W^\pm W^\pm$ and $H^\pm\rightarrow hW^\pm$.
\subsection{Discovery for small $v_\Delta$: $pp\rightarrow H^{++}H^{--}\rightarrow\ell^+\ell^+\ell'^-\ell'^-$}
The dominant discovery channel for the triplet model is $H^{++}H^{--}$ and the cleanest discovery process is $pp\rightarrow H^{++}H^{--}\rightarrow\ell^+\ell^+\ell'^-\ell'^-$. Several theoretical and experimental phenomenological studies of its LHC signatures have been performed\,\cite{Sui:2017qra, Ghosh:2017pxl,Dev:2017ouk,Perez:2008ha, Barger:1982cy, Gunion:1989in, Muhlleitner:2003me, Han:2007bk, Huitu:1996su, Dion:1998pw, Akeroyd:2005gt, Biswas:2017tnw, Aad:2012cg,Aaltonen:2011rta,ATLAS:2012mn,ATLAS:2012hi,Aad:2014hja,ATLAS:2014kca,Aad:2015oga,Sirunyan:2017ret,Aaboud:2017qph}. Recent related theoretical studies relevant to higher energy colliders include: (1) at a lepton collider with $\sqrt{s}=380\rm\,GeV$ and 3\,TeV, the production and decays of $H^{\pm\pm}$ were studied by Agrawal {\it et al.}\,\cite{Agrawal:2018pci}; (2) the $H^{++}H^{--}$ pair production cross section at the future 100\,TeV $pp$ collider was studied by Cai {\it et al.}\,\cite{Cai:2017mow}; (3) the $H^{++}H^{--}\rightarrow\tau^\pm\ell^\pm\ell^\mp\ell^\mp/\ell^+\tau^+\ell^-\tau^-$ processes were studied by Li\,\cite{Li:2018jns} at the high-luminosity and high-energy LHC as well as the future 100\,TeV circular $pp$ collider (FCC); (4) the multi-lepton final state of $H^{++}H^{--}$ at 13\,TeV LHC and FCC was studied by Mitra {\it et al.}\,\cite{Mitra:2016wpr} in the RMH by fixing $\lambda_{1}=0.13$ and $\lambda_2=\lambda_3=\lambda_4=1$. To the best of our knowledge, in the NMH this channel at the FCC has not yet been studied.

In what follows, we discuss our collider simulation for this channel  with a mass range from 40 GeV to 5000 GeV. The simulation is done by using {\tt MadGraph}\,2.3.3\,\cite{Alwall:2014hca} and the aforementioned pre-generated CTHM UFO file to generate events, and then each generated event undergoes parton shower and hadronization through {\tt Pythia-pgs}\,2.4.4\,\cite{Sjostrand:2006za} before arriving at the detector. The detector response is simulated by {\tt {\tt Delphes}}\,3.3.0\,\cite{deFavereau:2013fsa}, where the 100\,TeV FCC {\tt Delphes} card\,\cite{delphescard} is used at this step. To analyze the data collected by {\tt Delphes}, we use {\tt ROOT}\,6.06.02\,\cite{Antcheva:2009zz}. 

The dominant backgrounds for this channel are $ZW^\pm W^\mp$ and $ZZ$ as we are performing an exclusive analysis. In total, we generate 1,000,000 events for both the signal and the two backgrounds, and our preselection cuts for the signal and the backgrounds are: (1) transverse momentum $p_T>20$\,GeV for all final state particles; (2) absolute pseudorapidity $|\eta|<2.5$ for all final state particles. Since the Boosted Decision Trees (BDT)\,\cite{tmva} can maximize the cut efficiency and thus have better performance than a cut-based analysis\,\cite{Roe:2004na}, we will utilize this feature of BDT to train and test all the events that have passed the preselection cuts. We list the variables used during BDT training and test in Table \ref{bdtvar}: 
\begin{table}[!ht]
\caption{A list of BDT variables for the $p p \to H^{\pm\pm}H^{\mp\mp}\to\ell^+\ell^+\ell'^-\ell'^-$ signal and its backgrounds.}\label{bdtvar}
  \centering
  \begin{tabular}{c}\toprule[1.5pt]
    $\slashed{E}_T$: Missing transverse energy; $HT$: Scalar sum of transverse momentum\\
    $m_{H^{++}}$ : Positively doubly-charged Higgs mass, $m_{H^{--}}$ : Negatively doubly-charged Higgs mass\\
    $p_{T,\ell^{+}}^{\text{leading}}$, $p_{T,\ell^{+}}^{\text{sub-leading}}$: Transverse momentum of the $\ell^+$ with leading and sub-leading $p_T$ respectively\\
    $p_{T,\ell^{-}}^{\text{leading}}$, $p_{T,\ell^{-}}^{\text{sub-leading}}$: Transverse momentum of the $\ell^-$ with leading and sub-leading $p_T$ respectively\\
    $\Delta\phi_{\ell^+\ell^+}$, $\Delta R_{\ell^+\ell^+}$: $\Delta\phi$ and $\Delta R$ of the two positively charged leptons\\
    $\Delta\phi_{\ell^-\ell^-}$, $\Delta R_{\ell^-\ell^-}$: $\Delta\phi$ and $\Delta R$ of the two negatively charged leptons\\
    $m_{Z,1}$, $m_{Z,2}$: Two minimal combinations of the four leptons with same flavor and opposite charges\\
 \bottomrule[1.25pt]
\end{tabular}\par
\end{table}

\subsection{Discovery for large $v_\Delta$: $pp\rightarrow H^{++}H^{--}\rightarrow W^+W^+W^-W^-\to\ell^+\ell^+\ell'^-\ell'^-\slashed{E}_T$}

From the BR discussion in Sec.\,\ref{moddec}, we observe that the $H^{++}H^{--}\rightarrow\ell^+\ell^+\ell'^-\ell'^-$ channel can only cover the small $v_\Delta$ region, and we expect the large $v_\Delta$ region to be covered by the $pp\rightarrow H^{++}H^{--}\rightarrow W^+W^+W^-W^-$ channel. In this paper, we only focus on the $W^\pm\to\ell^\pm\nu_\ell$ mode for all the four $W$ bosons. In this case, the $4W$ channel has exactly the same backgrounds as the $H^{++}H^{--}\rightarrow\ell^+\ell^+\ell'^-\ell'^-$ channel considered in last sub-section. 

Repeating the same procedures as for the $H^{++}H^{--}\rightarrow\ell^+\ell^+\ell'^-\ell'^-$ channel, we generate 1,000,000 events for our signal and use the background data generated in last sub-section. We also use the same BDT training and test variables as those listed in Table\,\ref{bdtvar} to analyze this channel.

\subsection{Discovery for intermediate and large $v_\Delta$: $pp\rightarrow H^{\pm\pm}H^{\mp}\rightarrow\ell^\pm\ell^\pm hW^\mp\rightarrow \ell^\pm\ell^\pm b\bar{b}\ell^\mp\slashed{E}_T$ and $pp\rightarrow H^{\pm\pm}H^{\mp}\rightarrow W^\pm W^\pm hW^\mp\rightarrow \ell^\pm\ell^\pm b\bar{b}\ell^\mp\slashed{E}_T$}
While the $H^{++}H^{--}\rightarrow\ell^+\ell^+\ell'^-\ell'^-$ ($pp\rightarrow H^{++}H^{--}\rightarrow W^+W^+W^-W^-\to\ell^+\ell^+\ell'^-\ell'^-\slashed{E}_T$) only covers the small (large) $v_\Delta$ region, the $H^\pm H^{\mp\mp}$ can provide complementary discovery potential for the large and intermediate $v_\Delta$ region. To obtain information about $\lambda_{4,5}$, we require $H^{\pm}$ to decay into a $hW^\pm$ final state, while $H^{\mp\mp}$ can decay into either an $\ell^{\mp}\ell^{\mp}$ or a $W^\mp W^\mp$ final state. These two processes yield the same final state particles and, thus, share the same backgrounds. The backgrounds we consider for these two processes are: $hZW^\pm; t\bar{t}j$ and $W^\pm W^\mp b\bar{b}j$ with the light jet $j$ misidentified as a lepton with a fake rate of 0.01\%\,\cite{delphescard}; $t\bar{t}W^\pm$, $t\bar{t}Z$ and $ZZh$ with one lepton missing; $ZW^\pm jj$ with the two light jets misidentified as two $b$ quarks with a fake rate of 10\% for $c$ misidentified as $b$ and a 0.01\% fake rate for other light quarks\,\cite{delphescard}; and $ZW^\pm b\bar{b}$. The signals and the backgrounds are summarized below in Table \ref{table:s-bkg}. 

\begin{table}[htbp!]
\caption{Signals for intermediate and large $v_\Delta$ are listed in the first two rows. The two signals share the same backgrounds, which are listed in the following eight rows.}\label{table:s-bkg}
\centering
\begin{tabular}{c|c|c|c|c|c|c}
\hline
\multirow{2}{*}{\bf Signal} & \multicolumn{6}{|c}{$pp\rightarrow H^\mp H^{\pm\pm}\rightarrow hW^\mp \ell^\pm \ell^\pm\rightarrow b\bar{b}\ell'^\mp \ell^\pm \ell^\pm \slashed{E}_T$ (for intermediate $v_\Delta$)} \\
\cline{2-7}
& \multicolumn{6}{|c}{$pp\rightarrow H^\mp H^{\pm\pm}\rightarrow hW^\mp W^\pm W^\pm\rightarrow b\bar{b}\ell'^\mp \ell^\pm \ell^\pm \slashed{E}_T$ (for large $v_\Delta$)} \\
\hhline{=|=|=|=|=|=|=}
\multirow{8}{*}{\bf Background} & \multicolumn{6}{|c}{$pp\rightarrow hZW^\pm\rightarrow b\bar{b}\ell^+\ell^-\ell'^\pm \slashed{E}_T$}\\
\cline{2-7}
& \multicolumn{6}{|c}{$pp\rightarrow hZZ\rightarrow b\bar{b}\ell^+\ell^-\ell'^+\ell'^-$} \\
\cline{2-7}
& \multicolumn{6}{|c}{$pp\rightarrow ZW^\pm j j\rightarrow \ell^+\ell^-\ell'^\pm j j\slashed{E}_T$} \\
\cline{2-7}
& \multicolumn{6}{|c}{$pp\rightarrow t\bar{t}Z\rightarrow W^+bW^-\bar{b}\ell^+\ell^-\rightarrow b\bar{b}\ell'^+\ell''^-\ell^+\ell^- \slashed{E}_T$} \\
\cline{2-7}
& \multicolumn{6}{|c}{$pp\rightarrow ZW^\pm b\bar{b}\rightarrow b\bar{b}\ell^+\ell^-\ell'^\pm \slashed{E}_T$} \\
\cline{2-7}
& \multicolumn{6}{|c}{$pp\rightarrow W^+W^-b\bar{b}j\rightarrow b\bar{b}\ell^+\ell'^- j \slashed{E}_T$} \\
\cline{2-7}
& \multicolumn{6}{|c}{$pp\rightarrow t\bar{t}W^\pm\rightarrow W^+bW^-\bar{b}\ell^\pm\slashed{E}_T\rightarrow b\bar{b}\ell'^+\ell''^-\ell^\pm \slashed{E}_T$} \\
\cline{2-7}
& \multicolumn{6}{|c}{$pp\rightarrow t\bar{t}j\rightarrow W^+bW^-\bar{b}j \rightarrow b\bar{b}\ell'^+\ell''^-j \slashed{E}_T$} \\
\hline
\end{tabular}
\end{table}

\begin{table}[!ht]
\caption{A list of BDT variables for $W^\pm W^\pm hW^\mp$, $\ell^\pm\ell^\pm hW^\mp$ channels and their backgrounds. Since these two signals share the same backgrounds, we use the same BDT variables for both channels.}\label{sigbdtvar}
  \centering
  \begin{tabular}{c}\toprule[1.5pt]
    $\slashed{E}_T$: Missing transverse energy; $HT$: Scalar sum of transverse momentum\\
    $m_{H^{\pm\pm}}$ : Doubly-charged Higgs mass\\
    $m_h$, $m_Z$: SM Higgs and $Z$ boson mass; $m_{W,T}$: transverse mass of $W^\mp$ boson\\
    $\Delta\phi_{b\bar{b}}$, $\Delta R_{b\bar{b}}$: $\Delta\phi$ and $\Delta R$ of two $b$ quarks; $\Delta\phi_{\ell^\pm\ell^\pm}$, $\Delta R_{\ell^\pm\ell^\pm}$: $\Delta\phi$ and $\Delta R$ of two same-sign leptons\\
    $p_{T,b}^{\text{leading}}$, $p_{T,b}^{\text{sub-leading}}$: leading and sub-leading transverse momentum of the $b$ quark\\
    $\eta_{b}^{\text{leading}}$, $\eta_{b}^{\text{sub-leading}}$: pseudo-rapidity of the $b$ quark with leading and sub-leading $p_T$ respectively\\
    $p_{T,\ell^{\text{same}}}^{\text{leading}}$, $p_{T,\ell^{\text{same}}}^{\text{sub-leading}}$: leading and sub-leading transverse momentum of the same-sign leptons\\
    $\eta_{\ell}^{\text{leading}}$, $\eta_{\ell}^{\text{sub-leading}}$: pseudo-rapidity of the same-sign leptons with leading and sub-leading $p_T$ respectively\\
    $\eta_{\ell^{\text{oppo.}}}$, $p_{T,\ell^{\text{oppo.}}}$: pseudo-rapidity and transverse momentum of the opposite-sign lepton\\
 \bottomrule[1.25pt]
\end{tabular}\par
\end{table}

As for the $H^{++}H^{--}$ process, we use the same tools to generate events, the same preselection cuts to analyze the events. For the BDT training and test, the training variables we use for these two processes and the backgrounds are listed in Table \ref{sigbdtvar}. In addition, for the $t\bar{t}j$, $W^\pm W^\mp b\bar{b}j$ and $ZW^\pm jj$ backgrounds, {{we also add the following cuts at the generator level}}: (1) $p_T^{j,b}\ge10$\,GeV; (2) $|\eta^{j,b}|\le5$; (3)$\Delta R^{jj,bb,bj}\ge0.05$. With these requirements, in total, we generate 50,000,000 events for signal $\ell^\pm\ell^\pm hW^\mp$ and 1,000,000 events for signal $W^\pm W^\pm hW^\mp$; 4,579,172 events for $W^\pm W^\mp b\bar{b}j$; 5,000,000 events for $ZZh$ and $ZhW^\pm$; 29,000,000 events for $t\bar{t}Z$; 30,000,000 events for $t\bar{t}W^\pm$ and $t\bar{t}j$; 15,000,000 events for $ZW^\pm jj$ and $ZW^\pm bb$.

\subsection{Discovery potential at the 100\,TeV collider}\label{subsec:disbdt}
The significance is defined as $\frac{S}{\sqrt{S+B}}$ throughout the paper, with $S=\sigma_s\cdot\mathcal{L}$ and $B=\sigma_{\text{bkg}}^{\text{tot}}\cdot\mathcal{L}$ the total signal and background event number at the collider, where $\sigma_s$ and $\sigma_{\text{bkg}}^{\text{tot}}$ are the final signal and final total background cross section respectively, and $\mathcal{L}$ is the integrated luminosity, which we choose to be $30\,\text{ab}^{-1}$\,\cite{fcc1,fcc2} throughout the paper. By requiring the signal significance to be greater or equal to 5, the BDT based result for the discovery channels is given in Fig.\,\ref{bdtdis}. Several features of these results merit emphasizing:
\begin{itemize}
\item We see that at small $v_\Delta$ where the neutrino masses are naturally generated through the type-II seesaw mechanism, the CTHM can be discovered over a very wide mass range from tens of \,GeV to several \,TeV through the $pp\rightarrow H^{++}H^{--}\rightarrow\ell^+\ell^+\ell'^-\ell'^-$ channel.
We also recast the current LHC constraints for this channel at 8\,TeV and 13\,TeV, which is done by rescaling the production cross sections and the BRs in Refs.\,\cite{ATLAS:2014kca,Aaboud:2017qph}. We find that the current LHC constraints only exclude the relatively small  $m_\Delta$ and small $v_\Delta$ region of the CTHM parameter space for our benchmark point, which therefore motivates a future collider study as we have done above. 

\item For the benchmark point we use, 
\begin{align}
m_{H^{\pm\pm}}=m_\Delta^2+3001{\rm\,(GeV^2)}\Rightarrow m_{H^{\pm\pm}}\gtrsim54.78\rm\,GeV,
\end{align}
such that LEP constraints\,\cite{Acton:1992zp,Abbiendi:2001cr} are automatically satisfied. Note that our Fig.\,\ref{bdtdis} is plotted as a function of $m_\Delta$ such that $m_\Delta=0$ corresponds a minimal mass of $m_{H^{\pm\pm}}\simeq54.78$\,GeV.

\item For the large $v_\Delta$ region, the $pp\rightarrow H^{\pm\pm}H^\mp\rightarrow W^\pm W^\pm hW^\mp$ channel allows discovery of the CTHM up to about 1 TeV. The LHC constraints for this channel are currently absent, and the corresponding parameter space will be covered by the future 100\,TeV collider. {{In addition, for intermediate $v_\Delta$'s, the overlap among $W^\pm W^\pm hW^\mp$, $\ell^\pm \ell^\pm hW^\mp$ and $H^{++}H^{--}$ channels can also allow us to roughly determine $m_\Delta\in[400,1000]$\,GeV and $v_\Delta\in[10^{-4.4},10^{-3.9}]$\,GeV if all these three channels are observed with significance 5 or more. The redundancy among these models would provide an important cross check that the signals are due to the CTHM.} }

\item {\color{black}For large $v_\Delta$ and large $m_\Delta$ region where the $H^{\pm\pm}\rightarrow W^\pm W^\pm$ channel dominates as can be seen from left panel of Fig.\,\ref{decayregionplotHPP}, one would expect the $H^{++}H^{--}\rightarrow W^+W^+W^-W^-\to \ell^+\ell^+\ell'^-\ell'^-\slashed{E}_T$ channel to cover much of that parameter space. {\color{black} Although our present analysis is not optimized to extend beyond $m_\Delta\sim 1.6$ TeV for this channel, one might expect use of other $W$ decay modes (and a correspondingly different BDT training) to allow extension to higher masses. As an example, we note that the authors in Ref.\,\cite{kang:2014jia} have studied the channel $p p \to H^{++}(\to W^+\ell\nu_\ell)H^{--}(\to W^- jj)$ and concluded that $H^{\pm\pm}$ could be discovered at the 14\,TeV LHC with $\mathcal{L}=10\text{-}30\rm\, fb^{-1}$. It is worth exploring whether use of this channel (or others) may also afford greater coverage for $m_\Delta \gsim 1.6$ TeV. }}

\begin{figure}[thb!]
\captionstyle{flushleft}
\begin{tabular}{c}
\includegraphics[scale=0.4]{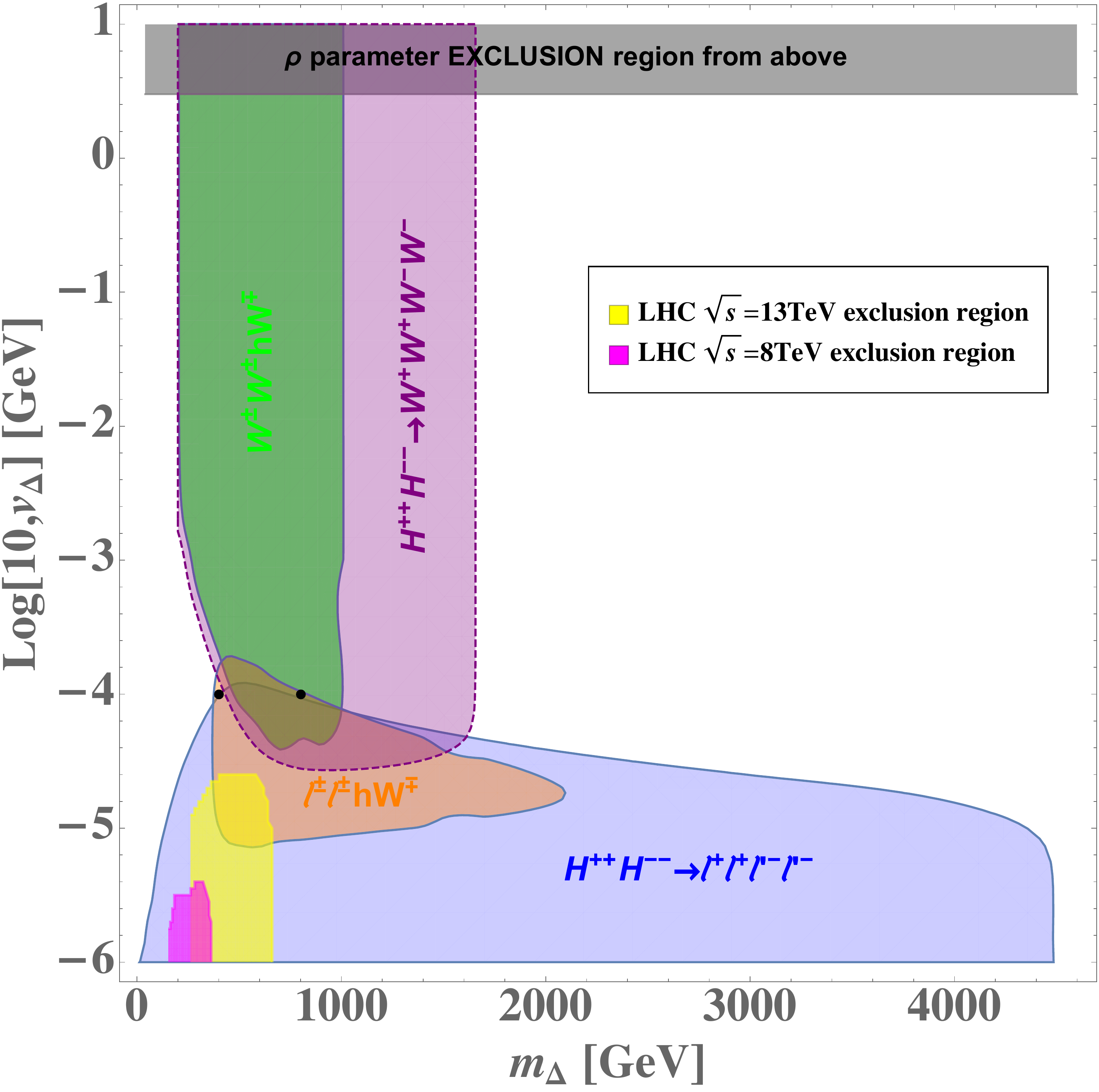}
\end{tabular}
\caption{{\color{black}Regions of significance $\ge 5\sigma $ in the $m_\Delta{-}v_\Delta$ plane with $m_{\nu_\ell}=0.01$\,eV ($\ell=e,\mu,\tau$), $\lambda_4=0$, $\lambda_5=-0.1$ and integrated luminosity of 30\,$\rm ab^{-1}$: The blue region corresponds to discovery using the $pp\rightarrow H^{++}H^{--}\rightarrow\ell^+\ell^+\ell'^-\ell'^-$ channel; the brown region is for the $H^{\pm\pm}H^{\mp}\rightarrow $ $\ell^\pm \ell^\pm$ $ hW^\mp$ channel ; the green region gives discovery using the  $H^{\pm\pm}H^{\mp}\rightarrow W^\pm W^\pm$ $hW^\mp$ mode. The yellow and magenta regions indicate the current LHC exclusion limits at $\sqrt{s}=13$ $\rm\,TeV$\,\cite{Aaboud:2017qph} and  $\sqrt{s}=8$\,\,TeV\,\cite{ATLAS:2014kca}, respectively. LEP constraints\,\cite{Acton:1992zp,Abbiendi:2001cr} are automatically satisfied since our benchmark point corresponds to $m_{H^{\pm\pm}}\gtrsim54.78$\,GeV. See the main text for a detail discussion. The black dots show two benchmark values of $m_\Delta$ used for Higgs portal coupling determination (see Section \ref{sec:lam45}).}}\label{bdtdis}
\end{figure}
\item {{One may also consider using the $H^{\pm\pm}H^{\mp\mp}\rightarrow W^\pm W^\pm\ell^\mp\ell^\mp$ channel to cover part of the parameter space. We note, however, that since the same-sign di-$W$ and the same-sign di-lepton decay channels are dominant only at large and small $v_\Delta$ respectively (as can be seen from the left panel of Fig.\,\ref{decayregionplotHPP}), we thus expect these channels to have enough significance only at $v_\Delta\sim(10^{-5},10^{-4})$\text{\,GeV}. The same region is already well covered by the $\ell^\pm\ell^\pm hW^\mp$ and $H^{++}H^{--}\rightarrow\ell^\pm\ell^\pm\ell'^\mp\ell'^\mp$ channels. 
}}

\item The $H^{++}H^{--}$ channel covers a very wide range over $m_\Delta$ at small $v_\Delta$ and the $W^\pm W^\pm$ $hW^\mp$ channel disappears around $m_\Delta=$1\,TeV. The reason for the ``long tail'' of the $H^{++}H^{--}$ channel can be understood from the blue region in Fig.\,\ref{bdtdisexplain}\,(a), from which we see that the $\rm BR(H^{\pm\pm}\rightarrow\ell^\pm\ell^\pm)$ decreases slowly with increasing $m_\Delta$ for $v_\Delta\lesssim10^{-4}$\,GeV, leading to a very slowly decreasing significance. In contrast, for the $W^\pm W^\pm$ $hW^\mp$ channel, the significance drops dramatically at $m_\Delta\approx$1\,TeV because of phase space suppression for heavier particles and decay BR suppression at smaller $v_\Delta$'s as can be seen from Fig.\,\ref{bdtdisexplain}(b).

\item {{We  remind the reader that we choose $m_\nu=0.01$\,eV for all the three light neutrinos generation throughout the paper. Since a larger (smaller) $m_\nu$ will correspond to a larger (smaller) Yukawa coupling and thus a larger (smaller) same-sign di-lepton decay BR of $H^{\pm\pm}$, we therefore expect the same-sign di-lepton decay regions in Fig.\,\ref{bdtdis} will shift upward (downward) for larger (smaller) $m_\nu$'s.

\item {\color{black}Finally, for our benchmark point, vacuum stability is not guaranteed at the Planck scale as discussed in Sec.\,\ref{secallconst}. In Ref.\,\cite{Chao:2012mx}, it was shown that vacuum stability up to the Planck scale actually prefers positive $\lambda_4$'s as indicated by the black arrow in the left panel of Fig.\,\ref{stability}. This difference is not, in general, problematic, as the stability region for our benchmark point amply covers the triplet mass range considered here. One could anticipate additional degrees of freedom modifying the behavior of the potential at larger scale, so as to ensure stability to the Planck scale. Nevertheless, it is interesting to ask how the reach indicated in Fig.~\ref{bdtdis} would evolve as we move along the black arrow in Fig.~\ref{stability} corresponding to higher stability scales.
We expect the discovery regions including the $H^\pm\to hW^\pm$ channel in Fig.\,\ref{bdtdis} to shrink for $0\lesssim\lambda_4\lesssim3$ as the $H^\pm\to hW^\pm$ decay BR decreases for $\lambda_4$'s in this region as can be seen directly from the upper right panel of Fig.\,\ref{slicedecay}. For $\lambda_4\gtrsim6$, one would expect the discovery regions including the $H^\pm\to hW^\pm$ chain to expand even though one needs to re-consider all the model constraints discussed in Sec.\,\ref{model:const}. For these larger values of the $\lambda_4$, however, we would expect to reach the limit of perturbativity well below the stability scale.}
}}

\end{itemize} 

\begin{figure}[thb!]
\captionstyle{flushleft}
\begin{tabular}{cc}
\includegraphics[scale=0.22]{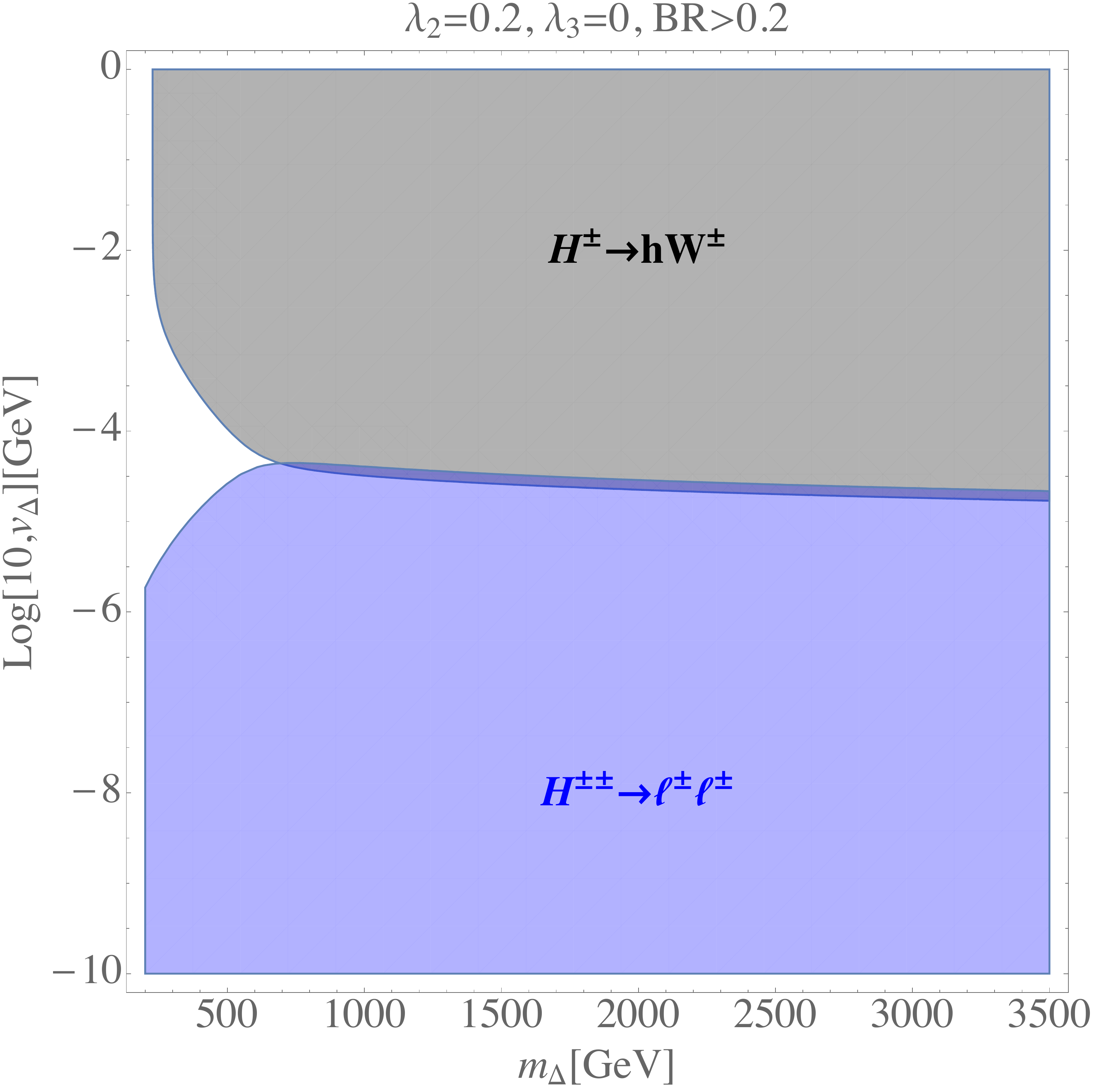} ~~&~~  \includegraphics[scale=0.22]{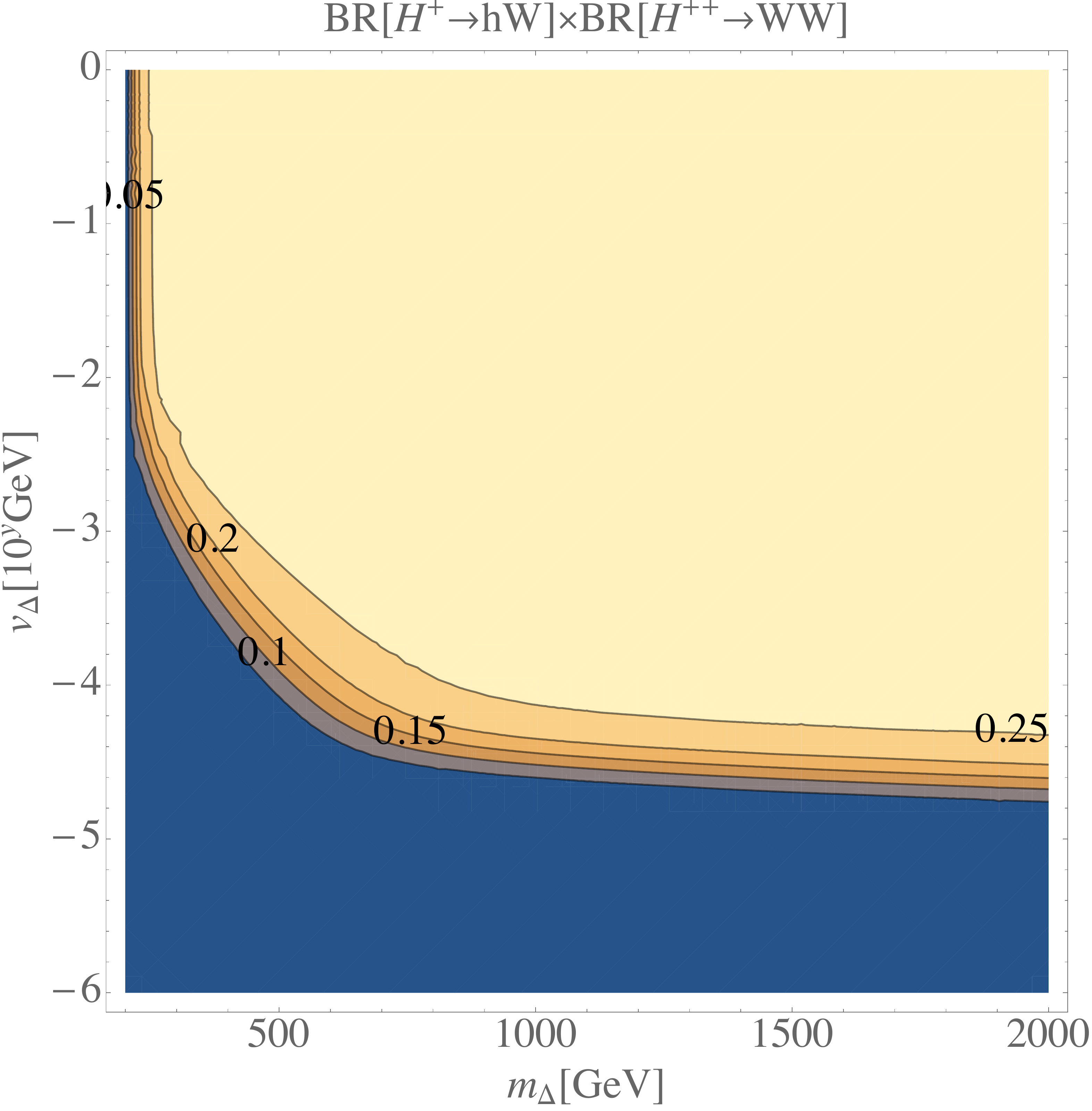}\\
(a) & (b)
\end{tabular}
\caption{Decay BRs for $\lambda_4=0$, $\lambda_5=-0.1$ and $m_\nu=0.01$\,eV. Figure (a): Decay BR$\ge20\%$ regions for $H^\pm\to hW^\pm$ and $H^{\pm\pm}\to\ell^\pm\ell^\pm$ channels. The slowly decreasing $\rm BR(H^{\pm\pm}\rightarrow \ell^\pm \ell^\pm)$ with increasing $m_\Delta$ explains the ``long-tail'' of the significance plot for $H^{++}H^{--}\rightarrow\ell^+\ell^+\ell'^-\ell'^-$ in Fig.\,\ref{bdtdis}. Figure (b): The solid lines indicate constant contours for $\rm BR(H^\pm\rightarrow hW^\pm)\times BR(H^{\mp\mp}\rightarrow W^\mp W^\mp)$. Product of the BRs is suppressed for small $v_\Delta$'s, which explains feature of the $W^\pm W^\pm$ $hW^\mp$ channel in Fig.\,\ref{bdtdis} in the small $v_\Delta$ region.}\label{bdtdisexplain}
\end{figure}

\section{Triplet Higgs potential determination and simulation}
\label{sec:lam45}
{\color{black}{From our result in the previous section, for $m_\Delta\lesssim4500$\,GeV, the $H^{++}H^{--}\to\ell^+\ell^+\ell'^-\ell'^-$, $W^\pm W^\pm hW^\mp$ and $\ell^\pm \ell^\pm hW^\mp$ channels can cover a significant portion of the parameter space of the CTHM except the region where $m_\Delta\gtrsim1$\,TeV and $v_\Delta\gtrsim10^{-4}$\,GeV. We expect  some of the latter region  to be covered by the $H^{++}H^{--}\to W^+W^+W^-W^-$ channel as discussed in last section. Therefore, the discovery potential for the CTHM at a  100\,TeV $pp$ collider is considerable. Assuming discovery of the doubly- and singly-charged scalars, we can fix $\lambda_5$ straightforwardly through the mass splitting as discussed in Sec.\,\ref{subsec:input}. However, to determine the important Higgs portal parameter $\lambda_4$, \color{black} additional information will be needed. For $v_\Delta$ larger than $\sim 10^{-5}$ GeV, the BR for $H^\pm\rightarrow hW^\pm$ is particularly useful
as discussed in Sec.\,\ref{lam4determ}\footnote{Note that, according to Fig.\,\ref{bdtdis} for $v_\Delta$ below $\sim10^{-5}$GeV, the $\ell^\pm \ell^\pm hW^\mp$ and the $W^\pm W^\pm hW^\mp$ channels will not be observable. In this region, one would need to explore other possible channels in order to determine $\lambda_{4,5}$.}. 

To investigate this possibility, we adopt the following strategy. First, we will carry out a detailed simulation for a choice of $\lambda_4+\lambda_5$ in the region where the BR$(H^\pm\rightarrow hW^\pm)$ is strongly-dependent on $\lambda_4+\lambda_5$, according to the top right panel of Fig.~\ref{slicedecay}. We will carry out this study for a choice of the $\lambda_j$ consistent the the stability and perturbativity considerations discussed above and for two different choices of the overall triplet mass scale, $m_\Delta$. Second, we will scan over the values of $\lambda_4$ and $m_\Delta$ for fixed $\lambda_5$, thereby varying the production cross section and BR from the values corresponding to our benchmark points. In doing so, we will rescale the significance of the signal accordingly. Third, we will repeat this analysis for different representative choices of $v_\Delta$ to indicate how the varying $H^{\pm\pm}$ BR affects the $\lambda_4$-sensitivity. Finally, we will compare the sensitivity with that of the observation of the rate for the SM Higgs boson to decay to a di-photon pair, as loop corrections from charged triplet scalars will affect the corresponding rate as functions of the Higgs portal couplings and $m_\Delta$. The results are plotted in Fig.\,\ref{haa}, where we show the corresponding regions of $5\sigma$ sensitivity to the model parameters. 

In what follows, we provide a more detailed discussion of the collider simulation and analysis than we provided for the results in Fig.~\ref{bdtdis}, given that we focus on the $H^\pm\rightarrow hW^\pm$ channel for coupling determination. 

}}

\subsection{Benchmark points}
\label{subsec:BMandS}

{\color{black} As discussed in Sec.\,\ref{moddec}, the $H^{++}H^{--}\rightarrow\ell^+\ell^+\ell'^-\ell'^-$ channel is powerful for the triplet model discovery at small $v_\Delta$, but it can not determine $\lambda_4$ as it is $\lambda_4$-independent. In contrast, $H^\mp H^{\pm\pm}\rightarrow hW^\mp \ell^\pm \ell^\pm/hW^\mp W^\pm W^\pm$ are promising for the determination of $\lambda_{4}$ at intermediate and large $v_\Delta$. In order to determine their collider signatures, we choose two representative benchmark points, taking into account vacuum stability, perturbative unitarity, perturbativity, neutrino masses and our result in Fig.\,\ref{bdtdis}:   $m_\Delta=800$\,GeV ($m_\Delta=400$\,GeV), $v_\Delta=10^{-4}$\,GeV, $m_h=125$\,GeV, $m_Z=91.1876$\,GeV, $m_\nu=0.01$\,eV, $\lambda_2=0.2$, $\lambda_3=0$, $\lambda_4=0$, $\lambda_5=-0.1$ for the $W^\pm W^\pm hW^\mp$ ($\ell^\pm \ell^\pm hW^\mp$) channel, which is a representative point of the large (small) $m_\Delta$ region. Note that although these benchmark parameter choices have $\lambda_4=0$, the sum $\lambda_4+\lambda_5$ differs from zero and lies in a region where BR($H^\pm\to hW^\pm)$ varies significantly with this combination of couplings. The choice of two the two different mass scales corresponds to the edges of various overlapping discovery regions, as indicated by the two black points in Fig.~\ref{bdtdis}.}

\subsection{Simulation: $pp\rightarrow H^\mp H^{\pm\pm}\rightarrow hW^\mp \ell^\pm \ell^\pm\rightarrow b\bar{b}\ell^\mp \ell^\pm \ell^\pm \slashed{E}_T$ for intermediate $v_\Delta$}
\label{sec:simu}
{\color{black}{In this section we will first generate data for $pp\rightarrow H^\mp H^{\pm\pm}\rightarrow hW^\mp \ell^\pm \ell^\pm\rightarrow b\bar{b}\ell^\mp \ell^\pm \ell^\pm \slashed{E}_T$ using {\tt MadGraph}, and then analyze the data by both a cut-based analysis and using the BDT method. In the former, we choose a set of ``hard cuts" by first comparing various signal and background distributions and endeavoring to optimize by hand the choice for greatest signal significance. As an alternative, we employ the BDT. As we show below, the BDT method generally provides a better signal efficiency and significance.
}}

\subsubsection{Cut based analysis: basic cuts}\label{basiccuts}
\begin{figure}[thb!]
\captionstyle{flushleft}
\begin{tabular}{cc}
 \includegraphics[width=90mm]{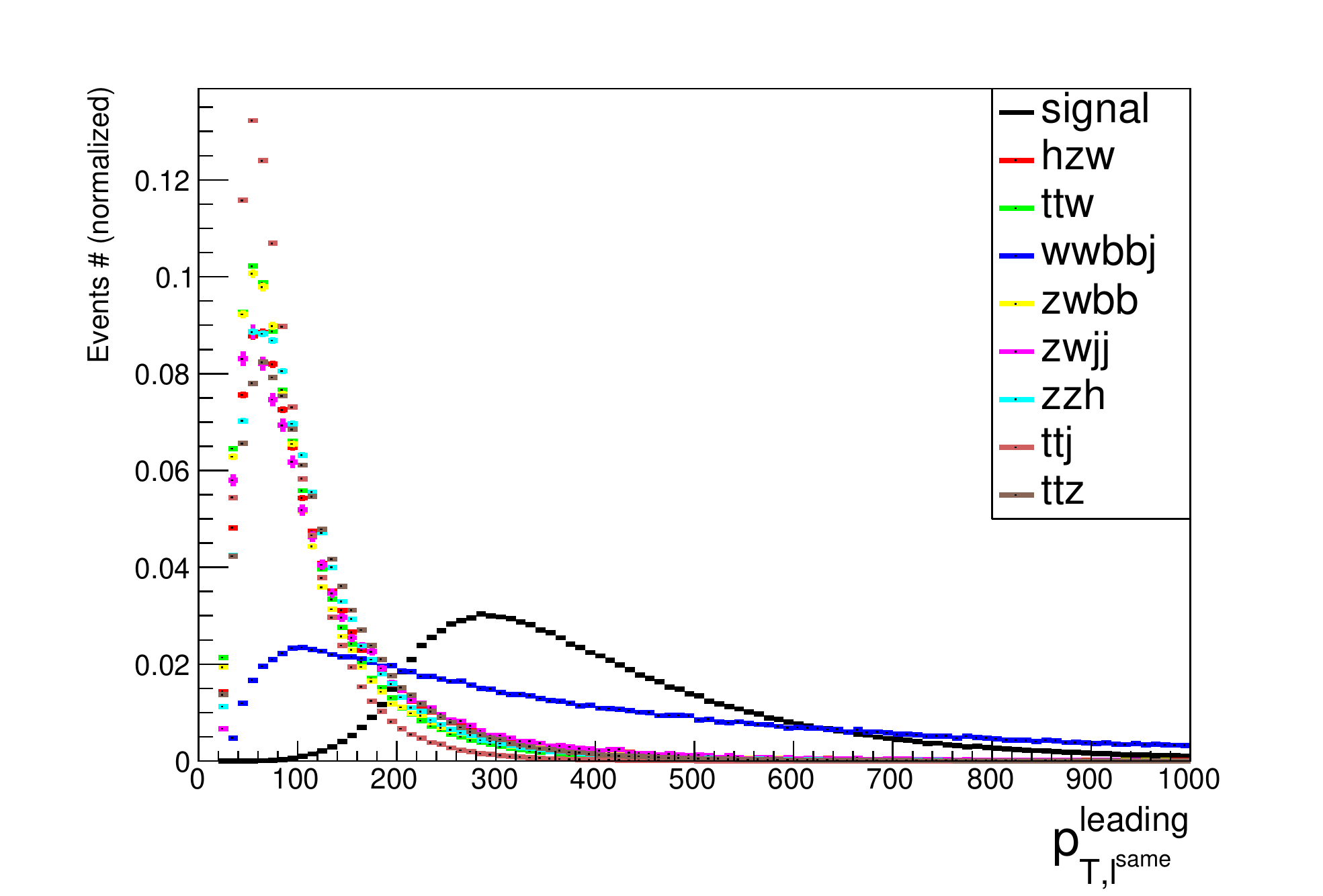} & \includegraphics[width=90mm]{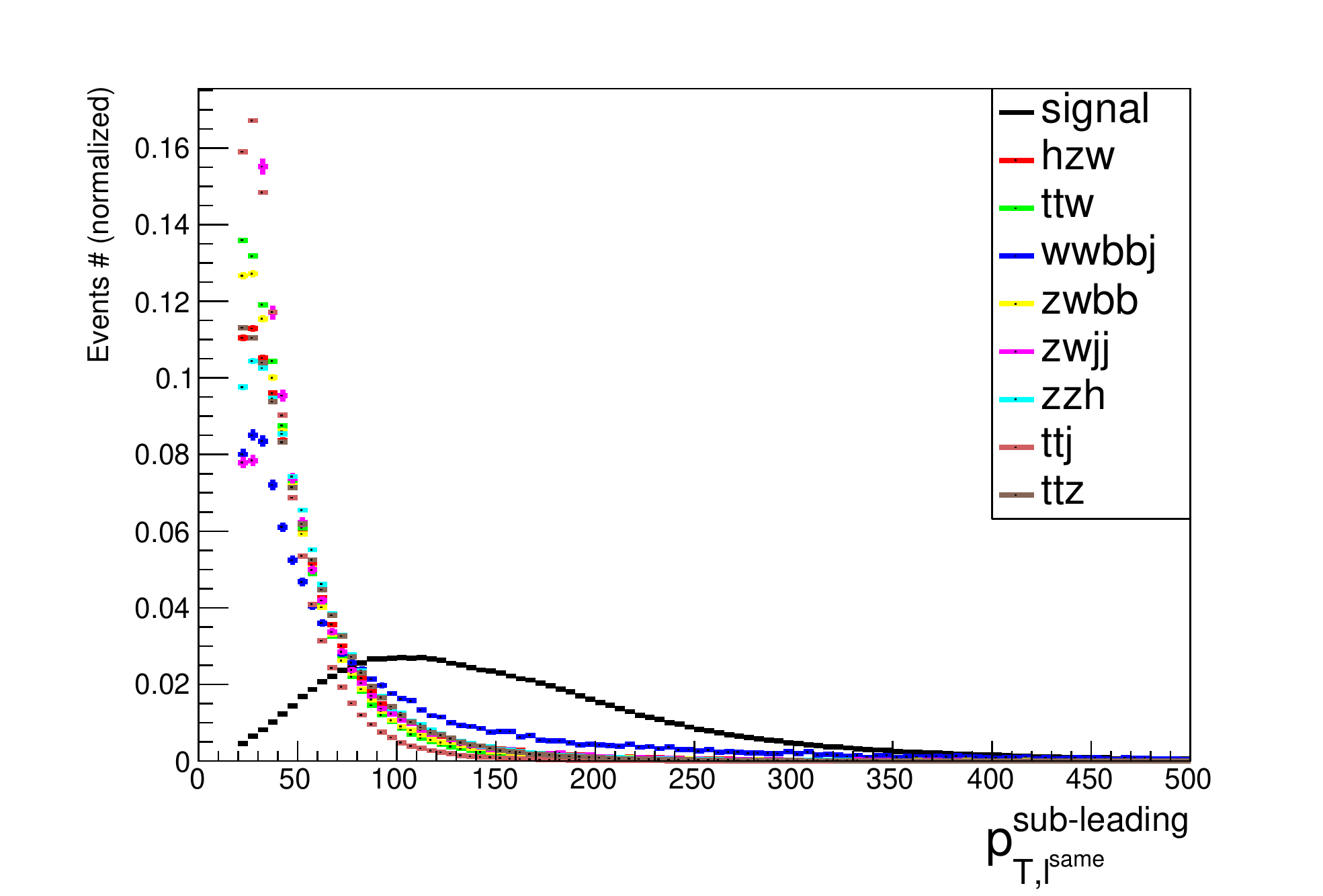} \\
 (a) same-sign lepton leading $p_T$ & (b) same-sign lepton sub-leading $p_T$ \\[6pt]
 \includegraphics[width=90mm]{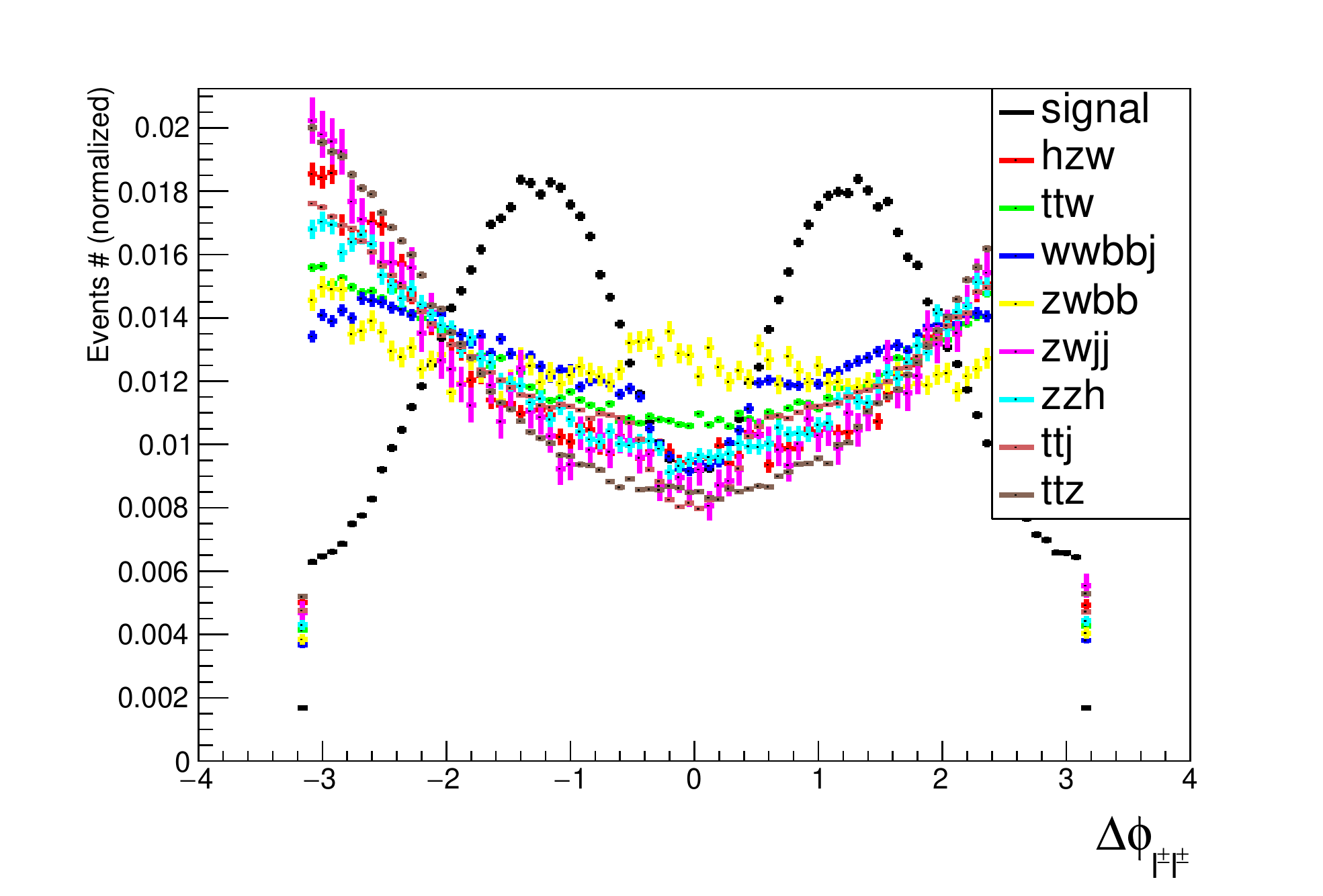} & \includegraphics[width=90mm]{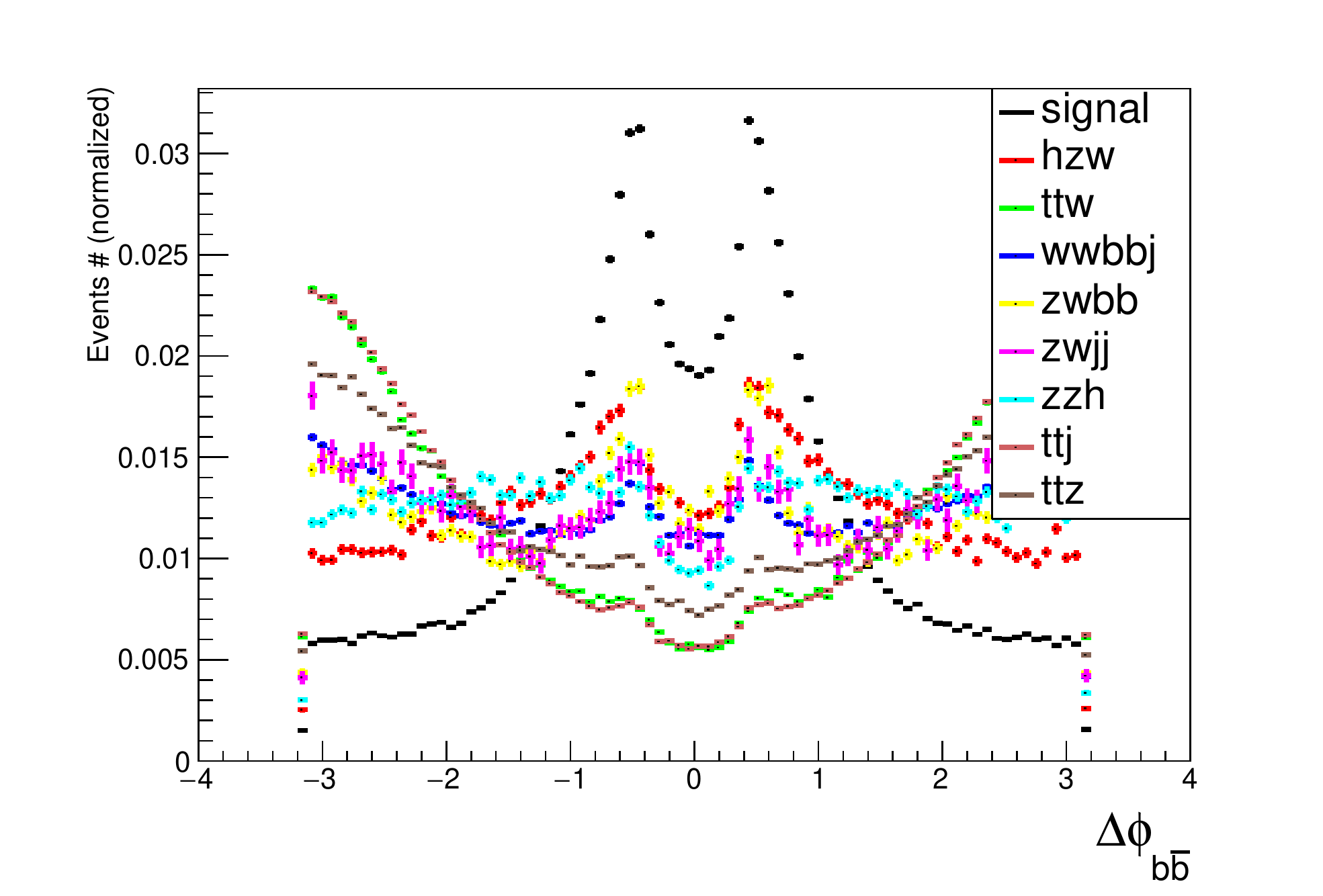}  \\
 (c) same-sign lepton $\Delta\phi$ & (d) $\Delta\phi$ of two $b$ quarks \\[6pt]
 \includegraphics[width=90mm]{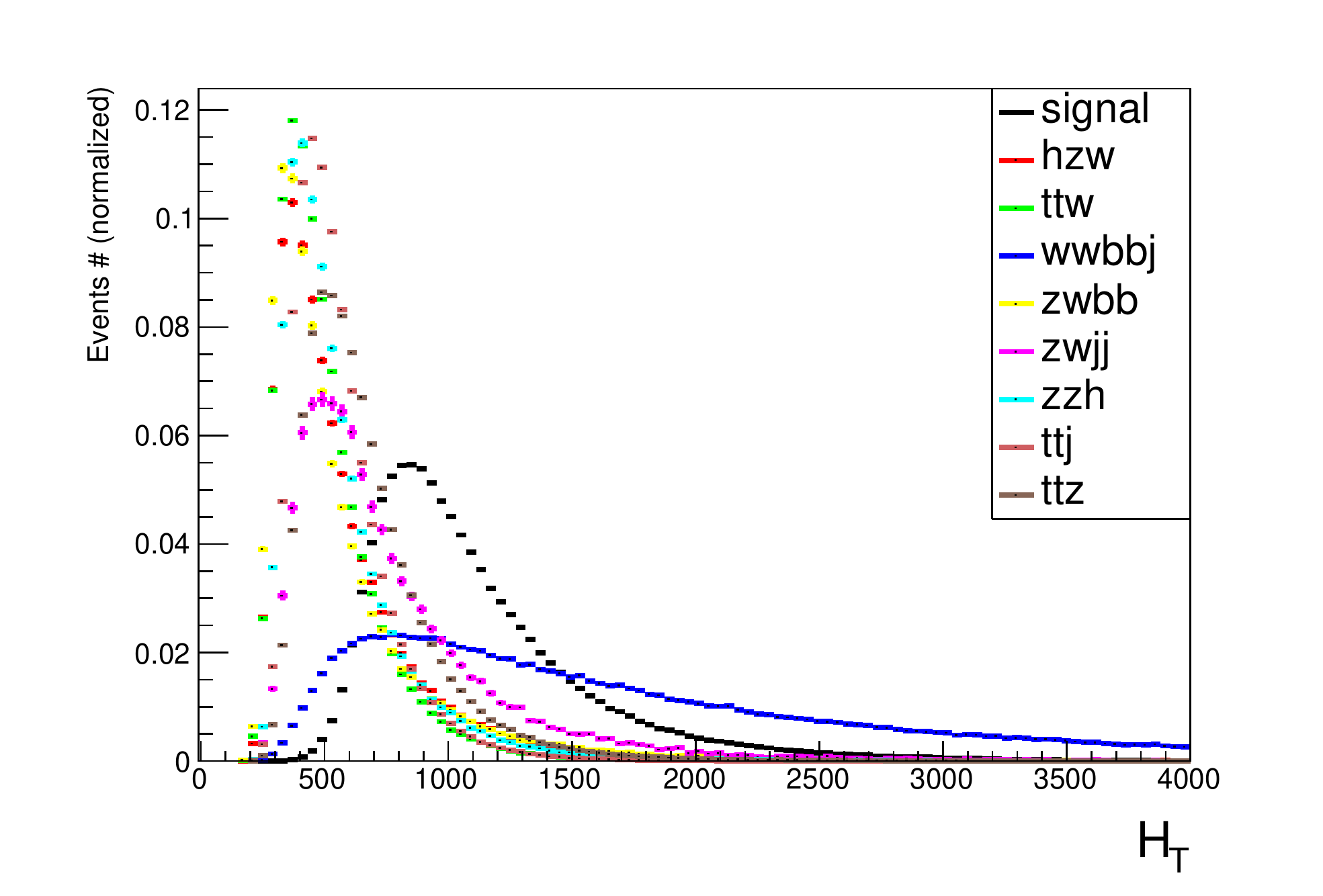} & \includegraphics[width=90mm]{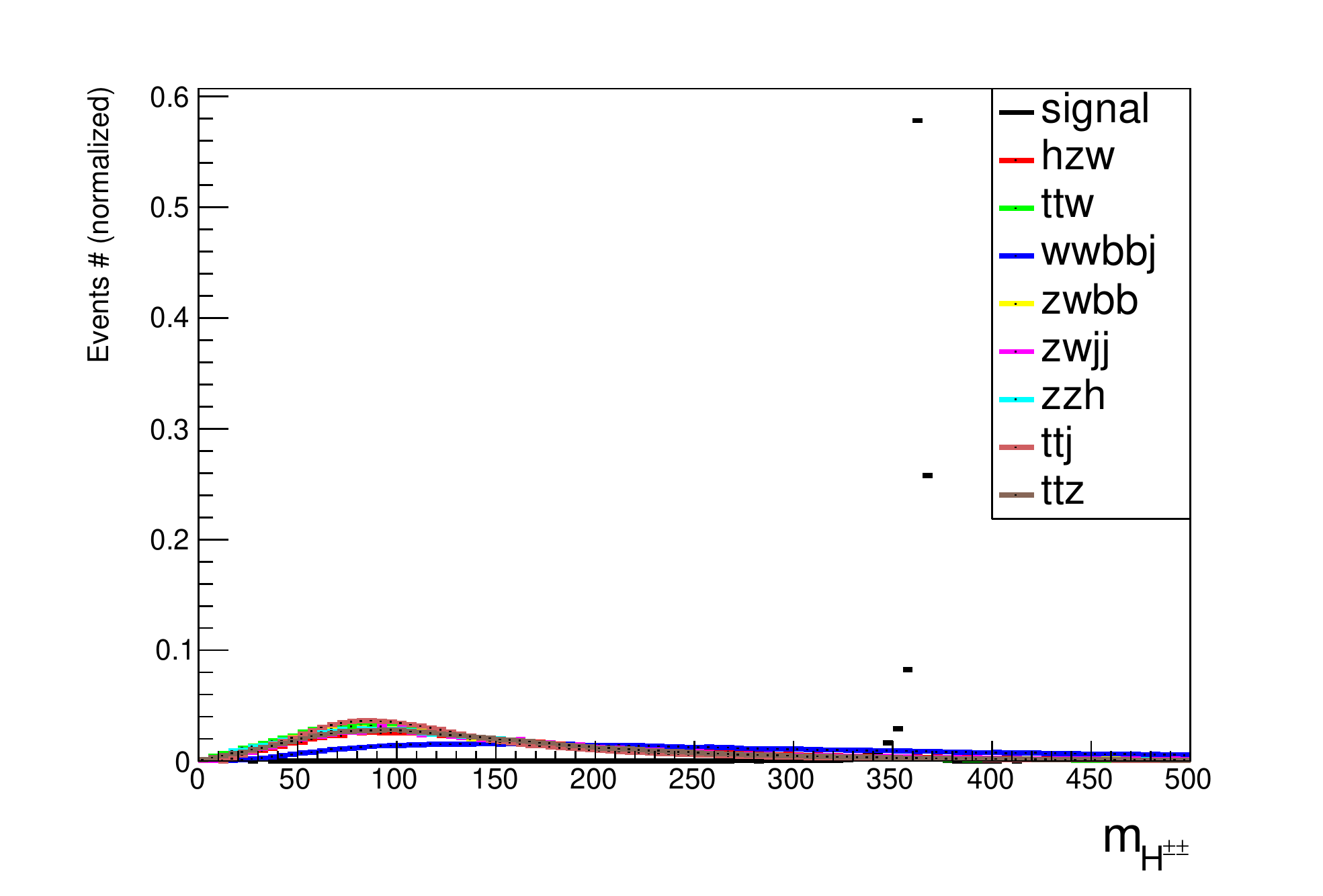} \\
 (e) $H_T$ & (f) Doubly-Charged Higgs invariant mass\\[6pt]
\end{tabular}
\caption{Representative reconstructed variables for the $\ell^\pm\ell^\pm hW^\mp$ channel after the basic cuts. We use the word ``signal'' to represent the $p p \to H^{\pm\pm}H^\mp\to \ell^\pm\ell^\pm hW^\mp$ channel in all histograms above.}\label{simu:bc}
\end{figure}

While analyzing the data by {\tt ROOT}\,6.06.02, we require all the final state particles have transverse momentum $p_T>20$\,GeV and pseudorapidity $|\eta|<2.5$; we also require exactly three leptons in the final state\footnote{with two of them are of same charge and of same flavor, and the third one with an opposite charge only.} and exactly two jets in the final state\footnote{with at least one of them being a $b$ quark.} for the signal and the $t\bar{t}W^\pm$, $t\bar{t}Z$, $hZW^\pm$, $ZW^\pm b\bar{b}$ and $hZZ$ backgrounds. For the $t\bar{t}j$ and $W^+W^-b\bar{b}j$ backgrounds, we require there are exactly two leptons and three jets\footnote{with at least one and at most two of the three jets are $b$ quarks. The light jet with the smallest $p_T$ among these three jets is taken to be a lepton with a 0.01\% fake rate\,\cite{delphescard}.} in the final state. For the $ZW^\pm jj$ background, when the light jet is a $c$ quark, we use a fake rate of $10\%$; and when the light jets are other light quarks, we use a fake rate of $0.01\%$\,\cite{delphescard}.

After the basic cuts, the result of reconstructed variables is given in Fig.\,\ref{simu:bc}, and the cut efficiency is given in Table \ref{tab:cft}. {\color{black}{By comparing the signal and the background distributions in Fig.\,\ref{simu:bc}, we find that $\Delta\phi$ and $\Delta R$ between the two $b$ quarks, scalar sum of the transverse momentum $H_T$, same-sign lepton leading and sub-leading $p_T$, same-sign lepton $\Delta\phi$ and $\Delta R$, $m_h$, $m_{H^{\pm\pm}}$ and $W$ boson transverse mass $m_{WT}$ have distinct features between our signal and the backgrounds, which can be exploited to reduce the backgrounds. These variables are the hard cuts we apply next.}}
\subsubsection{Cut based analysis: hard cuts}\label{hardcuts}
{\color{black}To improve the significance of the signal, we apply the following hard cuts in the same order as they are listed in Table \ref{llhctab}. After applying them}, the cut efficiency for each hard cut and significance of our signal are presented in Table \ref{tab:cft}. From the table, it is seen that the backgrounds are efficiently reduced and our signal has a final cross section about $7.3848\times10^{-4}$\,fb, with the significance being around 4; and the estimated event number for the signal after the basic cuts and the hard cuts is around 22 at the FCC with $\mathcal{L}=30\text{ab}^{-1}$.

\begin{table}[!htbp]
\centering
\caption{Cut flow table for $p p \to H^{\pm\pm}H^\mp\to \ell^\pm\ell^\pm hW^\mp$ under basic cuts (bc) and hard cuts (hc) with integrated luminosity of $30\,
\text{ab}^{-1}$. Here and in Table\,\ref{tab:wwhc}, we use the same abbreviations: ``proc.'' for ``processes''; ``E'' for ``base 10 exponential function''; ``cs'' for ``cross section'' with unit $fb$; ``eff.'' for ``efficiency'' in percent; ``signi.'' for ``significance'' and ``hci-j'' means ``applying hard cuts \text{i, $\cdots$, j}''}\label{tab:cft}
\begin{adjustbox}{max width = \textwidth}
\begin{tabular}{|c|c|c|c|c|c|c|c|c|c|}
\hline
 proc & original cs& - & bc & hc1 & hc1-2 & hc1-3 & hc1-4 & hc1-5 & hc1-6\\\hhline{|=|=|=|=|=|=|=|=|=|=|}
        &  & eff. &   2.94       &      5.78     &     5.84      &    14.86       &      95.95     & 45.07   & 6.25 \\\cline{3-10}
\bf hzw  & 0.6817 & cs   &   0.02       &   1.1584E-3      &    6.7652E-5       &    1.0053E-5      &   9.6460E-6        &  4.3474E-6  &  2.7268E-7    \\\hhline{|=|=|=|=|=|=|=|=|=|=|}
        & & eff. &    3.47     &     5.03    &     3.99      &     53.16     &     99.46      &  30.98  & 0  \\\cline{3-10}
\bf zzh  & 0.1107 & cs   &    3.8413E-3     &    1.9322E-4     &    7.7094E-6       &    4.0983E-6      &    4.0762E-6       &  1.2628E-6  &  0 \\\hhline{|=|=|=|=|=|=|=|=|=|=|}
        & & eff. &    0.25     &      5.04     &   3.34        &     48.39      &     100      &  46.67 & 14.29 \\\cline{3-10}
\bf zwjj  & 46.165 & cs   &   0.1133      &      5.7091E-3     &   1.9082E-4        &     9.233E-5      &     9.233E-5      &  4.3087E-5 & 6.1553E-6  \\\hhline{|=|=|=|=|=|=|=|=|=|=|}
       &  & eff. &   3.98  &  4.73     &   1.85      &    43.25    &   81.88   &   17.09  &  0 \\\cline{3-10}
\bf ttz  & 135.7 & cs   &  5.4044   &   0.2556     &    4.7167E-3     &   2.0402E-3     &   1.6705E-3   &  2.8544E-4   &  0 \\\hhline{|=|=|=|=|=|=|=|=|=|=|}
        & & eff. &     0.83    &    1.95     &   2.32     &    25      &   100        &   14.29  &   0   \\\cline{3-10}
\bf zwbb & 42.66 & cs   &    0.3521     &     6.8711E-3    &     1.5926E-4      &   3.9816E-5    &  3.9816E-5    &  5.688E-6  &  0 \\\hhline{|=|=|=|=|=|=|=|=|=|=|}
        & & eff. &   8.42     &     8.92      &    12.69       &   30.61        &    93.34       & 49.56  &  9.55   \\\cline{3-10}
\bf wwbbj  & 2.293 & cs   &    0.1932     &      1.7223E-2     &    2.1858E-3       &     6.6900E-4      &    6.2442E-4       & 3.0946E-4 & 2.9544E-5 \\\hhline{|=|=|=|=|=|=|=|=|=|=|}
&  & eff. & 2.74   &    19.40   &    1.18     &   39.94     &  81.03    &  27.33   &  12.57      \\\cline{3-10}
\bf ttw  & 68.7 & cs   & 1.8824   &  0.3652      &    4.3235E-3     &  1.7267E-3      &  1.3992E-3    &  3.8243E-4   & 4.809E-5   \\\hhline{|=|=|=|=|=|=|=|=|=|=|}
        & & eff. &  6.89  &   16.16     &    0.44     &   51.58     &    82.13      & 27.44    &  8.28 \\\cline{3-10}
\bf ttj  & 257 & cs   &    17.7094    &    2.8610     &    1.2456E-2     &   6.425E-3      &   5.2771E-3       &  1.4478E-3    &  1.1993E-4 \\\hhline{|=|=|=|=|=|=|=|=|=|=|}
\bf $\sigma_{\text{bkg}}^{\text{tot}}$  & 507.1454 & - & 25.6786 & 3.5130  & 2.4107E-2 & 1.1007E-2  & 9.1171E-3  & 2.4795E-3  & 2.0399E-4 \\\hhline{|=|=|=|=|=|=|=|=|=|=|}
        & & eff. &    16.15     &   62.03        &     58.30      &   87.20        &     96.94      & 78.43  & 98.50\\\cline{3-10}
\bf signal  & 0.0148 & cs   &   0.0024      &    1.4862E-3      &     8.6373E-4      &   7.5321E-4        &   7.3012E-4        & 5.7264E-4 & 7.3848E-4 \\\hhline{|=|=|=|=|=|=|=|=|=|=|}
\bf signi. & 0.1138 & - & 0.0820  &  0.1373  & 0.9467 & 1.2030  & 1.2744  & 1.7953  & 4.1664 \\ \hline
\end{tabular}
\end{adjustbox}
~\\~\\
\end{table}

\begin{table}[!ht]
\caption{A list of hard cuts for the $pp \to H^{\pm\pm}H^\mp\to\ell^\pm\ell^\pm hW^\mp$ channel.}\label{llhctab}
  \centering
  \begin{tabular}{c}\toprule[1.5pt]
    $m_Z\ge82$\,GeV or $m_Z\le 98$\,GeV, $80\text{\,GeV}\le m_h\le130$\,GeV\\
    $p_{T,b}^{\text{leading}}\ge80$\,GeV, $p_{T,\ell^{\text{oppo.}}}\ge40$\,GeV, $p_{T,\ell^{\text{same}}}^{\text{leading}}\ge200$\,GeV, $p_{T,\ell^{\text{same}}}^{\text{sub-leading}}\ge70$\,GeV, $H_T\ge700$\,GeV\\
    $0\le m_{WT}\le 90$\,GeV\\
    $-2\le\Delta\phi_{b\bar{b}}\le2$, $0\le\Delta R_{b\bar{b}}\le2$\\
    $-1.8\le\Delta\phi_{\ell^\pm\ell^\pm}\le1.8$, $0.6\le\Delta R_{\ell^\pm\ell^\pm}\le2.8$\\
    $340\text{\,GeV}\le m_{H^{\pm\pm}}\le390$\,GeV\\
 \bottomrule[1.25pt]
\end{tabular}\par
\end{table}

\subsubsection{BDT based analysis result}
To improve the cut efficiency, we also carry out a BDT based analysis as for analyzing model discovery at the 100\,TeV collider in Sec.\,\ref{sec:modeldis}. The result is shown in parallel with the cut-based result in Table\,\ref{tab:bdt} for comparison, and we find that the BDT method improves the signal significance by about a factor of 2 through optimizing the cut efficiency; in addition, the signal efficiency as well as the signal cross section are also improved by about a factor of 3.

\begin{center}
\begin{minipage}{\linewidth}
\centering
\captionof{table}{Comparison between BDT and cut-flow based results at $\mathcal{L}=30\,\text{ab}^{-1}$ for $pp \to H^{\pm\pm}H^\mp\to\ell^\pm\ell^\pm hW^\mp$.} \label{tab:bdt} 
\begin{tabular}{ C{2.5in} C{1.35in} C{1.35in} }\toprule[1.5pt]
& \bf BDT & \bf Cut based \\\midrule
\bf signal efficiency & 0.839 & 0.308 \\\midrule
\bf signal significance & 6.8922 & 4.1664 \\\midrule
\bf final signal cross section (fb) & $1.2417\times10^{-2}$ & $7.3848\times10^{-4}$ \\\midrule
\bf event number at detector & 60 & 22 \\
\bottomrule[1.25pt]
\end {tabular}\par
\bigskip
\end{minipage}
\end{center}

\subsection{Simulation: $H^\mp H^{\pm\pm}\rightarrow hW^\mp W^\pm W^\pm\rightarrow b\bar{b}\ell^\mp \ell^\pm \ell^\pm \slashed{E}_T$ process for intermediate and large $v_\Delta$}
The $H^\mp H^{\pm\pm}\rightarrow hW^\mp \ell^\pm \ell^\pm$ channel is helpful for the determination of $\lambda_4$ only at intermediate $v_\Delta$, for large $v_\Delta$'s, the $H^\mp H^{\pm\pm}\rightarrow hW^\mp W^\pm W^\pm$ channel can be used. Since it shares the same backgrounds as the $H^\mp H^{\pm\pm}\rightarrow hW^\mp \ell^\pm \ell^\pm$ channel in last sub-section, we generate 1,000,000 events for this signal and use the background data generated for the $H^\mp H^{\pm\pm}\rightarrow hW^\mp \ell^\pm \ell^\pm$ channel to study its collider phenomenologies. We still perform an exclusive analysis, and by using  the same basic cuts as for the $\ell^\pm\ell^\pm hW^\mp$ channel, we obtain the reconstructed variables under basic cuts for the $W^\pm W^\pm hW^\mp$ channel shown in Fig.\,\ref{wwbc1}. Note that $\Delta\Phi$ and $\Delta R$ between the two $b$ quarks and the two same-sign leptons, leading $p_T$ of the same-sign leptons, SM $h$, the doubly-charged Higgs and $Z$ boson masses and the transverse $W$ boson mass are the hard cuts that can be applied to further separate the signal from the backgrounds. Those hard cuts are applied in the same order as they are listed in Table \ref{wwhwhc}:
\begin{figure}[thb!]
\captionstyle{flushleft}
\begin{tabular}{cc}
 \includegraphics[width=90mm]{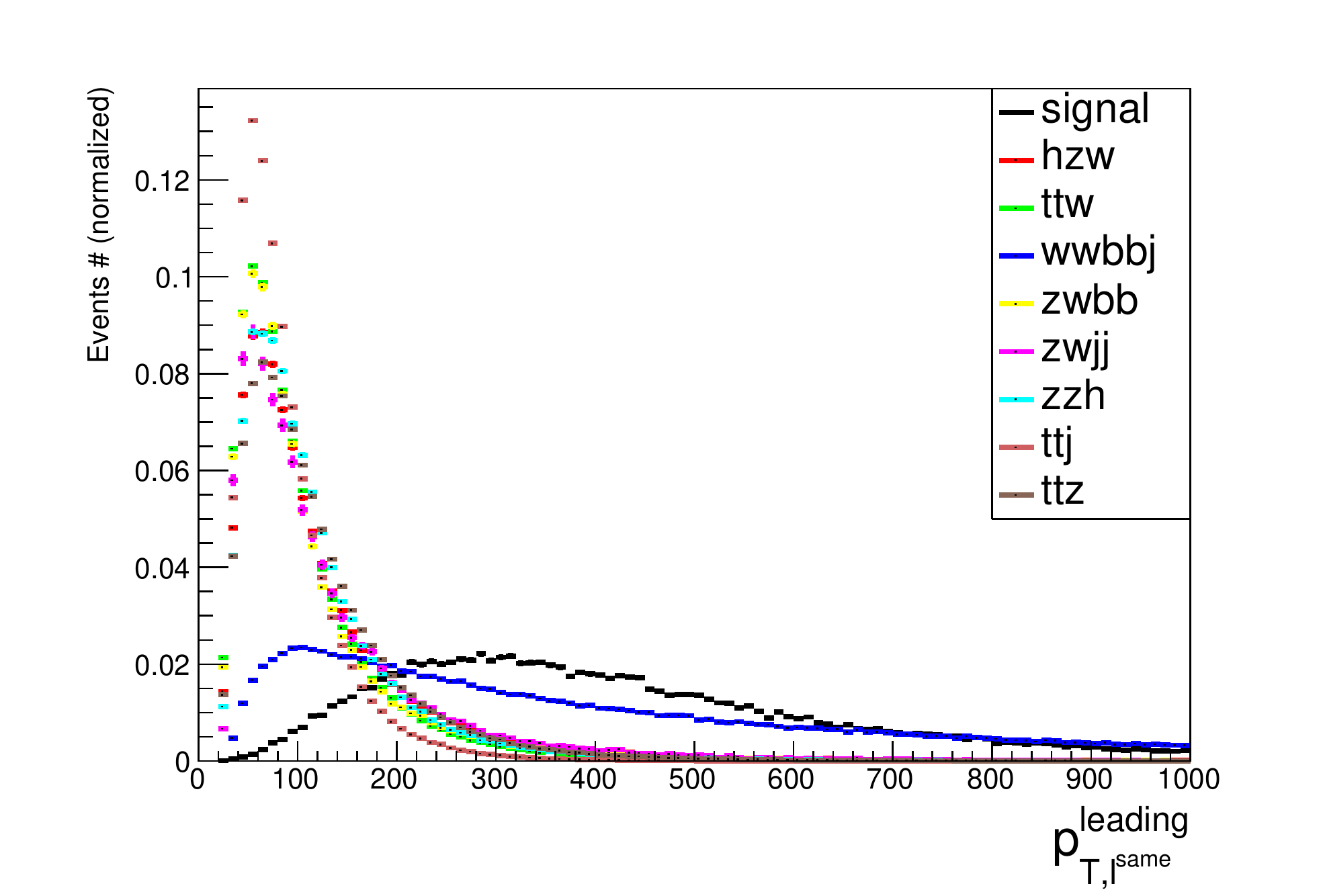} & \includegraphics[width=90mm]{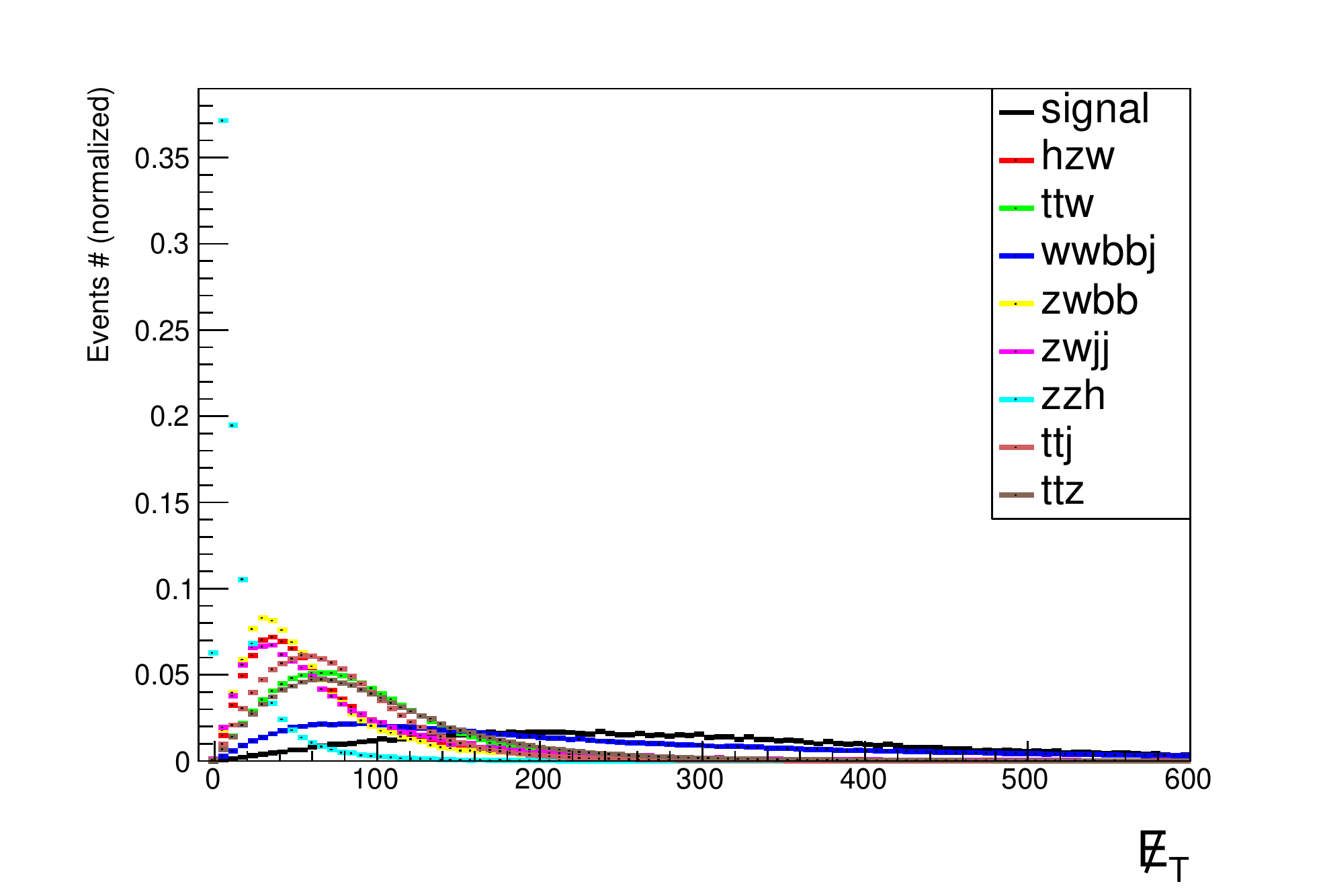} \\
 (a) same-sign lepton leading $p_T$ & (b) Missing $E_T$ \\[6pt]
  \includegraphics[width=90mm]{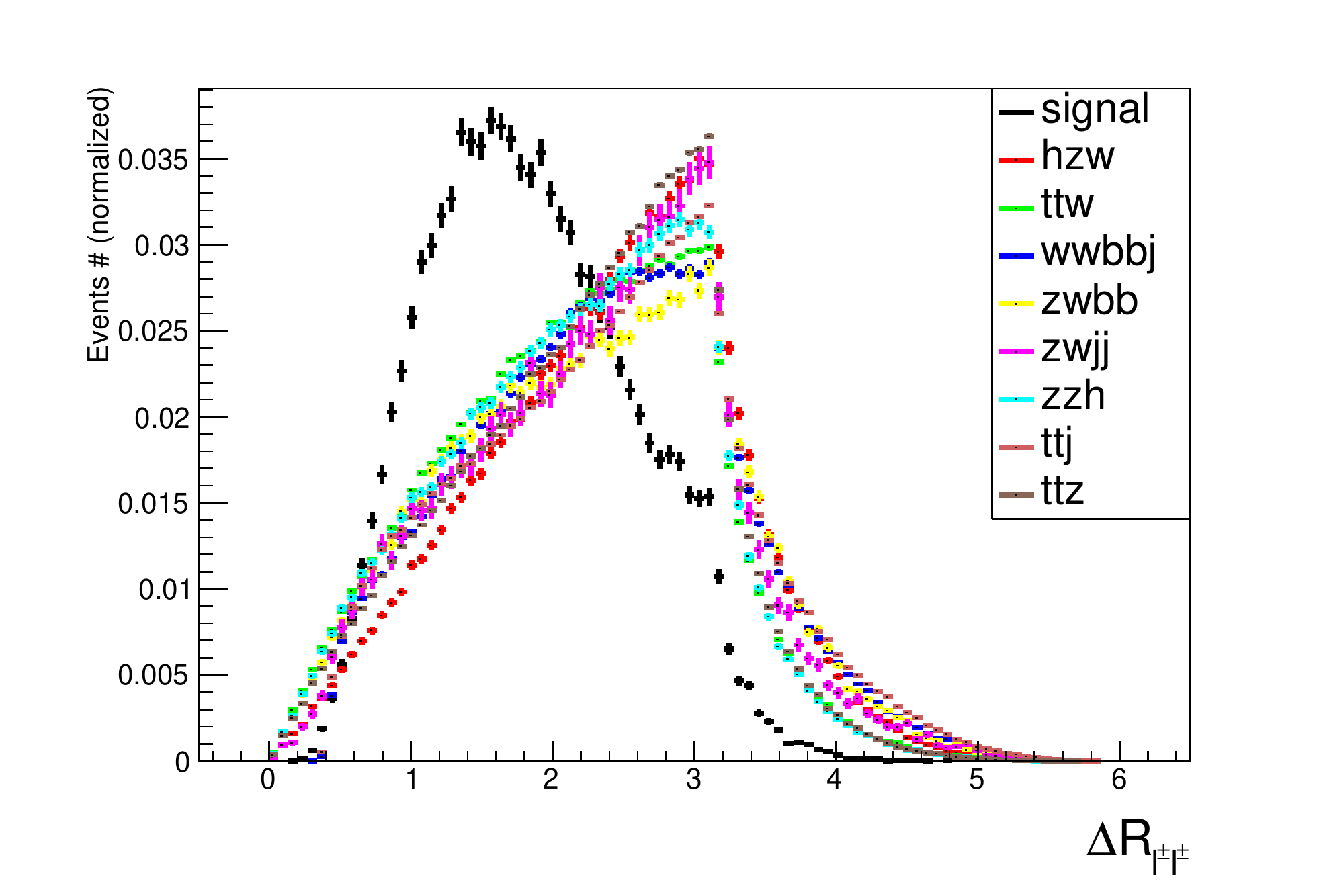} &   \includegraphics[width=90mm]{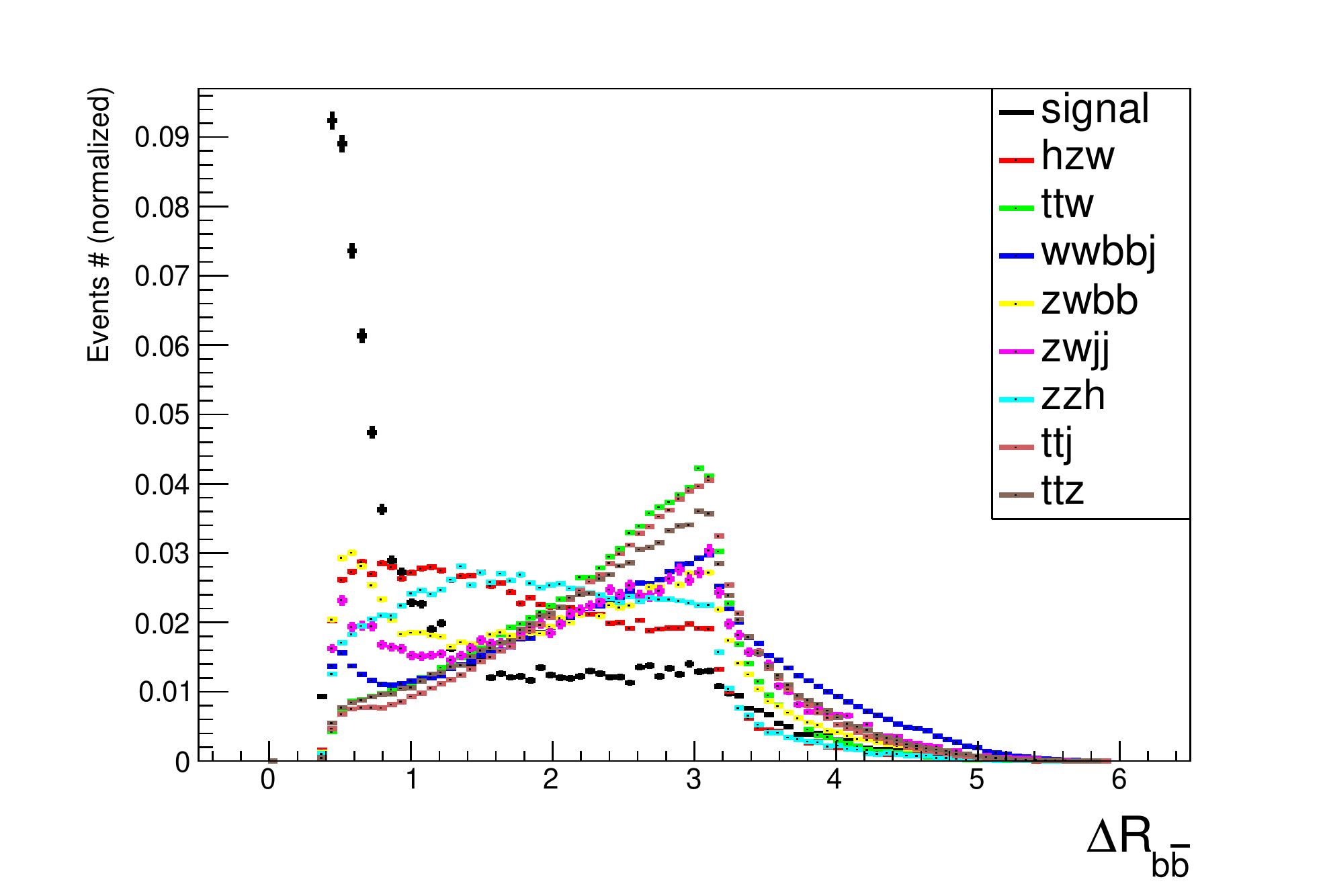}  \\
  (c) same-sign lepton $\Delta R$  & (d) $\Delta R$ of two $b$ quarks \\[6pt]
 \includegraphics[width=90mm]{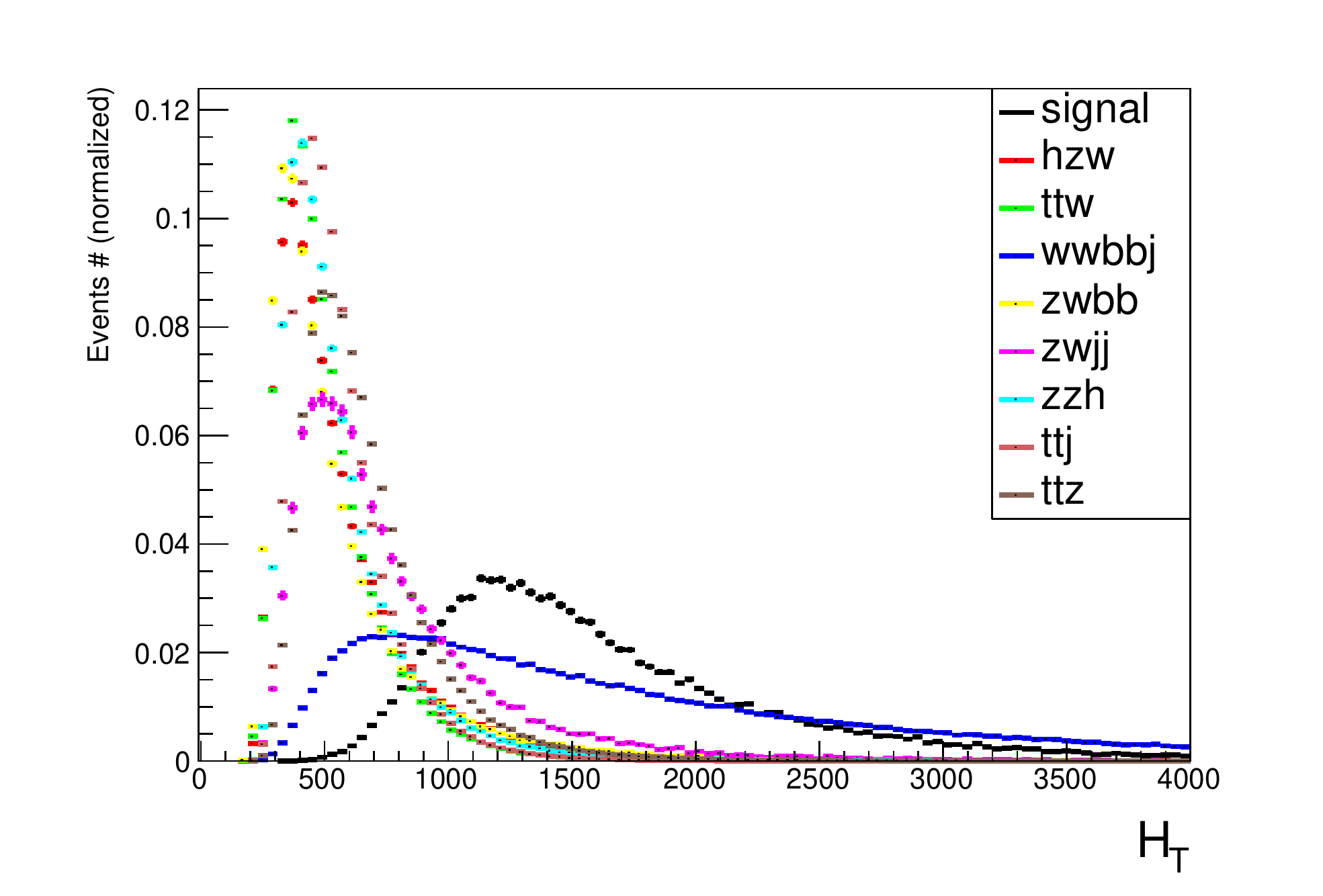}  & \includegraphics[width=90mm]{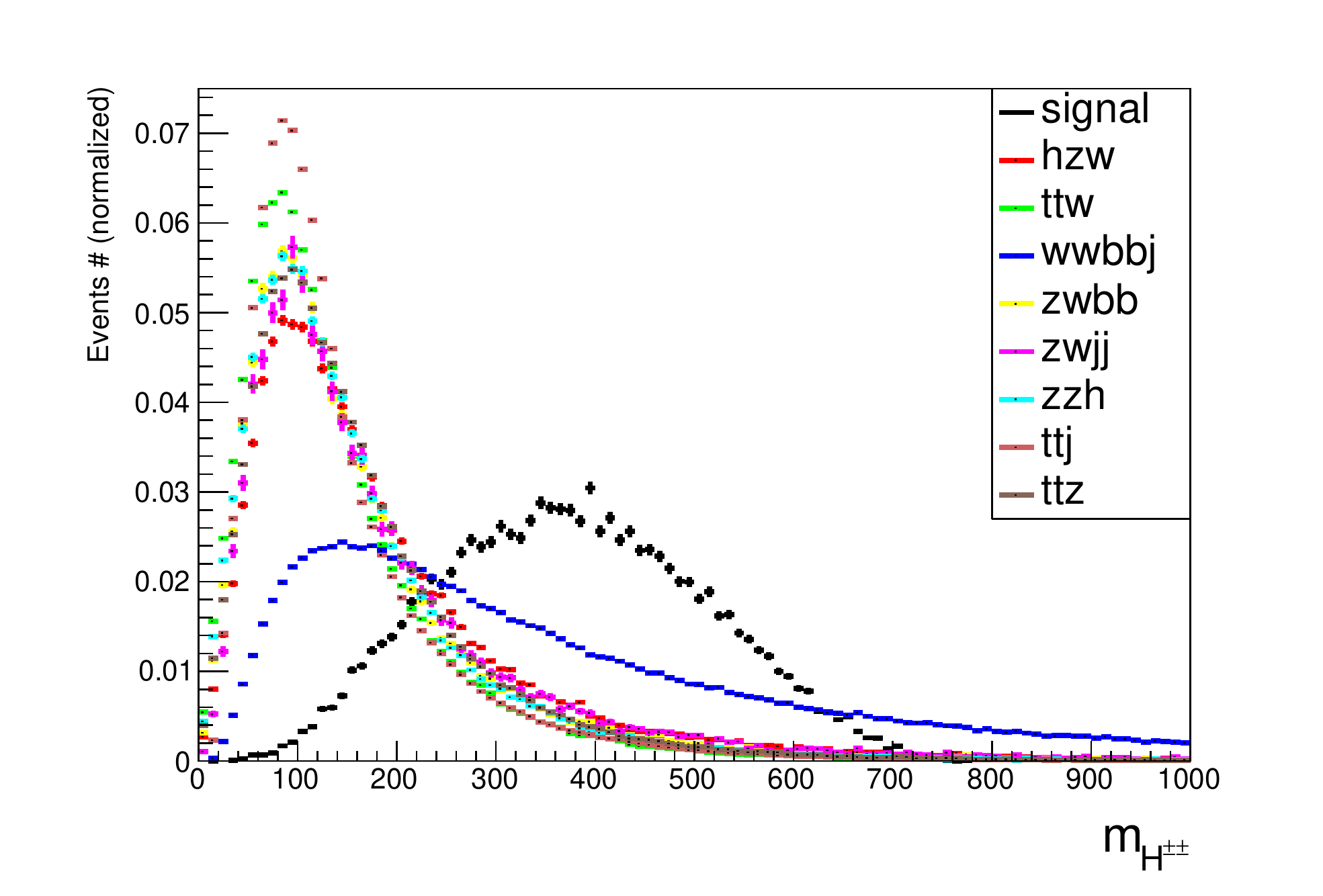}  \\
  (e) $H_T$ & (f) Doubly-charged Higgs invariant mass \\[6pt]
\end{tabular}
\caption{Reconstructed variables for the $W^\pm W^\pm hW^\mp$ channel under basic cuts. We use the word ``signal'' to represent the $p p \to H^{\pm\pm}H^\mp\to W^\pm W^\pm hW^\mp$ channel in all histograms above.}\label{wwbc1}
\end{figure}

\begin{table}[!htbp]
\centering
\caption{Cut flow table for $H^\mp H^{\pm\pm}\rightarrow hW^\mp W^\pm W^\pm$ under basic cuts (bc) and hard cuts (hc) with integrated luminosity of $30\, \text{ab}^{-1}$. Here we use the same abbreviations as in Table\,\ref{tab:cft}.}\label{tab:wwhc}
\begin{adjustbox}{max width = \textwidth}
\begin{tabular}{|c|c|c|c|c|c|c|c|c|}
\hline
proc.& original cs & - & bc & hc1 & hc1-2  & hc1-3 & hc1-4 & hc1-5\\\hhline{|=|=|=|=|=|=|=|=|=|}
        &  & eff. &   2.94       &  4.37   &  10.35   &  97.29  &  39.72  & 52.92   \\\cline{3-9}
\bf hzw  & 0.6817 & cs   &   0.02       &   8.741E-4  &  9.0474E-5  &  8.8025E-5  &  3.4965E-5  &  1.8503E-5  \\\hhline{|=|=|=|=|=|=|=|=|=|}
        & & eff. &    3.47     &  3.80   &  7.02  & 93.30   &  51.62  &  54.71  \\\cline{3-9}
\bf zzh  & 0.1107 & cs   &    3.8413E-3     &   1.4586E-4  & 1.0246E-5   & 9.5603E-6   &  4.9351E-6  &  2.6999E-6  \\\hhline{|=|=|=|=|=|=|=|=|=|}
        & & eff. &    0.25     &   5.40  &  9.20  &  89.62  &  61.59  &  55.45  \\\cline{3-9}
\bf zwjj  & 46.165 & cs   &   0.1133      &  6.1201E-3   &  5.6308E-4  &  5.0462E-4  & 3.1077E-4   &  1.7231E-4  \\\hhline{|=|=|=|=|=|=|=|=|=|}
       &  & eff. &   3.98  &  5.08    &  4.62  &  63.02  &  36.59  &  41.05  \\\cline{3-9}
\bf ttz  & 135.7 & cs   &  5.4044   &  0.2748   &  1.2704E-2  &  8.0062E-3  &  2.9292E-3  &  1.2026E-3  \\\hhline{|=|=|=|=|=|=|=|=|=|}
        & & eff. &     0.83    &  1.66   &  3.90  & 82.5 &  74.24  & 44.90   \\\cline{3-9}
\bf zwbb & 42.66 & cs   &    0.3521     &  5.8326E-3   &  2.2750E-4  &  1.8769E-4  &  1.3935E-4  &  6.2563E-5  \\\hhline{|=|=|=|=|=|=|=|=|=|}
        & & eff. &   8.42     &   10.72   &  24.12  &  83.77  & 52.14   &  65.83  \\\cline{3-9}
\bf wwbbj  & 2.293 & cs   &    0.1932     &   2.0719E-2   &  4.9978E-3  &  4.1865E-3  &  2.1826E-3  &  1.4369E-3  \\\hhline{|=|=|=|=|=|=|=|=|=|}
&  & eff. & 2.74   &  22.76   &  1.94  &  59.90  &  42.10  &  53.38  \\\cline{3-9}
\bf ttw  & 68.7 & cs   & 1.8824   &  0.4139   &  8.3198E-3  &  4.9832E-3  &  2.0977E-3  &  1.1198E-3  \\\hhline{|=|=|=|=|=|=|=|=|=|}
        & & eff. &  6.89  &  18.54   &  1.26  &  65.57  &  45.23  &  34.56  \\\cline{3-9}
\bf ttj  & 257 & cs   &    17.7094    &  3.2826   &  4.1454E-2  &  2.7182E-2  &  1.2293E-2  &  4.2491E-3  \\\hhline{|=|=|=|=|=|=|=|=|=|}
\bf $\sigma_{\text{bkg}}^{\text{tot}}$  & 507.1454 & - & 25.6786 & 4.0050  & 6.8367E-2 &  4.5148E-2   &  1.9993E-2    &  8.2645E-3   \\\hhline{|=|=|=|=|=|=|=|=|=|}
        & & eff. & 5.68 &  51.03  &  79.46  &  100  &  70.07  &  94.24  \\\cline{3-9}
\bf signal  & 0.0971 & cs   &  5.5079E-3     &  2.8104E-3    & 2.2331E-3   &  2.1615E-3  &  1.5146E-3  &  1.4273E-3  \\\hhline{|=|=|=|=|=|=|=|=|=|}
\bf sig. & 0.7467 & - & 0.1883 & 0.2433 & 1.4564  &  1.7220 & 1.7896 & 2.5123 \\ \hline
\end{tabular}
\end{adjustbox}
\end{table}

Results after applying the hard cuts are given in Table \ref{tab:wwhc}. And for comparison, the BDT based analysis is presented in parallel in Table \ref{bdtwwhc}, we see that BDT based analysis still gives a larger significance, which is about three times larger compared with cut-based result.

\begin{table}[!ht]
\caption{A list of hard cuts for the $pp\to H^{\pm\pm}H^\mp\to W^\pm W^\pm hW^\mp$ channel.}\label{wwhwhc}
  \centering
  \begin{tabular}{c}\toprule[1.5pt]
    $m_Z\ge80$\,GeV or $m_Z\le 100$\,GeV, $80\text{\,GeV}\le m_h\le140$\,GeV\\
    $p_{T,b}^{\text{leading}}\ge80$\,GeV, $p_{T,\ell^{\text{oppo.}}}\ge40$\,GeV, $p_{T,\ell^{\text{same}}}^{\text{leading}}\ge80$\,GeV, $p_{T,\ell^{\text{same}}}^{\text{sub-leading}}\ge50$\,GeV, $800\text{\,GeV}\le H_T\le2200$\,GeV\\
    $-1.4\le\Delta\phi_{b\bar{b}}\le1.4$, $0\le\Delta R_{b\bar{b}}\le2$\\
    $-2\le\Delta\phi_{\ell^\pm\ell^\pm}\le2$, $0\le\Delta R_{\ell^\pm\ell^\pm}\le2.8$\\
    $200\text{\,GeV}\le m_{H^{\pm\pm}}\le800$\,GeV\\
 \bottomrule[1.25pt]
\end{tabular}\par
\end{table}

\begin{center}
\begin{minipage}{\linewidth}
\centering
\captionof{table}{Comparison between BDT and cut-flow based results at $\mathcal{L}=30\,\text{ab}^{-1}$ for $pp\to H^{\pm\pm}H^\mp\to W^\pm W^\pm hW^\mp$.} \label{bdtwwhc} 
\begin{tabular}{ C{2.5in} C{1.35in} C{1.35in} }\toprule[1.5pt]
& \bf BDT & \bf Cut based \\\midrule
\bf signal efficiency & 0.6009 &  0.2591  \\\midrule
\bf signal significance &  6.8507   &  2.5123 \\\midrule
\bf final signal cross section (fb) & $3.3097\times10^{-3}$ & $1.4273\times10^{-3}$ \\\midrule
\bf event number at detector & 99 & 42 \\
\bottomrule[1.25pt]
\end {tabular}\par
\bigskip
\end{minipage}
\end{center}

\subsection{Determination of $\lambda_4$ upon discovery at the future 100\,TeV collider}\label{lam4haa}
As we have been addressing throughout the paper, the $H^\mp H^{\pm\pm}\rightarrow hW^\mp \ell^\pm \ell^\pm$ and the $H^\mp H^{\pm\pm}\rightarrow hW^\mp W^\pm W^\pm$ channels are important for the determination of $\lambda_4$, but our study above is done at only one benchmark point for both $H^\mp H^{\pm\pm}\rightarrow hW^\mp \ell^\pm \ell^\pm$ and $H^\mp H^{\pm\pm}\rightarrow hW^\mp W^\pm W^\pm$. To see how our result is sensitive to $\lambda_4$, we fix $\lambda_5=-0.1$ and perform a scan in the $\lambda_4\text{-}m_\Delta$ plane\footnote{Note that $\lambda_{2,3}$ are suppressed by $v_\Delta$, so their values do not matter here.}. Doing so, it is straightforward to rescale the signal and, thereby, obtain the variation in signal significance. 
The corresponding results are given in Fig.\,\ref{haa}\,(a), (b), (c) with $v_\Delta=10^{-1}$\,GeV, $v_\Delta=10^{-4}$\,GeV and $v_\Delta=10^{-5}$\,GeV respectively. There, we indicate the regions giving larger than $5\sigma$ significance for the two channels considered here.

In Fig.\,\ref{haa}(a), {\it i.e.}, at large $v_\Delta=10^{-1}\rm\,GeV$, only the $W^\pm W^\pm hW^\mp$ channel is useful, whereas the significance for $\ell^\pm \ell^\pm hW^\mp$ is less than 5 in the entire parameter space. The reason is that the rate for $H^{\pm\pm}\rightarrow\ell^\pm\ell^\pm$ is highly suppressed at large $v_\Delta$ as can be seen from left panel of Fig.\,\ref{decayregionplotHPP}. For $W^\pm W^\pm hW^\mp$, the appearance of the region at the upper-left corner is due to an increase of the decay BR for $H^\pm\rightarrow hW^\pm$ when $\lambda_4$ goes from negative to positive as can be seen from the upper right panel of Fig.\,\ref{slicedecay}. Therefore, at large $v_\Delta$, the $W^\pm W^\pm hW^\mp$ channel is more helpful for the determination of $\lambda_4$ at the FCC.

From Fig.\,\ref{haa}(b), {\it i.e.}, corresponding to intermediate $v_\Delta=10^{-4}\rm\,GeV$, both $W^\pm W^\pm hW^\mp$ and $\ell^\pm \ell^\pm hW^\mp$ can help to determine $\lambda_4$. The $W^\pm W^\pm hW^\mp$ channel covers a larger region at a higher mass scale while the $\ell^\pm \ell^\pm hW^\mp$ channel provides more coverage at a lower mass scale. The overlap between these two channels makes them useful as a cross check if the triplet scale is around $m_\Delta\in[400,900]$\,GeV. For $m_\Delta\in[900,1100]$\,GeV, the $W^\pm W^\pm hW^\mp$ channel can be used to determine $\lambda_4$; and for $m_\Delta\in[300,400]$\,GeV, we can use the $\ell^\pm \ell^\pm hW^\mp$ channel.

\begin{figure}[thb!]
\captionstyle{flushleft}
\begin{tabular}{ccc}
\includegraphics[width=55mm]{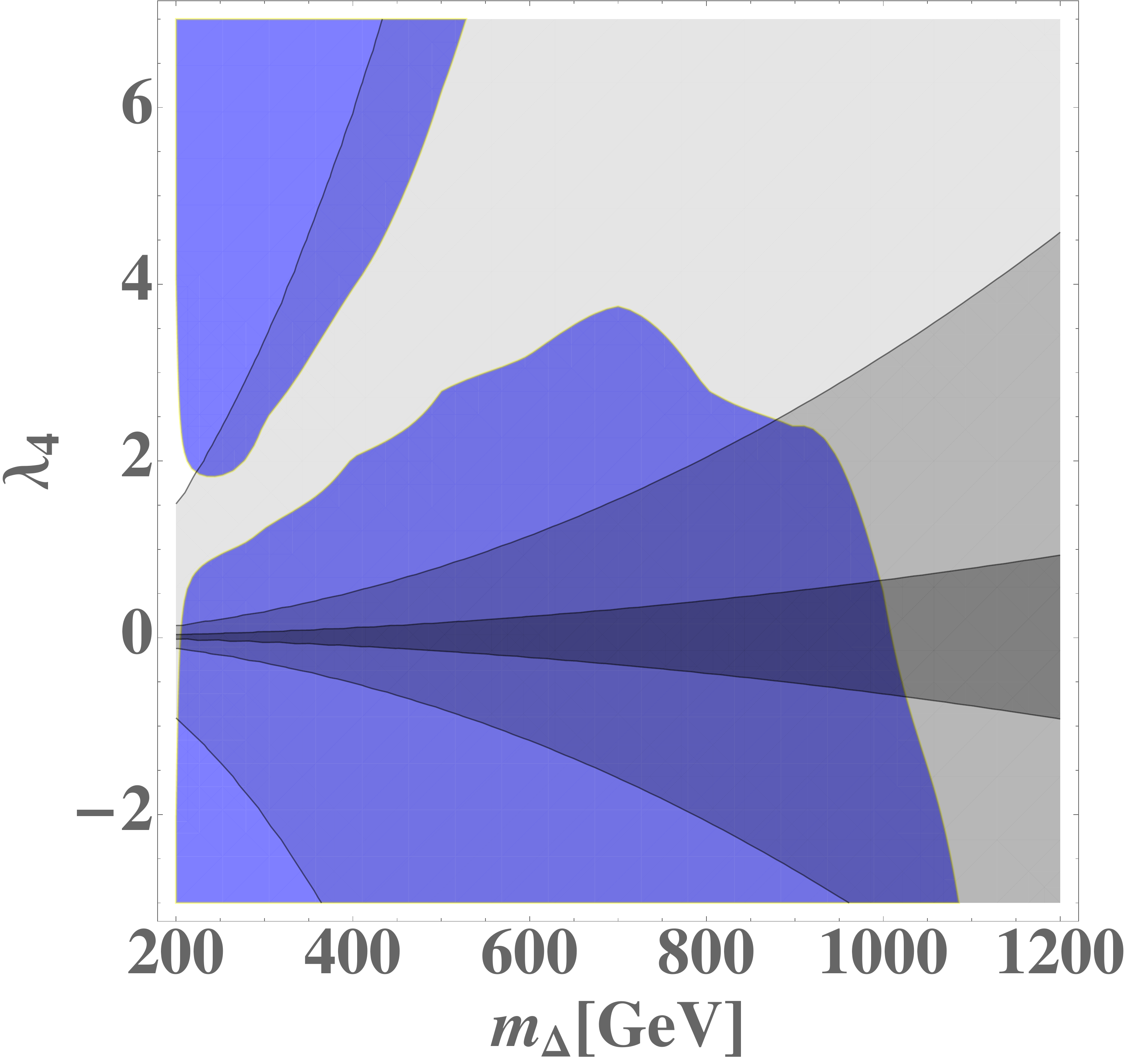} & \includegraphics[width=55mm]{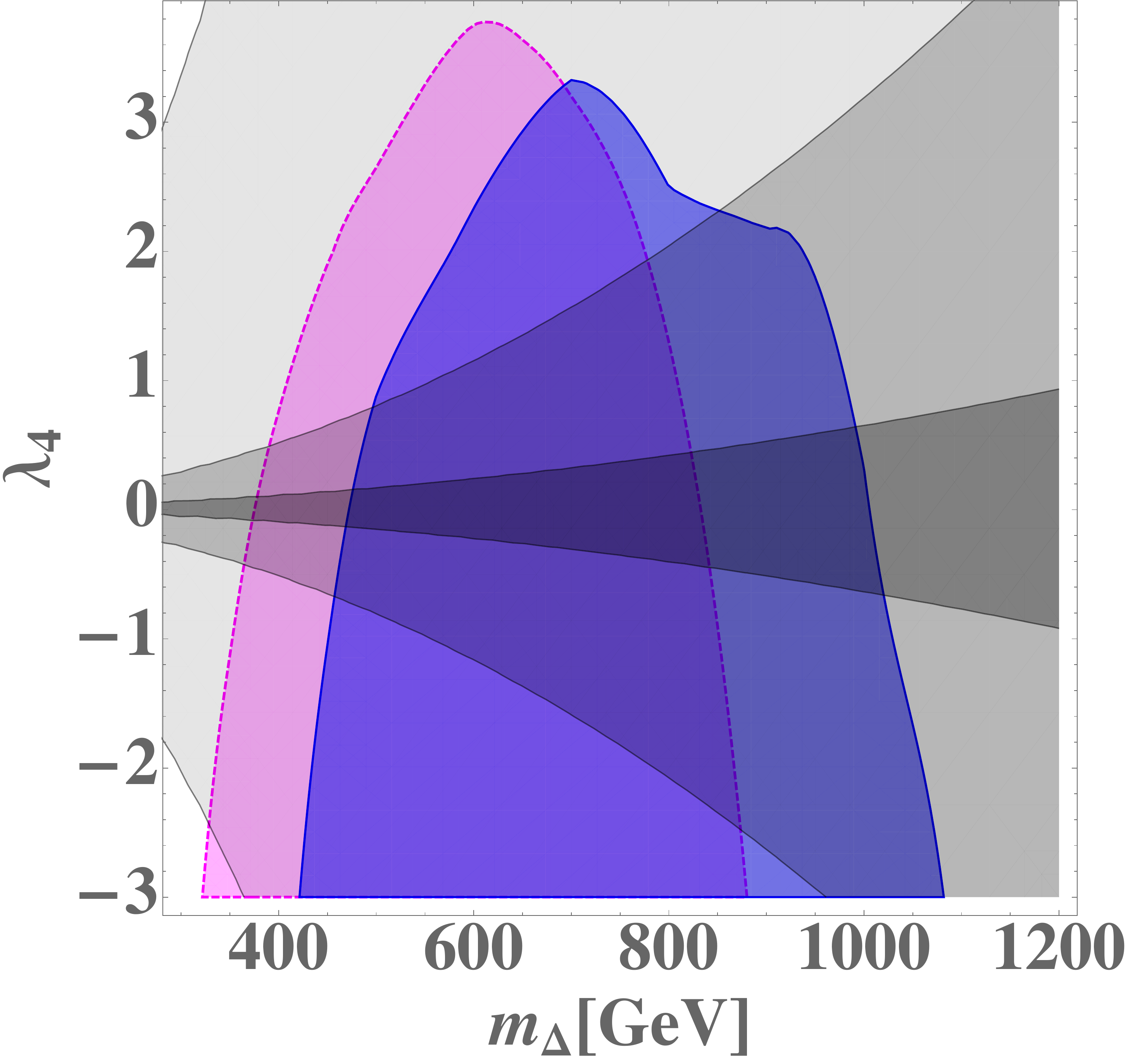} & \includegraphics[width=53mm]{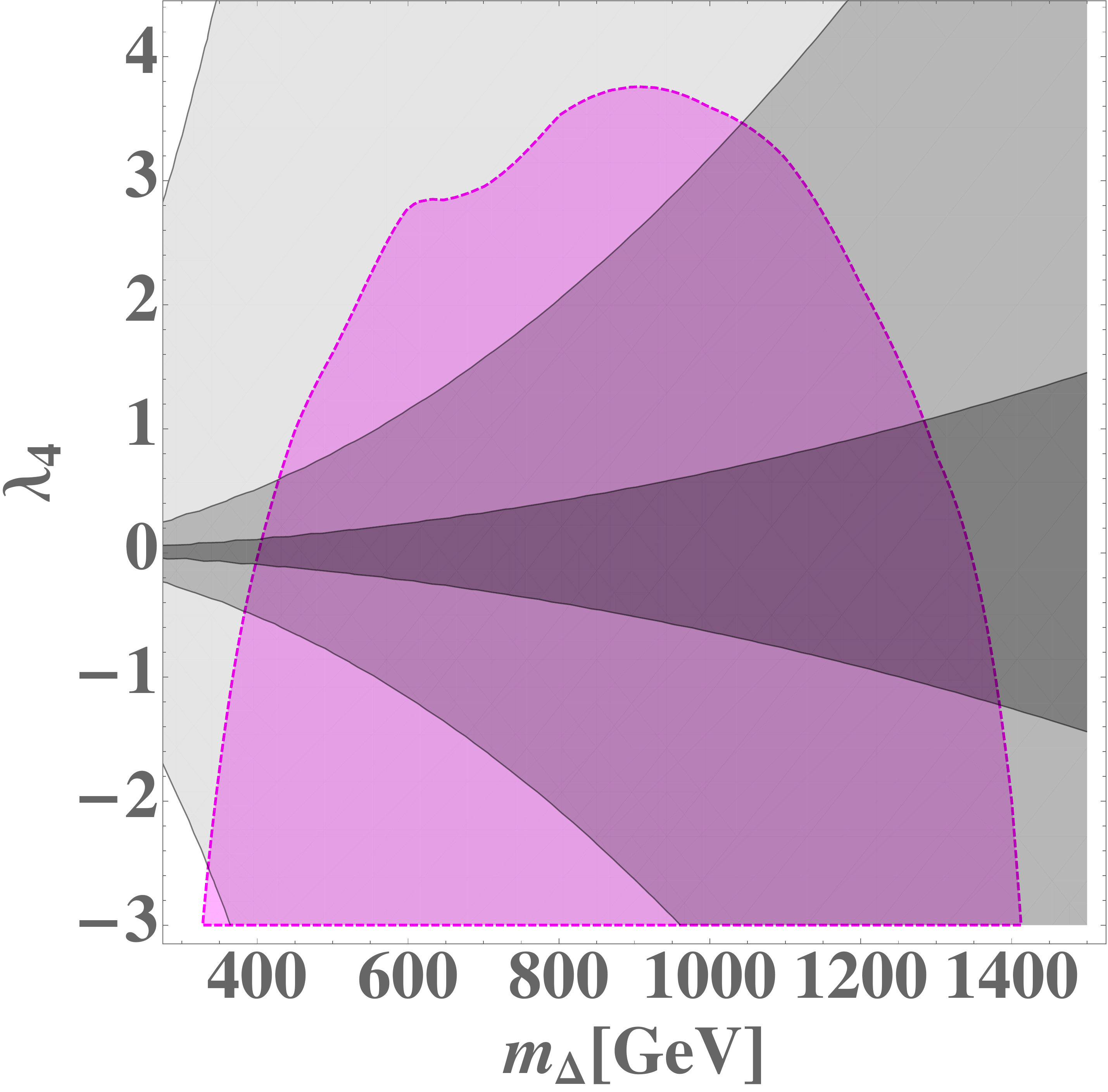}\\
(a) $v_\Delta=10^{-1}$\,GeV & (b) $v_\Delta=10^{-4}$\,GeV & (c) $v_\Delta=10^{-5}$\,GeV \\
\end{tabular}
\caption{Blue is significance $\ge5$ region for the $hW^\mp W^\pm W^\pm$ channel and magenta is that for the $hW^\mp \ell^\pm \ell^\pm$ channel. The outermost very light black region is the combined constraint on $R_{h\gamma\gamma}$ from ATLAS and CMS at 7\,TeV and 8\,TeV; the intermediate light black region is the planned FCC-ee constraint and the innermost black region is the planned FCC-ee+FCC-hh constraint on $R_{h\gamma\gamma}$.}\label{haa}
\end{figure}

And from Fig.\,\ref{haa}(c), {\it i.e.}, at small $v_\Delta=10^{-5}$\,GeV, only the $\ell^\pm \ell^\pm hW^\mp$ channel can be used to determine $\lambda_4$ since the $H^{\pm\pm}\rightarrow W^\pm W^\pm$ channel is highly suppressed as can be seen from the left panel of Fig.\,\ref{decayregionplotHPP}. Comparing this result with those at $v_\Delta=10^{-1}$\,GeV and $v_\Delta=10^{-4}$\,GeV, we see that at $v_\Delta=10^{-5}$\,GeV, the $\ell^\pm \ell^\pm hW^\mp$ channel covers the largest mass region up to about 1.4\,TeV.

{\color{black} It is now interesting to consider the possible complementarity between these direct probes of the Higgs portal coupling and mass with indirect tests.}
As has been studied in Refs.\,\cite{Arhrib:2011vc,Kanemura:2012rs}, the doubly-charged Higgs particle of the CTHM can give a sizable contribution to the $h\rightarrow \gamma\gamma$ decay rate especially for negative $\lambda_4$ and $\lambda_{45}$ due to a constructive interference\,\cite{Chun:2012jw}. We therefore expect the $h\rightarrow\gamma\gamma$ decay rate to provide an indirect determination of $\lambda_4$ by excluding some of the parameter space on the $\lambda_4\text{-}m_\Delta$ plane. {\color{black} In this context, we consider the ratio $R_{h\gamma\gamma}$ given
\begin{align}
R_{h\gamma\gamma}=\frac{\Gamma^{\text{NP}}(h\rightarrow\gamma\gamma)+\Gamma^{\text{SM}}(h\rightarrow\gamma\gamma)}{\Gamma^{\text{SM}}(h\rightarrow\gamma\gamma)},\label{haadef}
\end{align}
with $\Gamma^{\text{NP}}$ and $\Gamma^{\text{SM}}$ the new physics (NP) and pure SM contribution to the decay rate of $h\rightarrow\gamma\gamma$ respectively. From Eq.\,\eqref{haadef} we see that, if nature is completely described by SM, then this ratio will exactly be one; and any value that deviates from one might be a source of new physics.}
For the quark loop contributions, we retain only the dominant $t$ quark for the fermion loop contribution to $R_{h\gamma\gamma}$. The current LHC and the proposed FCC constraints on this ratio  is indicated in the $\lambda_4\text{-}m_\Delta$ plane in Fig.\,\ref{haa}\,(a), (b), (c)\footnote{The values we use for $R_{h\gamma\gamma}$ are: For the LHC, we use the current experimental value $1.16_{-0.18}^{+0.20}$\,\cite{Khachatryan:2016vau,Biswas:2017tnw}; For the FCC-ee collider, we use the proposed values, {\it i.e.}; $1\pm0.05$, and $1\pm0.01$ for FCC-hh collider\,\cite{Contino:2016spe}. }, where the lightest black region is the combined constraint on $R_{h\gamma\gamma}$ from ATLAS and CMS at 7\,TeV and 8\,TeV; the intermediate black region is the planned FCC-ee constraint and the darkest black region shows the combined planned FCC-ee+FCC-hh constraint on $R_{h\gamma\gamma}$.

From Fig.\,\ref{haa}\,(a), we see that the current LHC constraint on $R_{h\gamma\gamma}$ is almost ruled out the small $m_\Delta$ and large $\lambda_4$ region, but in other regions, the current LHC constraints on the $\lambda_4\text{-}m_\Delta$ plane are relatively weak. This situation, however, will be changed considerably by the future 100\,TeV collider as can be seen from the darker black region in Fig.\,\ref{haa}\,(a), (b), (c).  

{\color{black} Thus, combination of the direct and indirect probes of the CTHM would be advantageous in the determination of $\lambda_4$. If future precision measurements of the $h\to\gamma\gamma$ decay rate agree with the SM expectations, a substantial portion of the $\lambda_4\text{-}m_\Delta$ parameter space will be excluded, thereby assisting in the determination of $\lambda_4$. In the remaining regions of parameter space, 
$\lambda_4$ could eventually be determined by $H^\mp H^{\pm\pm}\rightarrow \ell^\pm \ell^\pm hW^\mp $ and $H^\mp H^{\pm\pm}\rightarrow W^\pm W^\pm hW^\mp $ based on our study above.   {{It is also possible that future experiments at the LHC, FCC-ee, or FCC-hh see a deviation of $R_{h\gamma\gamma}$  from the SM prediction. In this case, if $\lambda_5$ is determined from mass splitting (-0.1 in our case) 
we might also also conclude that: (1) If the deviation is detected through the $hW^\mp W^\pm W^\pm$ ($hW^\mp \ell^\pm \ell^\pm$) channel, the triplet will have a large (small) vev with $|\lambda_4|\sim 1$; (2) if the deviation is observed from both $hW^\mp W^\pm W^\pm$ and $hW^\mp \ell^\pm \ell^\pm$ channels, an intermediate triplet vev can be inferred with $|\lambda_4| \sim 1$.}}}
\section{Conclusion}
\label{sec:conclusion}
{\color{black} In this paper, we have investigated the model discovery and Higgs portal parameter determination of the Complex Triplet Higgs Model  at a prospective 100\,TeV $pp$ collider. The triplet with Y=2 has long been known as a key ingredient in generating non-zero neutrino masses through the type-II seesaw mechanism. The triplet interacts with the SM through its electroweak gauge interactions, its coupling to the leptons in the type-II see saw interaction, and to the Higgs doublet via the Higgs portal parameters $\lambda_4$ and $\lambda_5$. The latter modify the scalar potential and may enable a strong first order electroweak phase transition, as needed for electroweak baryogenesis.}

The CTHM parameter space is constrained by current experiments at the LHC in the region where the triplet is light ($\lesssim600$\,GeV) and its vev, $v_\Delta$, is small ($\lesssim10^{-4.6}$\,GeV). In this paper, we have analyzed the reach of a prospective 100\,TeV $pp$ collider by working in the Normal Mass Hierarchy (NMH) framework, wherein the doubly-charged Higgs particle $H^{\pm\pm}$ is the heaviest. Based on our study, we conclude that a large part of the CTHM parameter space will be covered by the 100\,TeV collider in the future as shown in our Fig.\,\ref{bdtdis}. More specifically, we find that :

\begin{enumerate}
\item The $H^{++}H^{--}$ and $H^{\pm\pm}H^\mp$ channels have the largest and the second largest cross section respectively, making them the dominant discovery channels of the CTHM. Importantly, the $H^{++}H^{--}\rightarrow\ell^+\ell^+\ell'^-\ell'^-$ channel is recognized as the smoking-gun signature of the CTHM, which can be used to discover the triplet up to a mass $\sim$4.5\,TeV when $v_\Delta\lesssim10^{-4}$\,GeV. In addition, for $v_\Delta\gtrsim10^{-4}$\,GeV, the triplet model can be discovered by the $H^{\pm\pm}H^\mp\rightarrow\ell^\pm\ell^\pm hW^\mp/W^\pm W^\pm hW^\mp$ channel when the triplet mass is below $\sim$1\,TeV.
\item For $v_\Delta\gtrsim10^{-4}$\,GeV, the triplet can also be discovered through the $H^{++}H^{--}\rightarrow W^+W^+W^-W^-\to\ell^+\ell^+\ell'^-\ell'^-\slashed{E}_T$ channel when the triplet mass is below $\sim$1.7\,TeV. In arriving at this conclusion, we use  the same BDT training and test variables as for the $H^{++}H^{--}\rightarrow\ell^+\ell^+\ell'^-\ell'^-$ channel. However, if one were to choose a different set of BDT training and test variables to optimize the cut efficiency, or if one were to study different final states like in Ref.\,\cite{kang:2014jia}, one might anticipate that the quartic-$W$ channel will also cover the upper right white corner in Fig.\,\ref{bdtdis}, such that the whole parameter space can be explored at the future 100\,TeV collider.
\item Upon discovery, Higgs portal parameter $\lambda_5$ can be determined straightforwardly from the mass splitting $\Delta m\approx\frac{|\lambda_5|v^2}{8m_\Delta}$ defined in Eq.\,\eqref{massspec}.
\end{enumerate}

While the triplet can be discovered over a wide range and $\lambda_5$ can be calculated straightforwardly from the mass splitting upon discovery, determination of the other Higgs portal parameter $\lambda_4$ is more complicated even after discovery. Fortunately, we can obtain $\lambda_4$ through precise measurements of the decay branching ratios. We find that only four decay vertices are helpful and summarize them in Table\,\ref{tab:1}. At the same time, to further narrow down the parameter space, precise measurements on the $h\to\gamma\gamma$ decay rate can help indirectly to the determination of $\lambda_4$ by excluding some of the parameter space, as shown in our Fig.\,\ref{haa}.

In this work, we only focus on the charged triplet Higgs particles in the NMH framework. However, the neutral triplet Higgs particles can also be used for model discovery and the Higgs portal parameter determination at the 100\,TeV collider. Looking ahead to future studies of the neutral states,  we comment that:
\begin{enumerate}
\item In the NMH framework, the $HA$ channel has the third largest cross section. We present the decay patterns of $H$ and $A$ in Fig.\,\ref{decayregionplotH} and Fig.\,\ref{decayregionplotA} respectively in Appendix\,\ref{HAdecay}. Recall from Table\,\ref{tab:1} that $A\to hZ$ is relevant for $\lambda_4$ determination, we find that the $pp\to HA\to hh\,hZ\to \gamma\gamma b\bar{b}b\bar{b}\ell^+\ell^-$ channel only has $\mathcal{O}(100)$ events at the future collider with $\sqrt{s}=100\rm\,TeV$ and $\mathcal{L}=30\rm\, ab^{-1}$ even without considering the backgrounds. Again, the event number can be improved by studying different final states or different decays chain including vertices in Table\,\ref{tab:1}.
\item For $\lambda_4$ determination, the $H^{\pm}\to hW^\pm$ channel has a larger branching ratio for $\lambda_{45}<0$. In comparison, $H\to ZZ$ has a larger branching ratio $\lambda_{45}>0$, which makes the vacuum stable to a higher scale compared with the benchmark point we use in this work. On the other hand, $H\to W^+W^-$/$A\to hZ$ channel dominates for both positive and negative $\lambda_{45}$ as can be seen from the right panel of Fig.\,\ref{decayregionplotH} and Fig.\,\ref{decayregionplotA}. Therefore, theoretically, the $HA$ channel also provides a way to for model discovery and $\lambda_{4,5}$ determination at the 100\,TeV collider.
\end{enumerate}

\acknowledgments{YD would like to thank Huai-Ke Guo and Hao-Lin Li for many useful discussions on MadGraph and Olivier Mattelaer for quickly answering several technical questions about MadGraph through launchpad. YD, MJRM, and JHY were supported in part under U.S. Department of Energy contract DE-SC0011095. JHY is also supported by the National Science Foundation of China under Grants No. 11875003 and the Chinese Academy of Sciences (CAS) Hundred-Talent Program.
}
\appendix

\section{Summary of current experimental constraints on the CTHM}
\label{app:expcon}
All upper/lower limits below are at 95\% confidence level unless otherwise specified.
\subsubsection{Singly charged Higgs particle $H^\pm$}
\begin{itemize}
\item For $pp$ collision at $\sqrt{s}=7$\,TeV, $\int\mathcal{L}dt=4.5\text{\,fb}^{-1}$, corresponding to the $m_h^\text{max}$ scenario of the Minimal Supersymmetric Standard Model (MSSM)\,\cite{Carena:2002qg}, $90\text{\,GeV} <m_H^\pm < 150\text{\,GeV}$ is excluded by assuming $\text{BR}(H^+\rightarrow \tau\nu)=100\%$\,\cite{Aad:2012tj}, where $\text{BR}$ stands for BR and same notation below.
\item For $pp$ collision at $\sqrt{s}=7$\,TeV $\int\mathcal{L}dt=4.5\text{\,fb}^{-1}$ and $\text{BR}(H^+\rightarrow \tau\nu)=100\%$, they find $\text{BR}(t\rightarrow b H^+)<\text{1\%-5\%}$ for $m_H^+\in[90,150]$\,GeV\,\cite{Aad:2012tj}. Later in the same year after the discovery of the Higgs particle, they improve their result to be $\text{BR}(t\rightarrow b H^+)<\text{0.8\%-3.4\%}$ for $m_H^+\in[90,160]$\,GeV\,\cite{Aad:2012rjx}. And assuming $\text{BR}(H^+\rightarrow c\bar{s})=100\%$ instead, they find $\text{BR}(t\rightarrow b H^+)<\text{1\%-5\%}$ for $m_H^+\in[90,150]$\,GeV\,\cite{Aad:2013hla}. While for $\sqrt{s}=8$\,TeV $\int\mathcal{L}dt=19.5\text{\,fb}^{-1}$, they find $\text{BR}(t\rightarrow H^+b)\times\text{BR}(H^\pm\rightarrow \tau^\pm\nu)<\text{0.23\%-1.3\%}$ for $m_H^+\in[80,160]$\,GeV. They also conclude that $\sigma(pp\rightarrow tH^\pm + X)\times\text{BR}(H^+\rightarrow \tau\nu)<\text{0.76\text{\,pb}-4.5}$\,pb for $m_H^+\in[180,1000]$\,GeV, which excludes the mass region $m_H^\pm\in[200,250]$\,GeV with large $\tan\beta$ in the context of MSSM\,\cite{Aad:2014kga}.
\item For $pp$ collision at $\sqrt{s}=8$\,TeV, $\int\mathcal{L}dt=20.3\text{\,fb}^{-1}$ and a VBF produced $H^\pm$, $\sigma(pp\rightarrow H^\pm+X)\times\text{BR}(H^\pm\rightarrow W^\pm Z)<\text{31\text{\,fb}-1020}$\,fb for $m_H^\pm\in(200,1000)$\,GeV\,\cite{Aad:2015nfa}.
\item For $pp$ collision at $\sqrt{s}=13$\,TeV and $\int\mathcal{L}dt=3.2\text{\,fb}^{-1}$, $\sigma(pp\rightarrow H^\pm t[b])\times\text{BR}(H^\pm\rightarrow \tau\nu)<\text{1.9\text{\,fb}-15}$\,fb for $m_H^\pm\in(200,2000)$\,GeV\,\cite{Aaboud:2016dig}.
\item For $pp$ collision at $\sqrt{s}=13$\,TeV and $\int\mathcal{L}dt=3.2\text{\,fb}^{-1}$ and a VBF produced $H^\pm$, $\sigma(pp\rightarrow H^\pm+X)\times\text{BR}(H^\pm\rightarrow W^\pm Z)<\text{36\text{\,fb}-573}$\,fb for $m_H^\pm\in(200,2000)$\,GeV\,\cite{Sirunyan:2017sbn}.
\end{itemize}
\subsubsection{doubly charged Higgs particle $H^{\pm\pm}$:}
\begin{itemize}
\item For $pp$ collision at $\sqrt{s}=7$\,TeV and $\int\mathcal{L}dt=1.6\text{\,fb}^{-1}$, $\sigma(H^{++}H^{--})\times\text{BR}(H^{\pm\pm}\rightarrow \mu^\pm\mu^\pm)<\text{1.7\text{\,fb}-11}$\,fb for $m_H^{\pm\pm}\in[100,400]$\,GeV. Interpreted in left-right symmetric models\,\cite{Pati:1974yy,Mohapatra:1974hk,Senjanovic:1975rk,Rizzo:1981xx}, $m_{H^{\pm\pm}}^L<355$\,GeV and $m_{H^{\pm\pm}}^R<251$\,GeV are excluded by assuming $\text{BR}(H^{\pm\pm}\rightarrow\mu^\pm\mu^\pm)=100\%$. For $\text{BR}(H^{\pm\pm}\rightarrow\mu^\pm\mu^\pm)=33\%$, $m_{H^{\pm\pm}}^L< 244$\,GeV and $m_{H^{\pm\pm}}^R<209$\,GeV are excluded\,\cite{Aad:2012cg}.
\item For $pp$ collision at $\sqrt{s}=1.96$\,TeV and $\int\mathcal{L}dt=6.1\text{\,fb}^{-1}$, $m_{H^{\pm\pm}} < \text{190-245}$\,GeV (depending on the decay modes and the couplings) are excluded\,\cite {Aaltonen:2011rta}.
\item For $pp$ collision at $\sqrt{s}=7$\,TeV and $\int\mathcal{L}dt=4.7\text{\,fb}^{-1}$, the cross section of a same-sign di-lepton pair in the fiducial region with $p_T^{e^\pm}>20$\,GeV, $p_T^{\mu^\pm}>25$\,GeV and $|\eta|<2.5$ is constrained to be between 1.7fb and 64fb\,\cite{ATLAS:2012mn}.
\item For $pp$ collision at $\sqrt{s}=7$\,TeV and $\int\mathcal{L}dt=4.7\text{\,fb}^{-1}$, assuming pair production of $H^{++}H^{--}$, $m_{H^{\pm\pm}}<409$\,GeV, $m_{H^{\pm\pm}}<375$\,GeV, $m_{H^{\pm\pm}}<398$\,GeV are excluded from $e^{\pm}e^{\pm}$, $e^\pm \mu^\pm$ and $\mu^\pm\mu^\pm$ final states respectively by assuming 100\% BR for each final state\,\cite{ATLAS:2012hi}.
\item For $pp$ collision at $\sqrt{s}=8$\,TeV and $\int\mathcal{L}dt=20.3\text{\,fb}^{-1}$, by assuming $\text{BR}(H^{\pm\pm}\rightarrow e\tau/\mu\tau)=100\%$, $m_{H^{\pm\pm}}<400$\,GeV is excluded\,\cite{Aad:2014hja}.
\item For $pp$ collision at $\sqrt{s}=8$\,TeV and $\int\mathcal{L}dt=20.3\text{\,fb}^{-1}$, by assuming $\text{BR}(H^{\pm\pm}\rightarrow e^\pm e^\pm/e^\pm \mu^\pm/\mu^\pm\mu^\pm)=100\%$, $m_{H^{\pm\pm}}^L<465\text{\,GeV}-550$\,GeV and $m_{H^{\pm\pm}}^R<\text{370\text{\,GeV}-435}$\,GeV are excluded\,\cite{ATLAS:2014kca}.
\item For $pp$ collision at $\sqrt{s}=8$\,TeV and $\int\mathcal{L}dt=20.3\text{\,fb}^{-1}$, for long-lived $H^{\pm\pm}$ pair produced through a Drell-Yan process (with only photon exchange included), $m_{H^{\pm\pm}}\in[50,660]$\,GeV is excluded\,\cite{Aad:2015oga}.
\item For $pp$ collision at $\sqrt{s}=13$\,TeV and $\int\mathcal{L}dt=35.9\text{\,fb}^{-1}$, for a VBF produced $H^{\pm\pm}$ particle, $s_H>0.18$ and $s_H>0.44$ are excluded for $m_{H^{\pm\pm}}=200$\,GeV and $m_{H^{\pm\pm}}=1000$\,GeV respectively in the GMM\,\cite{Sirunyan:2017ret}.
\item For $pp$ collision at $\sqrt{s}=13$\,TeV and $\int\mathcal{L}dt=36.1\text{\,fb}^{-1}$, by assuming $\text{BR}(H^{\pm\pm}\rightarrow e^\pm e^\pm/e^\pm \mu^\pm/\mu^\pm\mu^\pm)=100\%$, $m_{H^{\pm\pm}}^L<770\text{\,GeV}-870$\,GeV are excluded\,\cite{Aaboud:2017qph}.
\end{itemize}
\subsubsection{Electric charge neutral particles:}
\begin{itemize}
\item For $pp$ collision at $\sqrt{s}=8$\,TeV and $\int\mathcal{L}dt=19.5-20.3\text{\,fb}^{-1}$, $m_A=140$\,GeV and $\tan\beta>5.4$ in the $m_h^{\text{max}}$ scenario of the MSSM is excluded\,\cite{Aad:2014vgg}.
\item For $pp$ collision at $\sqrt{s}=8$\,TeV and $\int\mathcal{L}dt=20.3\text{\,fb}^{-1}$, $\sigma(gg\rightarrow A)\times\text{BR}(A\rightarrow Zh)\times\text{BR}(h\rightarrow \tau\bar{\tau}(b\bar{b}))<\text{0.098\text{\,pb}-0.013}\text{\,pb}~(\text{0.57fb-0.014\,pb})$ for $m_A\in[220,1000]$\,GeV\,\cite{Aad:2015wra}. Constraints on the 2HDM parameter space are also discussed therein.
\item For $pp$ collision at $\sqrt{s}=8$\,TeV and $\int\mathcal{L}dt=19.7\text{\,fb}^{-1}$, $\sigma(pp\rightarrow A)\times\text{BR}(A\rightarrow hZ\rightarrow b\bar{b}\ell^+\ell^-)\in[3,30]$\,fb (with $\ell=e,\mu$) is excluded for $m_A\in[250,600]$\,GeV\,\cite{Khachatryan:2015lba}. The result is used to reduce the parameter space of the 2HDM, see Figure 5 therein.
\item For $pp$ collision at $\sqrt{s}=8$\,TeV and $\int\mathcal{L}dt=20.3\text{\,fb}^{-1}$, $\sigma(pp\rightarrow H)\times\text{BR}(H \rightarrow ZZ)<\text{0.008\,pb-0.53\,pb}(\text{0.009\,pb-0.31\,pb})$ for a gluon-fusion (VBF) produced $H$ for $m_H\in[195,950]$\,GeV\,\cite{Aad:2015kna}, which is also used to constrain the 2HDM parameter space.
\item For $pp$ collision at $\sqrt{s}\text{=}8$\,TeV and $\int\mathcal{L}dt\text{=}20.3\text{\,fb}^{-1}$, the strongest limits are in the narrow-width: $\sigma_H\times\text{BR}(H\rightarrow W^+W^-)\text{<}830(240)$\,fb for a gluon-fusion (VBF) produced $H$ at $m_H=300$\,GeV. For $m_H=1500$\,GeV, $\sigma_H\times\text{BR}(H\rightarrow W^+W^-)\text{<}22(6.6)$\,fb\,\cite{Aad:2015agg}.
\item By studying $h\rightarrow (\gamma\gamma,ZZ^*\rightarrow4\ell, WW^*\rightarrow\ell\nu\ell\nu, Z\gamma, b\bar{b}, \tau^+\tau^-,\mu^+\mu^-)$ based on $pp$ collision data at $\sqrt{s}=7$\,TeV and $\int\mathcal{L}dt\text{=}4.7\text{\,fb}^{-1}$ and $\sqrt{s}=8$\,TeV and $\int\mathcal{L}dt=20.3\text{\,fb}^{-1}$, the authors in Ref.\,\cite{Aad:2015pla} set constraints on the parameter space of Minimal Composite Higgs Models (MCHM), additional electroweak singlet models and 2HDM. Especially, $m_A>370$\,GeV is constrained in hMSSM.
\item For $pp$ collision at $\sqrt{s}=8$\,TeV and $\int\mathcal{L}dt=19.7\text{\,fb}^{-1}$, in Ref.\,\cite{Khachatryan:2015tha} $\sigma(ggH)\times\text{BR}(H\rightarrow hh\rightarrow b\bar{b}\tau^+\tau^-)\text{<0.2fb-0.8fb}$ for $m_H\in[260,350]$\,GeV and $\sigma(ggA)\times\text{BR}(A\rightarrow hZ\rightarrow \tau^+\tau^-\ell^+\ell^-)\text{<20fb-40fb}$ for $m_A\in[220,350]$\,GeV. The results are also interpreted in the context of MSSM and type-II 2HDM.
\item For $pp$ collision at $\sqrt{s}=8$\,TeV and $\int\mathcal{L}dt=19.8\text{\,fb}^{-1}$, the lower limit on $\sigma(pp\rightarrow S' \rightarrow SZ)\times\text{BR}(S\rightarrow b\bar{b}(\tau^+\tau^-))\times\text{BR}(Z\rightarrow\ell^+\ell^-)$ ($S',~S$ are neutral Higgs bosons and $m_{S'}>m_S$.) is constrained to be 5fb for $\ell^+\ell^-\tau^+\tau^-$ final state, $m_{H/A}\in(500,1000)$\,GeV, $m_{A/H}\in(90,400)$\,GeV and 1-100fb for $\ell^+\ell^-b\bar{b}$ final state, $m_A\in[300,100000]$\,GeV respectively. While for the degenerate case, i.e., $m_A=m_H$, the parameter space is unexplored. The result is also explained in the context of 2HDM\,\cite{Khachatryan:2016are}.
\item For $pp$ collision at $\sqrt{s}\text{=}8$\,TeV and $\int\mathcal{L}dt\text{=}20.3\text{\,fb}^{-1}$, in the context of a type-II 2HDM, $m_A\lesssim500$\,GeV, $m_H\lesssim640$\,GeV, $m_A\text{=}m_H\lesssim620$\,GeV is excluded by considering only a pseudoscalar A, only a scalar H and the mass-degenerate scenario $m_A\text{=}m_H$ respectively\,\cite{Aaboud:2017hnm}.
\item For $pp$ collision at $\sqrt{s}=13$\,TeV and $\int\mathcal{L}dt=36.1\text{\,fb}^{-1}$, $\sigma(pp\rightarrow X\rightarrow W(Z)h)\times\text{BR}(W(Z)\rightarrow q\bar{q}^{(')})\times\text{BR}(h\rightarrow b\bar{b})\text{<83fb-1.6fb(77fb-1.1fb)}$ for $m_X\in[1.1,3.8]$\,TeV for a simplified model with a heavy vector triplet\,\cite{Aaboud:2017ahz}.
\item For $pp$ collision at $\sqrt{s}\text{=}13$\,TeV and $\int\mathcal{L}dt\text{=}36.1\text{\,fb}^{-1}$, the upper limit of $\sigma(pp\rightarrow X)\times\text{BR}(X\rightarrow ZV)\text{<1.7fb-1.4fb(0.42fb-1fb)}$ ($V\text{=}W,Z$, and $X$ a heavy resonance) for $m_X\in[300,3000]$\,GeV with a $X$ produced through a gluon-gluon-Fusion (VBF) process\,\cite{Aaboud:2017itg}.
\item For $pp$ collision at $\sqrt{s}=13$\,TeV and $\int\mathcal{L}dt=36.1\text{\,fb}^{-1}$, heavy neutral Higgs and gauge bosons in the ditau final state is studied and result is interpreted in hMSSM scenario, which excludes $\tan\beta\text{>}1.0(42)$ for $m_A\text{=}250(1500)$\,GeV\,\cite{Aaboud:2017sjh}.
\item For $pp$ collision at $\sqrt{s}=13$\,TeV and $\int\mathcal{L}dt=36.1\text{\,fb}^{-1}$, a heavy resonances ($Y$) decaying into a SM $h$ and another new particle $X$ ($X$ then decays into a light quark pair) is studied for $m_Y\in[1,4]$\,TeV and $m_X\in[50,1000]$\,GeV. $\sigma(pp\rightarrow Y \rightarrow Xh)\text{<}10^{-2}\text{\,pb-}10^{-3}$\,pb in the mass ranges under consideration\,\cite{Aaboud:2017ecz}.
\item For $pp$ collision at $\sqrt{s}=13$\,TeV, $\int\mathcal{L}dt=36.1\text{\,fb}^{-1}$, upper limit for $\sigma(pp\rightarrow A\rightarrow hZ)\times\text{BR}(h\rightarrow b\bar{b})$ is set to be from $5.5\times10^{-3}$\,pb to $2.4\times10^{-1}$\,pb for gluon-fusion production and $3.4\times10^{-3}$\,pb to $7.3\times10^{-1}$\,pb for associated production with $b$-quarks with $m_A\in[220,2000]$\,GeV\,\cite{Aaboud:2017cxo}.
\item For $pp$ collision at $\sqrt{s}=13$\,TeV, $\int\mathcal{L}dt=36.1\text{\,fb}^{-1}$, upper limits for $\sigma\times\text{BR}(H\rightarrow bb)$ are $14{-}830$\,fb for gluon-gluon fusion and $26{-}570$\, for b-associated production with $m_H\in[130,700]$\,GeV and $m_A\in[230,800]$\,GeV\,\cite{Aaboud:2018eoy}.
\item For $pp$ collision at $\sqrt{s}=13$\,TeV, $\int\mathcal{L}dt=35.9\text{\,fb}^{-1}$, $pp\to X\to ZZ \to 4\ell/2\ell2q/2\ell2\nu$, where $X$ is a heavy resonance, is studied in detail in Ref.\,\cite{Sirunyan:2018qlb} for $m_X\in[130,3000]$\,GeV. Limits on production cross section and the BR is set from their work. 
\item For $pp$ collision at $\sqrt{s}=13$\,TeV, $\int\mathcal{L}dt=35.9\text{\,fb}^{-1}$, an upper limit is set on the $t\bar{t}h$ production cross section relative to the SM expectation of $\mu=\sigma/\sigma_{\rm SM}$, the best fit value for which is $\mu=0.72\pm0.24(\rm stat)\pm0.38(\rm syst)$\,\cite{Sirunyan:2018mvw}.
\end{itemize}

\section{Decay rates of $h\rightarrow \gamma\gamma$}\label{haaapp}
Here we briefly review the computation of the ratio $R_{h\gamma\gamma}$. 
The current combined value from ATLAS and CMS at $\sqrt{s}=7$\,TeV and $\sqrt{s}=8$\,TeV is $R_{h\gamma\gamma}=1.16_{-0.18}^{+0.20}$\,\cite{Khachatryan:2016vau,Biswas:2017tnw}, and in the CTHM, the singly- and doubly-charged Higgs particles will contribute to $\Gamma^{\text{BSM}}$ through loop effects. There has been many literatures studying this contribution\,\cite{Biswas:2017tnw,Chun:2012jw,Arhrib:2011vc,Kanemura:2012rs,Akeroyd:2012ms}, and it was also shown that in the CTHM (\cite{Arhrib:2011vc,Kanemura:2012rs}), the doubly-charged Higgs particle will give a sizable contribution to the decay rate of $h\rightarrow \gamma\gamma$ especially for negative $\lambda_4$ and $\lambda_{45}$ due to a constructive interference\,\cite{Chun:2012jw}. Since we choose negative $\lambda_4$ and $\lambda_{45}$ in our cases, contributions to the rate from the CTHM can in turn be used to constrain the parameter space of the CTHM. To study this effect, we rewrite the result in Ref.\,\cite{Biswas:2017tnw,Djouadi:2005gj} by using our notation as follows:
\begin{align}
\Gamma(h\rightarrow \gamma \gamma) &=
\frac{G_{F}\alpha^{2}m_{h}^{3}}{128\sqrt{2}\pi^{3}}\left\vert\sum_{f}N_{c}Q_{f}^{2}g_{hff}A_{1/2}^{h}(\tau_{f})+g_{hW^+W^-}A_{1}^{h}(\tau_{W})\right.
\nonumber \\
 &\left. \quad + \frac{4m_W}{gm_h^2}g_{hH^{\pm}H^{\mp}}A_{0}^{h}(\tau_{H^{\pm}})-\frac{16m_Wv_\Phi}{gm_h^2}g_{hH^{\pm\pm}H^{\mp\mp}}A_{0}^{h}(\tau_{H^{\pm\pm}})\right\vert^{2}\,\,,
\label{Decay_width}
\end{align}
with $\alpha$ the fine structure constant, $g$ the U(1) coupling, $N_{c}^f$ the color factor ($N_{c}^f$=3 for quarks and 1 for leptons), $Q_{f}$ the fermion electric charge, $G_F$ the Fermi constant and $\tau_{i}=\frac{4m_{i}^{2}}{m_{h}^{2}} (i= f,W,H^{\pm},H^{\pm\pm})$. $g_{hW^\pm W^\mp}$, $g_{hH^\pm H^\mp}$ and $g_{hH^{\pm\pm}H^{\mp\mp}}$\footnotemark\footnotetext[16]{Note that these couplings are function of $\lambda_{4,5}$ as can be seen from Appendix\,\ref{frapp}.} are the couplings given in Appendix \ref{frapp}. And the loop functions $A_i$ are defined as:
\begin{eqnarray}
A_{1/2}(\tau_{x})&=& -2\tau_{x}\left\{1+(1-\tau_{x}){\cal  F}(\tau_{x})\right\}\,\,,\\
A_{1}(\tau_{x})&=& 2+3\tau_{x}+3\tau_{x}(2-\tau_{x}){\cal  F}(\tau_{x})\,\,,\\
A_{0}(\tau_{x})&=& 1-\tau_{x}{\cal  F}(\tau_{x})\,\,,\\ 
{\cal  F}(\tau_{x})&=& \left\{
\begin{array}{lr}
\left[\sin^{-1}(\sqrt{\frac{1}{\tau_{x}}})\right]^{2}&\,\,\,\,\,\,\text{for}\,\,\,\, \tau_{x}\geq 1,\\
-\frac{1}{4}\left[\ln (\frac{1+\sqrt{1-\tau_{x}}}{1-\sqrt{1-\tau_{x}}})-i\pi\right]^{2}&\,\,\,\,\,\, \text{for}\,\,\,\, \tau_{x}<1.
\end{array}
\right.
\end{eqnarray}

\section{$H$ and $A$ decays}\label{HAdecay}
In this section, we give the dominant decay channels of the neutral Higgs bosons. Note that $H\to ZZ/WW$ and $A\to hZ$ are relevant for $\lambda_{4,5}$ determination, we see that $A\to hZ$ and $H\to WW$ can be used for both positive and negative $\lambda_{45}$, while $H\to ZZ$ only works for positive $\lambda_{45}$. This scenario is different for the fourth channel related to the determination of $\lambda_{45}$, i.e., $H^\pm \to hW^\pm$, which works only for negative $\lambda_{45}$ .
\begin{figure}[thb!]
\captionstyle{flushleft}
\begin{tabular}{cc}
  \includegraphics[width=75mm,height=65mm]{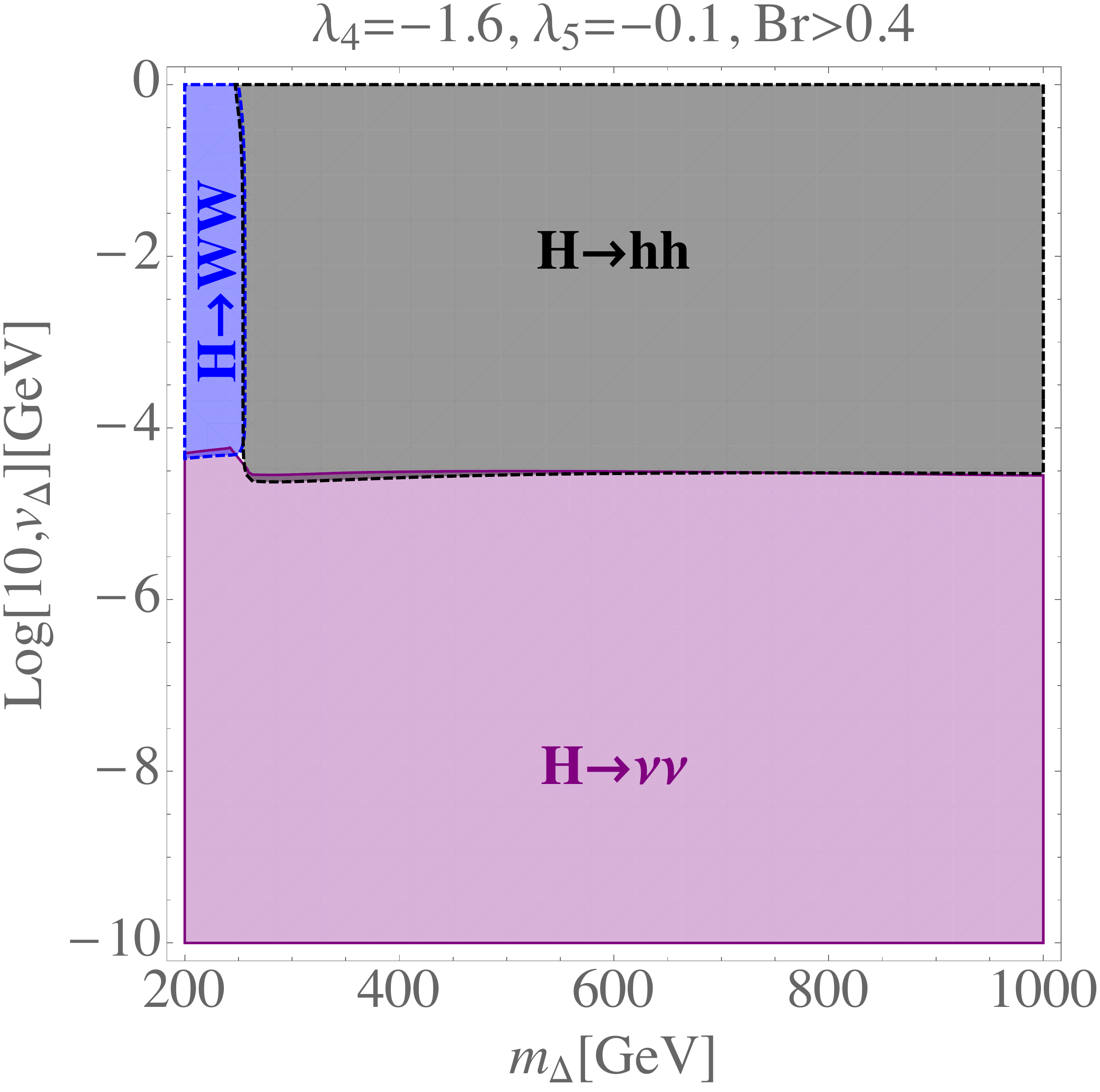} & \includegraphics[width=75mm,height=65mm]{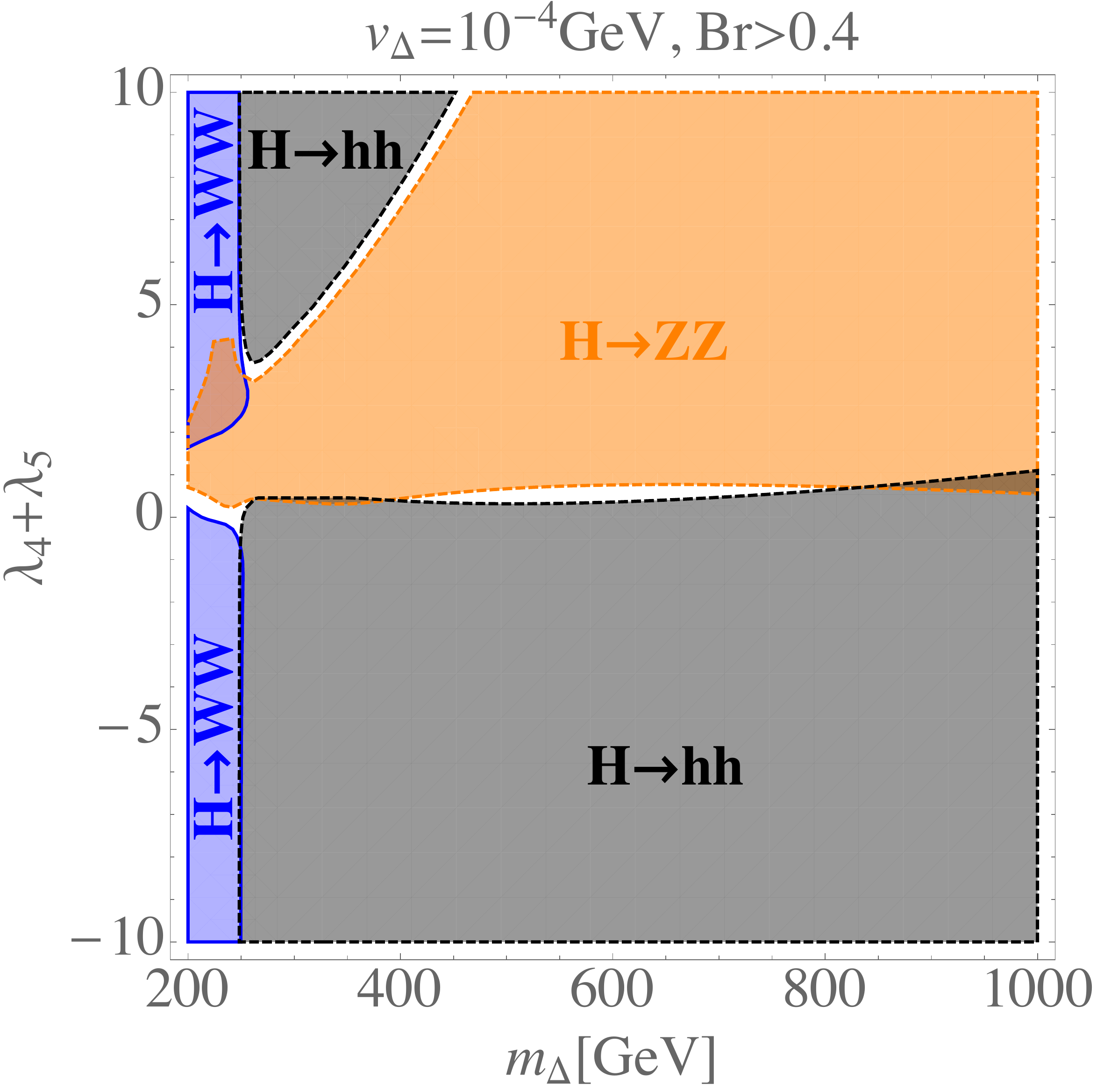}  \\
\end{tabular}
\caption{Decay region plots for $H$ with BR $\ge40\%$. Black region for the di-Higgs channel, blue region for the di-W boson channel and purple region for the di-neutrino channel. From the left panel, di-neutrino/di-$h$ channel dominates at small/large $v_\Delta$ respectively, and $W$-pair channel dominates at the large $v_\Delta$ and small $m_\Delta$ region. While from the right panel, we observe that di-$h$ (di-$Z$ boson) channel dominates for negative (positive) $\lambda_{45}$.}\label{decayregionplotH}
\end{figure}
\begin{figure}[thb!]
\captionstyle{flushleft}
\begin{tabular}{cc}
  \includegraphics[width=75mm,height=65mm]{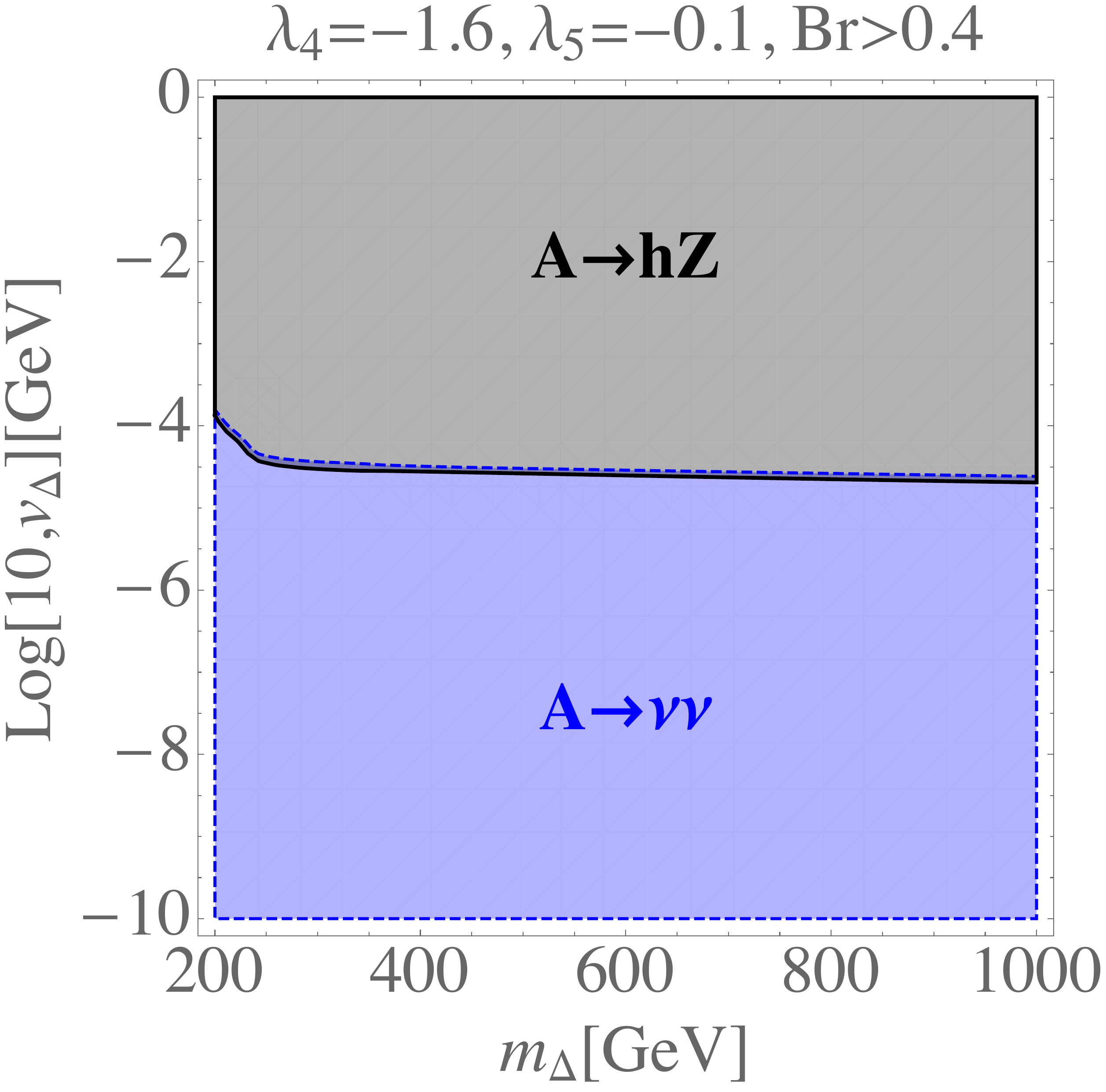} &   \includegraphics[width=75mm,height=65mm]{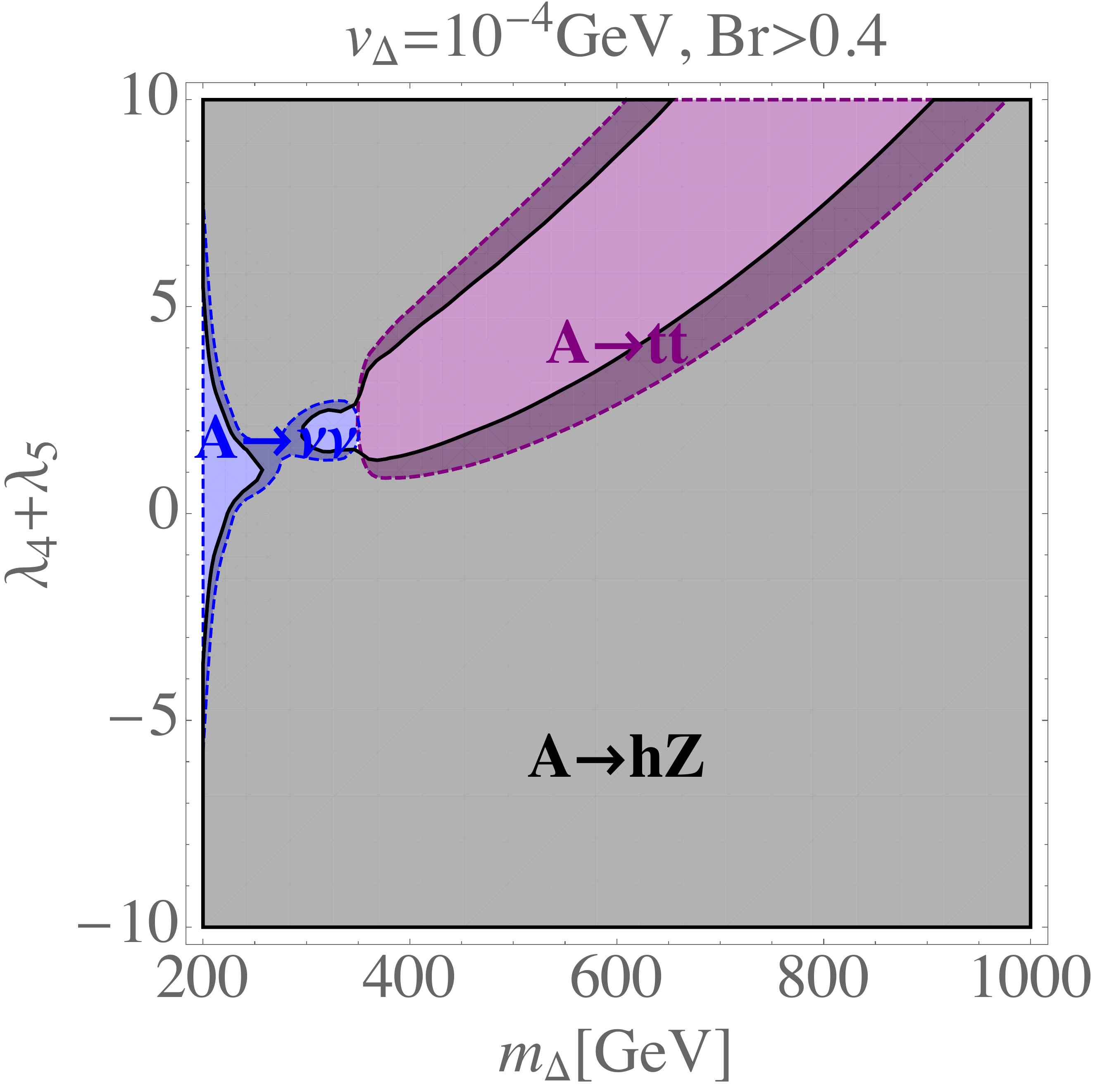} \\
\end{tabular}
\caption{Decay region plots for $A$ with $\rm BR\ge40\%$. Black region for the $hZ$ channel, purple region for the $t$ quark pair channel and blue region for the di-neutrino region.}\label{decayregionplotA}
\end{figure}

\section{Feynman rules for the CTHM}
\label{frapp}
\begin{center}
\begin{table}[h]
\begin{tabular}[t]{|c||c|c|}
\hline Interaction & Feynman Rule\footnote{Assuming all particles are incoming into the vertex, and to save ink, we use the following notations: $c_w\equiv\cos\theta_W$, $s_w\equiv\sin\theta_W$, $c_\alpha\equiv\cos\alpha$, $s_\alpha\equiv\sin\alpha$, $c_{\beta_{0,\pm}}\equiv\cos\beta_{0,\pm}$, $s_{\beta_{0,\pm}}\equiv\sin\beta_{0,\pm}$, $c_{2w}\equiv\cos(2\theta_W)$, $s_{2w}\equiv\sin(2\theta_W)$, $c_{2\alpha}\equiv\cos(2\alpha)$, $s_{2\alpha}\equiv\sin(2\alpha)$, $c_{2\beta_\pm}\equiv\cos(2\beta_\pm)$, $s_{2\beta_\pm}\equiv\sin(2\beta_\pm)$, $t_{\beta_0}\equiv\tan\beta_0$.}\\
\hline
$hW^\mp H^\pm$&$+\frac{ig}{2}(c_\alpha s_{\beta_\pm}-\sqrt{2}s_\alpha c_{\beta_\pm})(p_h-p_{H^\pm})^\mu$\\
$HW^\mp H^\pm$&$-\frac{ig}{2}(s_\alpha s_{\beta_\pm}+\sqrt{2}c_\alpha c_{\beta_\pm})(p_H-p_{H^\pm})^\mu$\\
$AW^\mp H^\pm$&$+\frac{g}{2}(s_{\beta_0} s_{\beta_\pm}+\sqrt{2} c_{\beta_0} c_{\beta_\pm})(p_A-p_{H^\pm})^\mu$\\
$hAZ$ & $-\frac{g}{2c_w}(c_\alpha s_{\beta_0}-2s_\alpha c_{\beta_0})(p_h-p_A)^\mu$ \\
$HAZ$ & $+\frac{g}{2c_w}(s_\alpha s_{\beta_0}+2c_\alpha c_{\beta_0})(p_H-p_A)^\mu$\\
$H^+ H^- Z$ & $\frac{ig}{2c_w}(c_{2w}s^2_{\beta^{\pm}}-2s^2_wc_{\beta^\pm}^2)(p_{H^-}-p_{H^+})^\mu$\\
$H^+ H^- \gamma$ & $ig s_w(p_{H^-}-p_{H^+})^\mu$\\
$H^{++} H^{--} Z$ & $\frac{ig}{c_w}c_{2w}(p_{H^{--}}-p_{H^{++}})^\mu$\\
$H^{++} H^{--} \gamma$ & $ig s_w(p_{H^{--}}-p_{H^{++}})^\mu$\\
$hZZ$ & $\frac{2iem_Z}{s_{2w}}(c_{\beta_0}c_\alpha+2s_{\beta_0}s_\alpha)g_{\mu\nu}$\\
$HZZ$ & $\frac{2iem_Z}{s_{2w}}(-c_{\beta_0}s_\alpha+2s_{\beta_0}c_\alpha)g_{\mu\nu}$\\
$hW^+W^-$ & $ig m_Z c_w(c_{\beta_0}c_\alpha+s_{\beta_0}s_\alpha)g_{\mu\nu}$\\
$HW^+W^-$ & $ig m_Z c_w(-c_{\beta_0}s_\alpha+s_{\beta_0} c_\alpha)g_{\mu\nu}$\\
$H^{\pm\pm}W^\mp W^{\mp}$ & $-\frac{i\sqrt{2}e^2v_\Delta}{s_w^2}g_{\mu\nu}$\\
$ZW^\pm H^\mp$ & $-\frac{ig^2v_\Phi s_{\beta^\pm}}{2c_w}g_{\mu\nu}$\\
$H^{\pm\pm}H^\mp W^\mp$ & $igc_{\beta^{\pm}}(p_{H^{\pm\pm}}-p_{H^{\mp}})^\mu$ \\
\hline\hline
$hH^{++}H^{--}$ & $-iv_\Phi(\lambda_4c_\alpha+\lambda_2t_{\beta_0}s_\alpha)$\\
$HH^{++}H^{--}$ & $-iv_\Phi(-\lambda_4s_\alpha+\lambda_2t_{\beta_0}c_\alpha)$\\
$hH^{+}H^{-}$ & $-\frac{i}{4}\left\{(8\lambda_1v_\Phi s_{\beta^\pm}^2+4\lambda_4 v_\Phi c^2_{\beta^\pm}+2\lambda_5v_\Phi c_{2\beta^\pm}+4\mu s_{2\beta^\pm})c_\alpha +(8\lambda_{23}v_\Delta c_{\beta^\pm}^2+4\lambda_4 v_\Delta s^2_{\beta^\pm}-4 v_\Delta \lambda_5 c^2_{\beta^\pm})s_\alpha \right\}$\\
$HH^{+}H^{-}$ & $-\frac{i}{4}\left\{-(8\lambda_1v_\Phi s_{\beta^\pm}^2+4\lambda_4 v_\Phi c^2_{\beta^\pm}+2\lambda_5v_\Phi c_{2\beta^\pm}+4\mu s_{2\beta^\pm})s_\alpha +(8\lambda_{23}v_\Delta c_{\beta^\pm}^2+4\lambda_4 v_\Delta s^2_{\beta^\pm}-4 v_\Delta \lambda_5 c^2_{\beta^\pm})c_\alpha \right\}$\\
$Hhh$ & $-\frac{i}{2}\left\{-2\sqrt{2}\mu c_\alpha(1-3s_\alpha^2)-2[6\lambda_1  c_\alpha^2+\lambda_{45}(1-3 c_\alpha^2)]v_\Phi s_\alpha+2[6 \lambda_{23} s^2_\alpha + \lambda_{45} (1-3s_\alpha^2)]v_\Delta c_\alpha\right\}$\\
$HAA$ & $-\frac{i}{2}\left\{-2\sqrt{2}\mu s_{\beta_0}(2s_\alpha c_{\beta_0}-c_\alpha s_{\beta_0})-2(2\lambda_1s_{\beta_0}^2 + \lambda_{45} c_{\beta_0}^2)v_\Phi s_\alpha+2(2\lambda_{23}  c^2_{\beta_0}+ \lambda_{45} s_{\beta_0}^2)v_\Delta c_\alpha\right\}$\\
$H^{\pm\pm}H^\mp H^\mp$ & $ -\frac{i}{2}\left(-2\sqrt{2}\lambda_3v_\Delta c_{\beta^\pm}^2+2\lambda_5v_\Phi s_{\beta^\pm}c_{\beta^\pm}+4\mu s_{\beta^\pm}^2\right)$\\
$hhh$ & $-6i\left[-\frac{s_\alpha c_\alpha^2\mu}{\sqrt{2}} + \frac{s_\alpha v_\Delta}{2}\left( \lambda_{45} c_\alpha^2 + 2 \lambda_{23} s_\alpha^2\right)+\frac{c_\alpha v_\Phi}{2}\left(2\lambda_1 c_\alpha^2 + \lambda_{45} s_\alpha^2\right)\right]$\\
$HHH$ & $-6i\left[-\frac{s_\alpha^2 c_\alpha \mu}{\sqrt{2}} + \frac{c_\alpha v_\Delta}{2}\left( \lambda_{45} s_\alpha^2 + 2 \lambda_{23} c_\alpha^2\right) - \frac{s_\alpha v_\Phi}{2}\left( 2\lambda_1 s_\alpha^2 + \lambda_{45} c_\alpha^2\right) \right]$\\
$hHH$ & $-\frac{i}{2}\left\{ 2\left[ 6\lambda_1  s_\alpha^2 + \lambda_{45} (1-3 s_\alpha^2) \right]v_\Phi c_\alpha +  2 \left[ 6\lambda_{23} c^2_\alpha + \lambda_{45} (1-3c_\alpha^2)\right] v_\Delta s_\alpha -2\sqrt{2}\mu s_\alpha(1-3c_\alpha^2) \right\}$\\
$hAA$ & $-\frac{i}{2}\left[2\sqrt{2}\mu s_{\beta_0}(2c_\alpha c_{\beta_0}+s_\alpha s_{\beta_0}) + 2 ( 2\lambda_1s_{\beta_0}^2 + \lambda_{45} c_{\beta_0}^2) v_\Phi c_\alpha + 2 ( 2\lambda_{23} c^2_{\beta_0} + \lambda_{45} s_{\beta_0}^2) v_\Delta s_\alpha\right]$\\
\hline
\end{tabular}
\label{tab:feynman}
\end{table}
\end{center}

\clearpage

\bibliography{ref}

\end{document}